\documentclass[12pt]{report}
\usepackage[latin5]{inputenc} 
\usepackage{amsmath}
\usepackage{graphicx}
\usepackage{dcolumn}
\usepackage{bm}
\usepackage{epsfig}
\usepackage{graphics}
\usepackage{xcolor}
\usepackage{bigints}
\usepackage{amssymb}
\usepackage{float}
\usepackage{theorem,latexsym}

\usepackage[a4paper, top=2.5cm, left=3.5cm, right=2.5cm, bottom=2.5cm]{geometry}
\usepackage{ifthen}
\usepackage{indentfirst} 
\usepackage{fancyhdr}    
\usepackage{times}       

\begin{document}





\begin{titlepage}
\thispagestyle{empty}
\vspace*{1.5cm}
\begin{center}
\Large{\textbf{BOSE-EINSTEIN CONDENSATION AND BLACK \\ HOLES
IN DARK MATTER AND DARK ENERGY}}
\end{center}
\vspace{3cm}
\begin{center}
\large{A Thesis Submitted to \\
the Graduate School of Engineering and Sciences of \\ \.{I}zmir Institute of Technology \\
in Partial Fulfillment of the Requirements for the Degree of \\
DOCTOR OF PHILOSOPHY
\\
in Physics
\\
by Kemal G{\"U}LTEK{\.I}N}
\end{center}
\vspace{3cm}
\begin{center}
\large{Supervisor: Prof. Dr. Recai ERDEM}
\end{center}

\vspace{3cm}
\begin{center}
\large{July 2023 \\ \.{I}zmir}
\end{center}
\end{titlepage}

\newpage
\chapter*{Acknowledgements}
\pagenumbering{roman}
I would like to express my deepest gratitude and endless respect to my supervisor Prof. Dr. Recai ERDEM for his inestimable support, guidance, patience, and encouragement in this thesis study. He was always welcoming me. It was an honour and a privilege to be supervised by him. 

I would like to thank Prof. Dr. Kadri YAKUT, Assoc. Prof. Dr. Suat DENG\.{I}Z, Prof. Dr. Alev Devrim G\"U\c{C}L\"U, and Assoc. Prof. Dr. Fatih ERMAN as the thesis defence jury members. They carefully read my thesis and provided useful comments that improved the quality of the thesis. It was a great pleasure for me to meet and discuss with them.

I also would like to thank Dr. Bet\"ul DEM\.{I}RKAYA for her contributions and supports. I am also very indebted to Hemza AZR\.{I} for his endless friendship and encouragement during my Ph.D period. Special thanks to my nearest friend Bahad{\i}r P\.{I}LTEN who was there when I needed help. I always feel lucky to have a friend like him.

Finally, my most important and heartfelt thanks to my family for their non-stop support, quidance, encouragements and patience. They have shown me unconditional love in all situations. I would like to dedicate this thesis to them.




\newpage
\chapter*{Abstract}
\begin{center}
BOSE-EINSTEIN CONDENSATION AND BLACK HOLES \\ IN DARK MATTER AND DARK ENERGY
\end{center}
The main aim of this study is to reveal curved space and particle physics effects on the formation of Bose-Einstein condensate scalar fields in cosmology and around a black hole. Cosmological scalar fields for dark energy and dark matter may be considered as a result of Bose-Einstein condensation. In this regard, our main attention will be devoted to Bose-Einstein condensates in curved space. By considering the dynamics of a scalar Bose-Einstein condensation at a microscopic level, we first study the initial phase of the formation of condensation in cosmology. To this end, we initially introduce an effective Minkowski space formulation that enables considering only the effect of particle physics processes, excluding the effect of gravitational particle production and enabling us to see cosmological evolution more easily. Then, by using this formulation, we study a model with a trilinear coupling that induces the processes. After considering the phase evolution of the produced particles, we find that they evolve towards the formation of a Bose-Einstein condensate if some specific conditions are satisfied. In principle, the effective Minkowski space formulation introduced in this study can be applied to particle physics processes in any sufficiently smooth spacetime. In this regard, we also analyse if a condensate scalar field is realized in the spacetime around a Reissner - Nordstr{\o}m black hole. We find that the produced particles of particle physics processes are localized in a region around the black hole and have a tendency toward condensation if the emerged particles are much heavier than ingoing particles. We also find that such a configuration is phenomenologically viable only if the scalars and the black hole have dark electric charges. Finally, we consider gravitational collapse around Schwarzschild black holes and form a first step towards a study in future about the effects of gravitational collapse on Bose-Einstein condensation.
 \newpage  
  \chapter*{Publications}
  \begin{enumerate}
\item  R.~Erdem and K.~G\"ultekin,
\textbf{A mechanism for formation of Bose-Einstein condensation in cosmology}, JCAP \textbf{10} (2019), 061 doi:10.1088/1475-7516/2019/10/061, [arXiv:1908.08784 [gr-qc]].
  \item R.~Erdem and K.~Gultekin,
\textbf{Particle physics processes in cosmology through an effective Minkowski space formulation and the limitations of the method}, Eur. Phys. J. C \textbf{81} (2021) no.8, 726 doi:10.1140/epjc/s10052-021-09524-8,
[arXiv:2102.05587 [gr-qc]].
\item R.~Erdem, B.~Demirkaya, and K.~G\"ultekin,
\textbf{Curved space and particle physics effects on the formation of Bose\textendash{}Einstein condensation around a Reissner\textendash{}Nordstr\o{}m black hole},
Eur. Phys. J. Plus \textbf{136} (2021) no.9, 972
[erratum: Eur. Phys. J. Plus \textbf{137} (2022) no.7, 822], doi:10.1140/epjp/s13360-021-01973-0, [arXiv:2110.00799 [gr-qc]].
\item R.~Erdem, B.~Demirkaya and K.~G\"ultekin,
\textbf{A metric for gravitational collapse around a Schwarzschild black hole}, Mod. Phys. Lett. A \textbf{38} (2023) no.07, 2350048, doi:10.1142/S0217732323500487, [arXiv:2007.04672 [gr-qc]].

\end{enumerate}

\newpage
 \tableofcontents

    \newpage       
	\listoffigures
 \newpage

\chapter{Introduction}
\pagenumbering{arabic}
The general theory of relativity, introduced by Albert Einstein in 1915, is the most comprehensive and standard framework for gravity. One of the main applications of general relativity (GR) is cosmology, i.e., the study of the universe at the largest (i.e. IR) scales. Observations suggest that the average distribution of matter at cosmological scales, i.e., at scales larger than 100 Mpc ($\approx 3.3 \times 10^{21} \ \text{km}$) is homogeneous and isotropic. The corresponding general relativistic description of the universe is formulated by Robertson-Walker-Friedmann-Lemaitre universe \cite{Liddle:1010476,Weinb. Cos}.

Observations imply that the amount of the ordinary matter called baryonic matter (i.e., stars, interstellar, intergalactic dust) is not enough to account for the whole matter in the universe. Therefore, there must be some non-luminous matter, called dark matter (DM), in addition to ordinary matter. Further, the observed accelerated expansion of the universe, when formulated in the Friedmann-Lemaitre-Robertson-Walker  model requires additional energy density called dark energy (DE). In this context, $\sim$ 95 $\%$ of the universe consists of non-luminous energy densities known as DE ($\sim$ 72 $\%$) and DM ($\sim$ 23 $\%$), and the remaining $\sim$ 5 $\%$ consists of visible matter known as baryonic matter ($\sim$ 4 $\%$) and radiation ($\sim 8\times 10^{-5} \ \%$). The simplest framework for DM is non-luminous cosmological dust, which is cold dark matter (CDM), while the simplest explanation for DE is the cosmological constant ($\Lambda$) of Einstein's general theory of relativity in addition to the baryonic matter and the radiation (photons and neutrinos) \cite{Sahni:2004ai}. This model is called the standard model of cosmology, namely the $\Lambda \text{CDM}$ model  \cite{Liddle:1010476,Weinb. Cos,Sahni:2004ai}.

$\Lambda \text{CDM}$ model agrees with observational data very well. However, this model has some crucial shortcomings. One of the main shortcomings of $\Lambda \text{CDM}$ is the cosmological constant (CC) problem \cite{Sahni:2004ai, Weinberg:1988cp, Nobbenhuis:2004wn}. The potential theoretical contributions to CC are huge when compared to DE's energy density. For example, the contribution of zero point energies to CC is about $10^{121}$ times the observed energy density of DE. This is called a CC problem. $\text{CDM}$ part of $\Lambda \text{CDM}$ also has some problems at a scale smaller than the galactic scale, such as the problem of predicting too dense cores for the galaxies (i.e., core-cusp or cuspy halo problem) and too many dwarf galaxies when compared with observations \cite{Sahni:2004ai, Weinberg:2013aya, Bull:2015stt}. There are also two potential problems for the $\Lambda \text{CDM}$ model, which are $H_0$ tension between its values determined locally and the ones derived at cosmological scales \cite{Riess_2021,2020A&A...641A...6P,Schoneberg:2021qvd}, and $\sigma_8 - S8$ problems \cite{Abdalla:2022yfr,Heymans:2020gsg,Amon:2022azi}.

To cope with these problems, alternative theories to $\Lambda \text{CDM}$ are considered. In literature, the most popular alternatives to $\Lambda \text{CDM}$ are scalar field models for DE and DM, e.g. quintessence models \cite{Amendola:2015ksp}, k-essence, phantom scalar field, and coupled scalar field models, etc. \cite{Amendola:2015ksp, Bose:2008ew, Scherrer:2004au}. In these types of models, the scalar fields are assumed to depend only on time because of the homogeneity and the isotropy of the universe at the background level. Such scalar fields are called cosmological scalar fields and may be considered to be the result of Bose-Einstein condensation \cite{Bose,Einstein}.

The dynamics of Bose-Einstein condensates are  described by the Gross-Pitaevskii (GP) equation \cite{Gross,Pitaevski}. GP equation may be considered the non-relativistic limit of $\phi^4$ theories \cite{Fukuyama:2007sx, Castellanos:2013ena, Erdem:2016hqw}. Many researches exist on Bose-Einstein condensate (BEC) scalar field models for DE and DM \cite{Fukuyama:2007sx, UrenaLopez:2008zh, Silverman:2001gx, Silverman:2002qx,Harko:2015nua, Nishiyama:2004ju, Besprosvany:2015ura, Das:2014agf, Takeshi:2009cy, Das:2015dca,Dymnikova:2000gnk,Dymnikova:2001jy}. They are usually studied at a macroscopic level (i.e., at the level of number densities or distribution functions) in cosmology, while its microscopic nature at the level of particle physics is not considered sufficiently. In this context, we have formulated a new mechanism for the formation of scalar field Bose-Einstein condensation in cosmology with particular emphasis on its microscopic description in particle physics in \cite{Erdem:2021mrw,Erdem:2019gpk}. 

Besides the cosmological framework, there are also some studies \cite{BALDESCHI1983221,Sin:1992bg,Goodman:2000tg,Arbey:2003sj,Boehmer:2007um,Harko:2011xw,Dwornik:2013fra,Robles:2012uy,Guzman:2013rua,DasGupta:2015qbz,Harko:2014vya,Harko:2019nyw,Chavanis:2016dab}, where the applicability of BEC at galactic scale, in particular around black holes, is considered. Galactic halos formed by bosons, either in the context of Bose-Einstein condensation or in an appropriate isothermal distribution, was first suggested by Baldeschi \textit{et al.} \cite{BALDESCHI1983221} in 1983. In 1994, this idea was improved by Sin \cite{Sin:1992bg}, who explained the problem of galaxy rotation curves by self-gravitating BECs. In 2000, Goodman suggested that a BEC interacting with gravity and with itself via an appropriate potential may play the role of dark matter \cite{Goodman:2000tg}. Arbey \textit{et al.} \cite{Arbey:2003sj} in 2003 showed that the BEC of a self-interacting charged scalar field may account for the dark matter and explain the problem of the rotation curve of the dwarf spiral galaxies. In 2007, B\"ohmer and Harko \cite{Boehmer:2007um} reinforced the idea that dark matter can be a BEC and they compared the predictions of their model with the observational data. There are also some recent studies pursuing the idea of BEC dark matter. The problem of predicting DM distributions at cores of the galaxies may be well-explained in the context of BEC in \cite{Harko:2011xw}. The observed rotational curves of galaxies is realised in studies \cite{Dwornik:2013fra,Robles:2012uy} if the BEC plays the role of DM halos. The dynamics of rotating BEC dark matter halos consisting of ultralight spinless bosons around a galaxy and the effect of the rotating halo on the galactic rotation curves are examined in \cite{Guzman:2013rua}. The formation of supermassive black holes via the collapsing BEC dark matter distributions is studied in \cite{DasGupta:2015qbz}. In \cite{Harko:2014vya,Harko:2019nyw}, the mechanisms of the gravitational collapse of the BEC dark matter halos with repulsive self-interactions of the particles are examined. The  collapse  of a self-gravitating BEC with  attractive  self-interaction is studied in \cite{Chavanis:2016dab}. In a wholly general relativistic framework, we have studied in \cite{Erdem:2021trk} if condensation could be formed around a Reissner-Nordstr{\o}m (RN) black hole \cite{Chandrasekhar:1985kt}, where we consider the curved space and particle physics effects on the formation of the condensate.


Besides, to set up a framework toward a future investigation into how gravitational collapse can affect Bose-Einstein condensation, we derive a new metric in \cite{Erdem:2020jgi} that describes the gravitational collapse of a cosmological fluid, in particular, dust around a primordial Schwarzschild black hole. The mechanism for forming stars, galaxies, and clusters of galaxies out of small inhomogeneities is based on gravitational collapse. In this regard, there are some significant studies \cite{McVittie,Tolman,Oppenheimer,Bondi,generalized-Bondi,Kaloper,Vaidya,generalized-Vaidya,general-Oppenheimer-Snyder,Thakurta,McClur-Dyer,Thakurta-not-BH,Thakurta-McClur-not-BH,Sultana-Dyer,Faraoni-Jacques} examining gravitational collapse. One of them is the McVittie spacetime \cite{McVittie} that studies, in 1933, Schwarzschild black holes embedded in a cosmological background. It studied the effects of cosmological expansion on the dynamics of black holes and considers the case of non-accretion, i.e.,  $G^0_{\;1}=0$ which corresponds to the zero radial power flux of the fluid. Tolman \cite{Tolman} in 1934 studied the effect of inhomogenities on cosmological models that consist of dust particles distributed around a central object without an accretion, i.e., the pressure of the dust is negligible. Another work by Oppenheimer and Synder in 1939 \cite{Oppenheimer} examined a star that collapses by the effect of its gravitational field. In 1947 Bondi \cite{Bondi} proposed a metric to understand some characteristics of the spherically symmetric systems with zero pressures. This metric can also be reduced to other type of metrics (such as Schwarzschild, Robertson-Walker, Oppenheimer and Synder) except the McVittie metric after various coordinate transformations specifying some particular value \cite{generalized-Bondi}. However, all these models we mentioned do not consider the accretion of the black holes \cite{generalized-Bondi,Kaloper}. On the other hand, in 1951, Vaidya \cite{Vaidya,generalized-Vaidya} proposed a completely different metric that allows the absorption or radiation of massless dust by a black hole so that the black hole is accreting in case of null dust. All the other metrics that describe the gravitational collapse are proposed as extensions of the above metrics. For instance, a generalized metric of Oppenheimer-Snyder \cite{general-Oppenheimer-Snyder} explains the way of the gravitational collapse of stars into regular black holes. Thakurta \cite{Thakurta} and McClur-Dyer \cite{McClur-Dyer} extend the McVittie metric. Still it has been shown \cite{Thakurta-not-BH,Thakurta-McClur-not-BH} that this metric does not correspond to the cosmological black holes. In contrast, Sultana-Dyer \cite{Sultana-Dyer} metric, as an extension to McVittie metric, studies a black hole in the presence of cosmological constant, i.e., a black hole in de Sitter space \cite{generalized-Bondi,Kaloper}. As another example, Faraoni-Jacques \cite{Faraoni-Jacques} extend the McVittie metric to explain an accreting black hole in the presence of an imperfect fluid with heat flow \cite{generalized-Bondi}. In other words, only Faraoni-Jacques black holes among McVitie type metrics can define accreting black holes. The Faraoni-Jacques solution in that form, however, cannot represent accreting black holes in common fluids, such as cosmic dust.



The present thesis is organised as follows: In Chapter 2, we briefly review the $\Lambda$CDM model and its potential problems. In Chapter 3, we briefly examine cosmological scalar field models, particularly the quintessence scalar field model. We also mention other types of scalar fields. In Chapter 4, we study the BEC scalar field at the levels of condensed matter physics and relativistic scalar field models. We conclude the section after examining some cosmological BEC scalar field models. In Chapter 5, we formulate a mechanism for the formation of the BEC scalar field in cosmology. To this end, we first introduce an effective Minkowski space formulation and then we proceed to realize the condensation. In Chapter 6, we study the curved space and particle physics effects on the formation of Bose-Einstein condensation around an RN black hole. In Chapter 7, we derive a new metric for the gravitational collapse around a black hole. This metric may lead to a condensate around an accreting black hole, which is the more realistic point to be addressed in future studies. We finally give concluding remarks in Chapter 8.
\newpage
\chapter{The Standard Model of Cosmology}
\section{A Brief Overview}
At the beginning of modern cosmology, without any data, the scientists (just for mathematical simplicity) came up with the idea  that we are not at the center of the universe, the universe looks the same at every point and in all directions, and we do not have a distinguished position in the universe. This hypothesis is called the `cosmological principle' (CP). CP implies that the universe is homogeneous (i.e., the universe appears the same at every point) and isotropic (i.e., the universe appears the same in all directions) on large scales $>$ 100Mpc. The cosmic microwave background (CMB) data \cite{1965ApJ...142..419P,Bennett_2013,Fixsen_2009} in the last decades seem to confirm CP.

\ \ \ The homogeneity and the isotropy of the universe on large scales imply that the universe should be described by Friedman-Robertson-Walker (FRW) metric \cite{Liddle:1010476,Weinb. Cos}:
\begin{equation}
ds^2=-dt^2+a(t)^2\bigg[\frac{dr^2}{1-kr^2}+r^2\big(d\theta^2+\text{sin}^2\theta d\varphi^2\big)\bigg].\label{met.}
\end{equation}
Here $ds$ measures the proper distance between two distinct locations in spacetime when they are separated by $dx^\mu$. $a(t)$ is the cosmic scale factor and measures how physical distances are expanding by time. $k$ is the constant measures the curvature of the spatial part of the spacetime and it may be rescaled to take the values +1 , 0, -1 for the spherical, flat, and hyperbolic geometries, respectively. 

The metric tensor of the spacetime $g_{\mu\nu}$ (in $ds^2=g_{\mu\nu}dx^\mu dx^\nu$), namely the gravitational field, is considered as the fundamental dynamical variable in the general theory of relativity, and it satisfies Einstein's field equation:
\begin{equation}
R_{\mu\nu}-\frac{1}{2}Rg_{\mu\nu}-g_{\mu\nu}\Lambda=8\pi G_N T_{\mu\nu},\label{G}
\end{equation}
where  both $R$ (Ricci curvature scalar), $R_{\mu\nu}$ (Ricci curvature tensor), and $\Lambda$ (cosmological constant) are the geometrical objects that describe the curvature of the spacetime and the right-hand side is related to the energy-momentum tensor  $T_{\mu\nu}$ of any energy distribution. (\ref{G}) tells us how the curvature of the spacetime and the matter affect each other. $g_{\mu\nu}\Lambda $ term may be absorbed in the redefinition of $T_{\mu\nu}$.


For the homogeneity and isotropy of the universe, the appropriate energy-momentum tensor is that of a perfect fluid, namely, $T^\mu_\nu=\text{diag}(-\rho, p, p, p)$ \cite{Liddle:1010476,Weinb. Cos}, where $\rho$ is the mass density and $p$ is the pressure of the fluid (A perfect fluid is a fluid with no heat flow nor viscosity).
By considering the metric in Eq.(\ref{met.}) together with the perfect fluid form for the energy-momentum tensor, the Einstein field equations (\ref{G}) result in the following Friedmann equations:
\begin{equation}
H^2\equiv \bigg(\frac{\dot{a}}{a}\bigg)^2=\frac{8\pi G}{3}\rho-\frac{k}{a^2}\label{Fr1}
\end{equation}
\begin{equation}
\frac{\ddot{a}}{a}=-\frac{4\pi G}{3}(\rho+3p),\label{Fr2}
\end{equation}
where $\dot{a}=\frac{da}{dt}$, $\ddot{a}=\frac{d^2a}{dt^2}$, and the rate of the expansion of the universe is described by the Hubble parameter $H= \frac{\dot{a}}{a}$. The first Friedmann equation (\ref{Fr1}) describes how the universe's energy drives its expansion. The second Friedmann equation (\ref{Fr2}) gives the rate of the acceleration of the expansion. These two Friedmann equations can be combined to obtain the continuity equation 
\begin{equation}
\dot{\rho}+3H(\rho+p)=0,\label{Cont.}
\end{equation}
which can be also derived by the conservation law of the energy-momentum tensor, i.e., $\nabla_\nu T^{\mu\nu}=0$. $\rho$ and $p$ may be related by the equation of state parameter $w=\frac{p}{\rho}$ which is constant for each energy density in $\Lambda\text{CDM}$. For a constant $w$, Eq.(\ref{Cont.}) can be solved to obtain the relation 
\begin{equation}
\rho \propto a^{-3(1+w)},
\end{equation}
which after plugging into Eq.(\ref{Fr1}) with $k\simeq 0$ by the observations \cite{Barnett:1996hr}, gives the explicit time dependence of the scale factor: 
\begin{align}
a(t)\propto t^{2/(3(1+w))} \ \ \ \text{for} \ \ \ w\neq 1
\\
a(t)\propto e^{Ht} \ \ \ \text{for} \ \ \ w=-1.
\end{align}

The universe consists of radiation, non-relativistic matter, and DE \cite{Liddle:1010476,Weinb. Cos}. 
For each fluid, the equation of state parameter can be written separately as $w_i=\frac{p_i}{\rho_i}$. Extremely relativistic matter is included in radiation, which at this time consists of only photons and neutrinos and we have $w_{\text{rad}}=1/3$ for radiation. Hence $\rho_{\text{rad}} \propto a^{-4} $ and the scale factor evolves as $a_{\text{rad}} \propto t^{1/2} $ for the radiation dominated universe. For non-relativistic matter consisting of baryons and DM, the pressure may be negligible compared to their mass density so that we have $w_{\text{mat}}=0$. Thus $\rho_{\text{mat}} \propto a^{-3} $ and the scale factor evolves as $a_{\text{mat}} \propto t^{2/3} $ for the matter dominated universe. For the cosmological constant (i.e., vacuum energy), we have $w_{\Lambda}=-1$. Thus, $\rho_{\Lambda}$ is constant and the scale factor grows as $a_{\Lambda} \propto e^{Ht}$, which implies an exponentially accelerated expansion for the universe. Radiation controlled the dynamics of the early universe until around 47,000 years after the Big Bang. The energy density of matter overtook those of radiation (and vacuum energy) about 47,000 years after the Big Bang. The cosmological constant dominated epoch is regarded as the final of the known universe's three eras and it started after the matter dominated epoch, when the universe was approximately 9.8 billion years old. 


If we introduce a parameter $\Omega_i=\rho_i / \rho_c$ where $\rho_c=3H^2 / 8\pi G$ is the critical density, Eq.(\ref{Fr1}) can be rewritten as 
\begin{equation}
\Omega=\Omega_{rad}+\Omega_{mat}+\Omega_{\Lambda}=1+\frac{k}{(aH)^2}.
\end{equation}
The observed present values of the quantities for each type fluid are \cite{Liddle:1010476,Weinb. Cos}
\begin{equation}
 \Omega_{\text{rad}} \sim 8\times 10^{-5}, \ \ \ \ \Omega^{\text{Baryonic}}_{\text{mat}} \sim 0.04, \ \ \ \ \Omega^{\text{Dark}}_{\text{mat}} \sim 0.23, \ \ \ \ \Omega_{\Lambda} \sim 0.72\label{obs}
\end{equation}
with $k \simeq 0$ (i.e., approximately flat). As clearly can be seen from Eq.(\ref{obs}), the present universe is dominated by the cosmological constant $\Lambda$ adopted as dark or vacuum energy.
\section{Problems of the Standard Model}
Having introduced the standard model of cosmology ($\Lambda$CDM) briefly, we are now able to mention its problems in short. The problem of cosmological constant arises from corrections to the vacuum energy density of the energy-momentum tensor in Eq.(\ref{G}), e.g. as arising from zero point energies of quantum theory. The contribution of zero point energies to vacuum energy density is \cite{Mukhanov:2007zz,Itzykson:1980rh,Peskin:1995ev,Maggiore:2005qv}
\begin{equation}
\rho_\text{vac}= \ <0\mid \delta\hat{T}_{00}\mid 0> \ =\frac{1}{2}\int_0^\infty \frac{d^3\textbf{k}}{(2\pi)^3}\sqrt{k^2+m^2},
\end{equation}
which consists of all the vacuum energy (i.e., lowest energy state) contributions of quantum fields of mass $m$. Because there is an upper limit for the energy scale of quantum field theory, the infinity in the upper limit of the integral can be replaced by some cut-off scale $k_\text{max}$ \cite{Mukhanov:2007zz,Itzykson:1980rh,Peskin:1995ev,Maggiore:2005qv}. This gives
\begin{equation}
\rho_\text{vac}\approx\frac{k_\text{max}^4}{16\pi^2}.
\end{equation}
Since the General Relativity still hold true slightly below the Planck scale, one may use for the cut-off $k_\text{max}=m_\text{pl}=1,22\times10^{19}\text{GeV}$ \cite{Weinberg:1988cp, Nobbenhuis:2004wn,COPELAND_2006}, which results in the vacuum energy density 
\begin{equation}
\rho_\text{vac}\approx 10^{74}\text{GeV}^4.
\end{equation}
This contributes to the bare energy density of $\Lambda$ and the effective vacuum energy density is obtained as
\begin{equation}
\rho^{\text{theo}}_{\text{eff}}\sim \rho_\Lambda+10^{74}\text{GeV}^4.
\end{equation}
However, the present cosmological observations based on the standard model of cosmology shows that \cite{Weinberg:1988cp, Nobbenhuis:2004wn,COPELAND_2006}
\begin{equation}
\rho^{\text{obs}}_{\text{eff}}\lesssim 10^{-47}\text{GeV}^4.
\end{equation}
Therefore, $\rho^{\text{theo}}_{\text{eff}}$ predicted from the theory is about $10^{121}$ times larger than $\rho^{\text{obs}}_{\text{eff}}$ based on observations. This huge difference gives rise to the so-called cosmological constant problem and this extreme discrepancy is needed to be explained and resolved.

For DM, the problems arise from the discrepancies between the observations and the simulations based on the $\Lambda$CDM, at small scales. One of the well-known problems is the rotational curves of galaxies. The observations show that in a wide variety of distances around a galaxy, the rotating stars orbit in circles at approximately constant tangential speeds as opposed to the theory where these tangential speeds are expected to be decreased. Therefore, there must be additional matter distributions to keep the rotational speeds of the stars constant. Another disagreement is called a core-cusp problem. The simulations based on $\Lambda$CDM form DM distributions in galaxies with inner radial density $\rho \propto r^{-1}$, which shows the cuspy density profile at small radii. However, observations show the constant-density DM profiles at cores as $\rho \propto r^0$. Another problem is the missing satellites problem (or dwarf galaxy problem). Compared to the measured number of satellite galaxies around the Milky Way, the simulations forecast excessive dwarf galaxies orbiting Milky Way-sized galaxies. There is also an issue with the local voids at small scales, which consist of only a few galaxies in their vast spaces. However, simulations predict more dense regions in these voids. A detailed review of these and additional problems at small scales are given in \cite{Weinberg:2013aya, Bull:2015stt,Famaey:2011kh}. Understanding these problems is needed for the applicability of standard cosmology at small scales.


The Hubble tension is another intriguing issue in cosmology at the moment \cite{Riess_2021,2020A&A...641A...6P,Schoneberg:2021qvd}. Local measurements of the universe's current expansion rate that don't substantially rely on cosmological model assumptions, such as those obtained from supernovae, provide numbers clustering around $73 \ \text{km}\ \text{s}^{-1} \ \text{Mpc}^{-1}$. $\Lambda$CDM, which examines the early universe through the CMB, offers a second, indirect method for measuring what we assume to be the same amount. Using this approach, the Hubble constant is estimated to be approximately $67 \ \text{km}\ \text{s}^{-1} \ \text{Mpc}^{-1}$. Thus the difference between the readings taken locally and at cosmological scales is around 4$\sigma$, the standard deviation between the two measurements.

In addition to the Hubble tension, there is another main anomaly called $\sigma_8$ tension \cite{Abdalla:2022yfr,Heymans:2020gsg,Amon:2022azi}. On a scale of $8 \ \text{h}^{-1} \ \text{Mpc}$ ($ \text{h} = H_0 / (100 \ \text{km} \ \text{s}^{-1} \ \text{Mpc}^{-1})$), $\sigma_8$ represents the amount of current matter fluctuation. There is a discrepancy in this amount between the study of recent data on the CMB observation and observations of galaxy clusters. According to CMB estimates $\sigma_8 \sim  0.82$ in contrast to lower estimates $\sigma_8 \sim 0.75$ based on the detection of galaxy clusters.

\newpage
\chapter{Cosmological Scalar Field Models as Alternatives to the Standard Model of Cosmology}
\vspace{-0.5 cm}

As discussed in the previous chapter, the evolution of the universe is described by the Einstein field equations (\ref{G}). These equations are obtained by considering the Einstein-Hilbert action \cite{A1, H1}
\begin{equation}
S_{EH}=\int d^4 x \sqrt{|g|}\bigg[\frac{c^4}{16\pi G_N}\big(R(g)-2\Lambda \big)+ \mathcal{L}_M \bigg],\label{EH}
\end{equation}
where $g \equiv \text{Det}[g_{\mu\nu}]$ is the determinant of the metric tensor, $R(g)=g^{\mu\nu}R_{\mu\nu}(g)$ is
the curvature scalar of metric, $\Lambda$ is the cosmological constant, $\mathcal{L}_M$ is the Lagrangian density describing matter fields, $G_N$ is the gravitational Newtonian constant and $c$ is the speed of light in vacuum. The energy-momentum tensor is given by $T_{\mu\nu}\equiv \frac{-2}{\sqrt{|g|}}\frac{\delta S_M}{\delta g^{\mu\nu}}$, where $S_M=\int d^4 x \sqrt{|g|} \mathcal{L}_M$ is the matter part of the action (\ref{EH}). As we pointed out, $\Lambda \text{CDM}$ has some problems related to DE and DM. Therefore, some alternative models to $\Lambda \text{CDM}$ are proposed in the literature. These models may be classified broadly as `modified gravity models' and  the `modified energy-momentum models'. In this chapter, some alternative scalar field models to $\Lambda \text{CDM}$ are discussed in the context of `modified energy-momentum models'. In these models, DE and DM are provided by scalar fields with a negative pressure for DE and zero pressure for DM. This may be done by replacing $\Lambda$ and/or CDM by the contribution of a scalar field. In this chapter, we shall be interested in the  modified $T_{\mu\nu}$ models, particularly cosmological scalar models. Although we particularly consider some pure scalar field models in this chapter, there are also some recent intriguing models such as \cite{Dengiz:2011ig,Dengiz:2016eoo,Tanhayi:2012nn} concerning Weyl's local invariant gravity theories where the local conformal gauge symmetry is achieved with the help of additional extra scalar and gauge fields, and the masses of the excitations are generated through a specific spontaneous symmetry breaking mechanism as in the standard model. It may be interesting in future studying these extra fields in the context of dark matter and/or dark energy.

In the following section, we shall start with the quintessence model, the most preferred scalar field model. Then, in the following sections, we shall continue with  brief discussions on the other preferred scalar field models, such as k-essence, phantom, tachyon and coupled scalar field models \cite{Amendola:2015ksp}.

\section{Quintessence}
Quintessence model is associated with a canonical scalar field, say $\phi$, which has the following form of the Lagrangian density \cite{Caldwell:1997ii,Carroll:1998zi}
\begin{equation}
\mathcal{L}_{\phi}=X-V(\phi),\label{X1}
\end{equation}
where $X=-\frac{1}{2}g^{\mu\nu}\partial_{\mu}\phi\partial_{\nu}\phi$ and $V(\phi)$ are the kinetic energy and the potential of the field, respectively. The corresponding action is obtained by replacing the cosmological constant $\Lambda$ in Eq.(\ref{EH}) with the lagrangian density of the quintessence field, $\mathcal{L}_{\phi}$:
\begin{equation}
S=\int d^4 x \sqrt{|g|}\bigg[\frac{c^4}{16\pi G_N}R(g)+\mathcal{L}_{\phi}+ \mathcal{L}_M \bigg],\label{S}
\end{equation} 
where the main dynamical fields are $g_{\mu\nu}$ and $\phi$. The condition of stationarity of the action under variations of $g_{\mu\nu}$ results in
\begin{equation}
R_{\mu\nu}-\frac{1}{2}Rg_{\mu\nu}=8\pi G_N \big(T^{M}_{\mu\nu}+T^{\phi}_{\mu\nu}\big).\label{fe1}
\end{equation}
Here $T_{\mu\nu}\equiv \frac{-2}{\sqrt{|g|}}\frac{\delta S_M}{\delta g^{\mu\nu}}$, where $S_M=\int d^4 x \sqrt{|g|} \mathcal{L}_M$  and $T^{\phi}_{\mu\nu}\equiv \frac{-2}{\sqrt{|g|}}\frac{\delta S_\phi}{\delta g^{\mu\nu}}= \partial_{\mu}\phi\partial_{\nu}\phi-g_{\mu\nu}\big[\frac{1}{2}g^{\alpha\beta}\partial_{\alpha}\phi\partial_{\beta}\phi+V(\phi)\big]$, where $S_\phi=\int d^4 x \sqrt{|g|} \mathcal{L}_\phi$. The equation of state parameters for $T^{M}_{\mu\nu}$ (for a general perfect fluid) and for $T^{\phi}_{\mu\nu}$ (for quintessence field) are $p_M=w_M \rho_M$ and $p_\phi=w_\phi\rho_\phi$, respectively. Here $w_M$ is introduced for the cases of the radiation and non-relativistic matter and $w_\phi$ is introduced in place of $w_\Lambda$. Bu using the FRW metric with the above form for $T^{\phi}_{\mu\nu}$, the energy density and the pressure of the quintessence field are obtained as
\begin{eqnarray}
&\rho_\phi=\frac{\dot{\phi}^2}{2}+V(\phi),
\\ &
p_\phi=\frac{\dot{\phi}^2}{2}-V(\phi).
\end{eqnarray}
Then, the corresponding equation of state of the field is given by
\begin{equation}
w_\phi=\frac{p_\phi}{\rho_\phi}=\frac{\dot{\phi}^2-2V(\phi)}{\dot{\phi}^2+2V(\phi)},\label{eqs1}
\end{equation}
where we have assumed that $\phi=\phi(t)$, i.e. $\nabla \phi\rightarrow 0$,  by considering the average homogeneity and the isotropy of the universe. Eq.(\ref{eqs1}) implies that the equation of state parameter of $\phi$ may change dynamically with time for the different epochs of the universe in contrast to the constant equation of state parameter of $\Lambda$. The field equations (\ref{fe1}) corresponding to the FLRW universe are given by
\begin{eqnarray}
&H^2=\frac{8\pi G}{3}\bigg[\frac{\dot{\phi}^2}{2}+V(\phi)+\rho_M \bigg],
\\
&\dot{H}=-\frac{8\pi G}{2}\bigg[\dot{\phi}^2+\rho_M+p_M\bigg].
\end{eqnarray}
By use of these equations, we have two continuity equations for both the fluid and the quintessence field, as given below:
\begin{eqnarray}
&\dot{\rho}_M+3H(\rho_M+p_M)=0, 
\\
&\dot{\rho}_\phi+3H(\rho_\phi+p_\phi)=0.
\end{eqnarray}

During the radiation and matter dominated epochs of the universe, it is assumed that 
$\rho_\phi \ll \rho_M$ if we assume that the only function of $\phi$ is to accelerate the expansion, i.e., $\phi$ plays the role of DE. The first condition for this is to make 
$\dot{\phi}^2/2\gg V(\phi)$ in Eq.(\ref{eqs1}). In this case, we obtain $w_\phi\simeq 1$ for the quintessence field in these eras and hence the continuity equation results in the density evolution of the field $\rho_\phi \propto a^{-6}$. This result is enough for $\phi$ to decrease much faster than the background fluid density, namely $\rho_{\text{rad}} \propto a^{-4}$ and $\rho_{\text{mat}} \propto a^{-3}$ for the radiation and matter dominated epochs, respectively. However, for the explanation of the late-time acceleration of the universe caused by DE density, $\rho_\phi$ should track $\rho_M$. This requires $\dot{\phi}^2 < V(\phi)$ and equivalently $w_\phi<-1/3$. Therefore, the evolution of the field $\phi$ by time should be slow enough in comparison with the potential $V(\phi)$, 
which is called the slow roll condition \cite{Bassett:2005xm}. The expected present DE dominated epoch is provided by the condition $\dot{\phi}^2\ll V(\phi)$ for which $w_\phi\simeq -1 $ as in the case of $\Lambda$. 

To make $\phi$ play the role of cosmological constant, it must mimic cosmological constant at late times. This may be provided by considering slowly varying potentials \cite{Caldwell:1997ii}. In this respect, some quintessence models are depending on the different kinds of quintessence potentials such as freezing models, thawing models, etc. Detailed discussions on these models are given in \cite{Amendola:2015ksp}.Using the quintessence potentials, the source of DE is then provided by the dynamical quintessence field $\phi$ in place of the constant $\Lambda$. 

Further, by imposing the condition $\dot{\phi}^2\approx 2V(\phi)$, which gives the zero pressure for $\phi$, i.e., $w_\phi=0$, one may assume that $\phi$ may play the role of DM. An example of this model is given in \cite{Sahni:1999qe}, where a physically reasonable more general potential for $\phi$ is introduced as $V(\phi)=V_0(\text{cosh}\ \lambda\phi -1 )^{\kappa_\phi}$, where $\kappa_\phi=\frac{1+w_\phi}{1-w_\phi}$. The choice of this potential makes the dynamics of $\phi$ more adaptable for cosmology so that one may employ it for DM and DE. Here the field may play the role of only DM for $w_\phi=0$, i.e., $\kappa_\phi=1$ while it may play the role of only DE for $w_\phi \leq -1/3$, i.e., $\kappa_\phi \leq 1/2$. In the same study, the possibility of a unified model of DE and DM is also discussed by considering the potential $V(\phi,\psi)=V_\phi(\text{cosh}\ \lambda_\phi \phi -1 )^{\kappa_\phi}+V_\psi(\text{cosh}\ \lambda_\psi \psi -1 )^{\kappa_\psi}$, where $\phi$ and $\psi$ are employed for DE (with $\kappa_\phi \leq 1/2$) and DM (with $\kappa_\psi=1$  ), respectively.
The quintessence field obeys the relation $-1<w<-1/3$ at the cosmological scale if the field plays the role of DE. Another study \cite{Guzman:2000zba} shows that $-1<w<-1/3$ also holds at a galactic scale if the field plays the role of DM, which is called quintessence-like or exotic DM. More examples of DM in the context of quintessence are given in \cite{Bi:2003qa,Fasiello:2016yvr,Mandal:2021xqp,Boehmer:2009tk}.


\section{k-essence}
Besides the quintessence models, there are also other modified matter models known as k-essence as alternatives to the DE model of $\Lambda$CDM \cite{ArmendarizPicon:2000dh, ArmendarizPicon:2000ah}. In these models, the Lagrangian density contains a non-canonical kinetic term of the field, which is a function of both the kinetic energy of the field  and the field itself. 
The typical action for these models is
\begin{equation}
S=\int d^4 x \sqrt{|g|}\bigg[\frac{c^4}{16\pi G_N}R(g)+P({\phi}, X)+ \mathcal{L}_M \bigg],\label{kessence}
\end{equation}
where $P({\phi}, X)$ is a function of the k-essence field $\phi$ and the kinetic energy $X=-\frac{1}{2}g^{\mu\nu}\partial_{\mu}\phi\partial_{\nu}\phi$. 
The corresponding energy-momentum tensor of the field is given by 
\begin{equation}
T^\phi_{\mu\nu}=P_{,X}\partial_{\mu}\phi\partial_{\nu}\phi + g_{\mu\nu}P,
\end{equation}
where $P_{,X}\equiv \frac{\partial P}{\partial X}$. Then the most general equation of state for such models is obtained as
\begin{equation}
w_\phi=\frac{p_\phi}{\rho_\phi}=\frac{P}{2X P_{,X}-P}.\label{k1}
\end{equation}
In the case of $|2X P_{,X}|\ll |P| $, the requirement $w_\phi \simeq  -1$ for DE is realized. There are some models belonging to the k-essence family such as the ghost condensate, tachyon, and Dirac-Born-Infeld (DBI) models \cite{Amendola:2015ksp}; the Lagrangian densities for these models are given  by $P=-X+X^2/M^4$ (with constant $M$ having  dimension of mass) for ghost condensate, $P=-V(\phi)\sqrt{-\text{det}(g_{\mu\nu}+\partial_\mu \phi \partial_\nu \phi)}$ for tachyon, and $P=-f(\phi)^{-1}\sqrt{1-2f(\phi)X}+f(\phi)^{-1}-V(\phi)$ for Dirac-Born-Infeld (DBI). With some reasonable choices of $f(\phi)$ and $V(\phi)$ \cite{Amendola:2015ksp}, the fields of the models behave as the sources of the DE.

There are also some studies on employing the non-canonical scalar field for both DE and DM \cite{Mishra:2018tki,Bose:2008ew,Brax:2020tuk,Scherrer:2004au}. As an example, one may consider a Lagrangian density $P({\phi}, X)=X\left(\frac{X}{M^4}\right)^{\alpha-1}-V(\phi)$ \cite{Mishra:2018tki},
which is the generalised form of the Lagrangian density (\ref{X1}) of the canonical scalar field. Then the energy density and the pressure are obtained as $\rho_{\phi}=(2\alpha-1)X\left(\frac{X}{M^4}\right)^{\alpha-1}+V(\phi)$ and $p_{\phi}=X\left(\frac{X}{M^4}\right)^{\alpha-1}-V(\phi)$, respectively. Here with some reasonable choices for the potential $V(\phi)$ \cite{Mishra:2018tki}, the equation of state of the kinetic term, $\rho_{X}$, of the density $\rho_{\phi}$, is given by $w_{X}=\frac{1}{2\alpha-1}$, where $w_{X}\simeq 0$ for $\alpha\gg 1$. We therefore conclude that for the proper potentials and high values of $\alpha$, the kinetic term $\rho_{X}$ may behave as DM while the potential term $V({\phi})$ may play the role of DE.



\section{Phantom}
The recent reported values \cite{Komatsu_2011,Alam:2003fg} for the equation of state of DE lead to the possibility that $w_{\text{DE}}<-1$, generally called phantom DE. Scalar field models of this type may be considered a specific k-essence case . Considering the model with a positive energy density $\rho_{\phi}$, the only condition to obtain $w_{\text{DE}}<-1$ is to get $P_{,X}<0$ in Eq.(\ref{k1}). This may be provided with the simplest choice of the Lagrangian density with a negative kinetic energy \cite{Caldwell:1999ew,Caldwell:2003vq}:
\begin{equation}
P(X,\phi)=-X-V(\phi).
\end{equation}
The corresponding energy density and the pressure of a phantom scalar field are given by $\rho_{\phi}=-\dot{\phi}^2/2+V(\phi)$ and $p_{\phi}=-\dot{\phi}^2/2-V(\phi)$. The equation of state of the field is then obtained as
\begin{equation}
w_{\phi}=\frac{-\dot{\phi}^2/2-V(\phi)}{-\dot{\phi}^2/2+V(\phi)},
\end{equation}
which results in $w_{\phi}<-1$ for $\dot{\phi}^2/2<V(\phi)$. Some detailed discussion on the cosmological dynamics of phantom field with different choices of the potential $V(\phi)$ are given in the papers \cite{PhysRevD.68.023509,Singh:2003vx,Sami:2003xv}.



\section{Tachyon}
The string theory tachyon model is associated with the action \cite{Garousi:2000tr,Sen:2002nu}
\begin{equation}
S_{\phi}=-\int d^4x \ V(\phi)\sqrt{-\text{det}(g_{\mu\nu}+\partial_\mu \phi \partial_\nu \phi)},
\end{equation}
where $V(\phi)$ is the potential of the field. The corresponding energy density and pressure by use of the FRW metric is then obtained as
\begin{equation}
\rho_\phi=\frac{V(\phi)}{\sqrt{1-\dot{\phi}^2}},
\end{equation}
\begin{equation}
p_\phi=-V(\phi)\sqrt{1-\dot{\phi}^2},
\end{equation}
which result in the equation of state parameter
\begin{equation}
w_\phi=-1+\dot{\phi}^2.
\end{equation}
If we here assume that the tachyon field plays the role of DE, we then expect that $w_\phi \approx -1$, which requires the condition $\dot{\phi}^2\ll 1$. Tachyon field is designated in k-essence family because it has an action of a type Eq.(\ref{kessence}). However, this differs from k-essence in the sense that the kinetic energy of the tachyon must be here suppressed in order to achieve accelerated expansion of the universe. Some examples on tachyon models are given in \cite{Padmanabhan:2002cp,Abramo:2003cp,Aguirregabiria:2004xd,Copeland:2004hq}.

\section{Coupled Scalar Field Models}
As mentioned in Eq.(\ref{obs}), DM and DE have the same order of densities. Therefore, there may be a relationship between them. In this regard, some models such as coupled quintessence field, coupled k-essence field, etc. \cite{Wetterich:1994bg,Amendola:1999er,Gumjudpai:2005ry}, where the scalar fields adopted as DE couple to the non-relativistic matter, are studied in this context.

As an example of these types of models, we consider the coupled quintessence field \cite{Amendola:2015ksp,Amendola:1999er}. Here, the Lagrangian density given in Eq.(\ref{S}) is modified with the interaction Lagrangian density $\mathcal{L}_{\text{int}}$:
\begin{equation}
S=\int d^4 x \sqrt{|g|}\bigg[\frac{c^4}{16\pi G_N}R(g)+\mathcal{L}_{\phi}+ \mathcal{L}_M + \mathcal{L}_{\text{int}} \bigg],
\end{equation}
where $ \mathcal{L}_{\text{int}}$ causes interaction between the quintessence field and the non-relativistic matter through a coupling between the energy-momentum tensors of quintessence and matter \cite{Amendola:2015ksp,Amendola:1999er}:
\begin{equation}
\nabla_\mu T^\mu_{\nu(\phi)}=-QT_M\nabla_\nu\phi, \ \ \ \ \ \nabla_\mu T^\mu_{\nu(M)}=QT_M\nabla_\nu\phi.
\end{equation}
Here $T_M=-\rho_M + 3p_M$ is the trace of the energy-momentum tensor for the matter fluids (radiation and non-relativistic matter) and $Q$ is the coupling constant. Since $w_{\text{rad}}=1/3$, the trace vanishes for the radiation. Consequently, only the non-relativistic matter, i.e., dark and baryonic matters, couple to the quintessence field $\phi$. This coupling with the suitable choice of the potential $V(\phi)$ \cite{Amendola:2015ksp,Amendola:1999er,Fuzfa:2007sv} brings about the desired explanations for DM and the universe's accelerated expansion.


\newpage
\chapter{Bose-Einstein Condensation}

Bose-Einstein condensation (BEC) is one of the most interesting physical phenomena explored by Bose \cite{Bose} and Einstein \cite{Einstein} in the 1920's. It was theoretically shown that under some conditions, large numbers of particles obeying the Bose statistics could collapse into the lowest quantum state to realize a macroscopic quantum phenomenon known as BEC, which was, after 75 years, experimentally verified \cite{Anderson:1995gf,Davis:1995pg,Bradley:1995zz}. The macroscopic dynamics of a BEC is described by the GP equation, which was independently derived by Gross \cite{Gross} and Pitaevskii \cite{Pitaevski} in 1961. In the first section of this chapter, we shall study the conditions to form a BEC and derive the GP equation in detail.




Besides, BEC can also be considered of the cause of spatial independence cosmological scalar fields given in Chapter 3. If a scalar field forms a Bose-Einstein condensate, the field then naturally depends only on time, as in the case of the scalar fields for DE and DM. In this regard, the formation of a condensate scalar field has physical importance at the cosmic level. Further, to explore the implications of a Bose-Einstein scalar field condensation in cosmology, one should also study the relations between the $\phi^4$ theories and the GP equation. To this end, in the Section 4.2 we shall derive the GP equations from $\phi^4$ theories by considering both flat and curved spacetimes. 

In the last section, we shall study some cosmological Bose-Einstein condensate scalar field models.

\section{A Brief Overview of Bose-Einstein Condensation in Condensed Matter Physics}
To understand BEC, the simplest case is to consider an ideal non-relativistic Bose gas, where we shall assume N undistinguishable, non-interacting, non-relativistic quantum particles confined in a box of volume $L^3$ \cite{GP,Schmitt:2014eka}. To this end, we need the quantum statistics of bosons represented by the grand-canonical ensemble, where the system is considered in thermal equilibrium at temperature $T$ with a reservoir with chemical potential $\mu$. Thus, the total occupation number of particles over all states, denoted by s, is given by
\begin{equation}
N=\sum_{s\geq0}\frac{1}{e^{\beta(\epsilon_s-\mu)}-1}=\frac{1}{e^{\beta(\epsilon_0-\mu)}-1}+\sum_{s>0}\frac{1}{e^{\beta(\epsilon_s-\mu)}-1}\equiv N_0+N_{\text{excited}},\label{N1}
\end{equation}
where $\beta=(k_B T)^{-1}$ with Boltzmann's constant $k_B$, and $N_0$ and $N_{\text{excited}}$ are the occupation numbers for the ground state with the energy $\epsilon_0$ and the excited states with the energy $\epsilon_s$, respectively. For the ideal gas confined in the box, the allowed eigenvalues $\epsilon$ of  the Hamiltonian $H^{(1)}=\hat{\textbf{p}}^2/2m$ (where the momentum operator $\hat{\textbf{p}}=-i\hbar\nabla$) for the particles in the gas are given by $\epsilon_n=p^2_n/2m$, where $\textbf{p}_n=2\pi\hbar \textbf{n}/L$ by the de Broglie relation with $\textbf{n}$ being a vector with either $0$ or $\pm$ integer components $n_x$, $n_y$, $n_z$. In this regard, Eq.(\ref{N1}) can be rewritten as 
\begin{align}
N=\frac{1}{e^{\beta(\epsilon_0-\mu)}-1}+ V\bigg(\frac{2\pi}{\hbar}\bigg)^3 \int \frac{1}{e^{\beta(\frac{p^2}{2m}-\mu)}-1}d\textbf{p}&=\frac{1}{e^{\beta(\epsilon_0-\mu)}-1}+\frac{V}{\lambda^3}g_{3/2}(e^{\beta\mu}) \nonumber
\\&
\equiv N_0+N_{\text{excited}},\label{N2}
\end{align}
where 
\begin{equation}
\lambda =\sqrt{\frac{2\pi \hbar^2}{mk_B T}}
\end{equation}
is the thermal wavelength and $g_{3/2}(e^{\beta\mu})=\frac{2}{\sqrt{\pi}}\int dx x^{1/2}\frac{1}{e^{(x-\beta\mu)}-1}$. Here, the chemical potential of the ideal Bose gas, either in the grand canonical ensemble or confined in the box, has the physical constraint $\mu<\epsilon_0$ \cite{GP} and  $N_0$ of the lowest energy state becomes increasingly large as $\mu \rightarrow \epsilon_0$. For the ideal Bose gas confined in the box, the lowest energy is $\epsilon_0=0$ (i.e., $\mu$ must be always negative) \cite{GP}. In the derivation of Eq.(\ref{N2}), we have used the integral $V\bigg(\frac{2\pi}{\hbar}\bigg)^3 \bigintss d\textbf{p}$ in place of summation $\sum_\textbf{p}$ with the transormation $p^2=2mk_B Tx$. By using the total number density $n=N/V$, Eq.(\ref{N2}) can be written as 
\begin{equation}
\lambda^3 n_0=\lambda^3 n-g_{3/2}(e^{\beta\mu}).
\end{equation}
As clearly seen from the last equation, its left-hand side should always be positive, i.e., $\lambda^3 n > g_{3/2}(e^{\beta\mu})$. The transition between a gas phase and a condensed phase starts when $\mu \rightarrow 0$ \cite{GP} (since $\epsilon_0 \rightarrow 0$ in this case), which implies 
\begin{equation}
\lambda > \lambda_c\equiv n^{-1/3} \ (2.612)^{1/3} \ \ \ \ \text{or} \ \ \ \ T<T_c\equiv \frac{2\pi \hbar^2}{mk_B \lambda^2_c}\label{BEC1},
\end{equation}
where $g_{3/2}(1)=2.612$. The interpretation of BEC may be better clarified by considering the fraction of the particle numbers as follows. Using (\ref{BEC1}) for Eq.(\ref{N2}), we get
\begin{equation}
\frac{N_0}{N}=1-\bigg(\frac{\lambda_c}{\lambda}\bigg)^3  \ \ \ \ \text{or} \ \ \ \ \frac{N_0}{N}=1-\bigg(\frac{T}{T_c}\bigg)^{3/2}.
\end{equation}
Here, when $\lambda > \lambda_c$ or $T<T_c $ (which implies that $N_0\neq 0$), the bosons start to form a BEC and when $\lambda \gg \lambda_c$ or $T\ll T_c $ (which implies that $N_0\approx N$), approximately all the particles accumulate on the ground state with the energy $\epsilon_0=0$ or equivalently with the momentum $\textbf{p}=0$. The latter case is the realization of BEC and the pure condensation of the gas phase occurs at $T=0$. Thus the state of BEC can be described by a unique condensate or macroscopic wave function.

The method applied above is for the ideal Bose gas. For a more general and realistic case, one should, however consider the interacting Bose gases \cite{GP,Schmitt:2014eka}. To understand the interacting picture, we use dilute Bose gas approximation and it is given by the condition $r_0\ll n^{-1/3}$, where $r_0$ is the range of the interatomic forces. By this condition, the interaction among particles can be described to be a two-body scattering process. We also impose $r_0\ll \lambda$ which states the very low momenta of the incoming Bose particles. In this case, we can consider s-wave ($l=0$) scattering length $a$ with $a\ll n^{-1/3}$; that is, the gas is weakly interacting. Thus the corresponding Hamiltonian operator for this configuration is given by
\begin{equation}
\hat{H}=\int \bigg[\frac{\hbar^2}{2m}|\nabla \hat{\Psi}(\textbf{r})|^2+V_\text{ext}(\textbf{r})|\hat{\Psi}(\textbf{r})|^2+\frac{1}{2}\int \hat{\Psi}^\dagger(\textbf{r}^\prime)\hat{\Psi}^\dagger(\textbf{r})V(\textbf{r}^\prime-\textbf{r})\hat{\Psi}(\textbf{r}^\prime)\hat{\Psi}(\textbf{r})d\textbf{r}^\prime \bigg]d(\textbf{r}),\label{H}
\end{equation}
where $V_\text{ext}(\textbf{r})$ and $V(\textbf{r})$ are the external and the two-body potentials, 
respectively.

The study of BEC in the case of the interacting approach is grounded on the Bogoliubov approximation as follows. The field operator, which is given in the form of $\hat{\Psi}(\textbf{r})=\sum_i \varphi_i \hat{a}_i$, can be separated into two as $\hat{\Psi}(\textbf{r})=\hat{\Psi}_0(\textbf{r})+  \hat{\Psi}_{\text{NC}}(\textbf{r})$, where the first term represents the ground or the condensate state and the second term is for the non-condensate states of the system. The macroscopic nature of the ground state of the system in the case of BEC implies $N_0=|\hat{a}_0|^2\gg 1$. In this respect, Bogoliubov considers the macroscopic component of the field operator $\hat{\Psi}_0(\textbf{r})$ a classical field and the other component $\hat{\Psi}_{\text{NC}}(\textbf{r})$ a perturbative field, i.e., $\hat{\Psi}(\textbf{r})=\Psi_0(\textbf{r})+  \delta\hat{\Psi}(\textbf{r})$. Further, in the case of very low temperatures, we can also ignore the second term and consider the field operator completely classical, i.e., $\hat{\Psi}(\textbf{r})=\Psi_0(\textbf{r})$, which means that the system completely behaves as a classical object in the case of BEC.

However, the Bogoliubov approximation is not applicable in the case of the realistic potentials $V(\textbf{r})$. In this regard, we additionally reconsider the potential in the third term of $\hat{H}$ as the effective or pseudo one $V_{\text{eff}}(\textbf{r})$ in the sense that these two potentials reproduce the same low-energy scattering properties. In the same term, we can also assume that most of the fields are in a condensate state, so $\Psi(\textbf{r}^\prime,t)\approx \Psi_0(\textbf{r},t)$ at the order of the range of the interatomic force. 

With the assumptions introduced above, the form of the Hamiltonian (\ref{H}) changes to 
\begin{equation}
\hat{H}=\int \bigg[\frac{\hbar^2}{2m}|\nabla \Psi_0(\textbf{r},t)|^2+V_\text{ext}(\textbf{r})|\Psi_0(\textbf{r},t)|^2+\frac{1}{2}g |\Psi_0(\textbf{r},t)|^4 \bigg] d\textbf{r},\label{HH}
\end{equation}
where we consider the time dependence of the wave function and time-independent external potential and $g=\int V_{\text{eff}}(\textbf{r})d\textbf{r}=4\pi \hbar^2 a/m$ with the s-wave scattering length $a$. Thus, using the Hamiltonian (\ref{HH}) in the Heisenberg equation of motion, $i\hbar\frac{\partial}{\partial t}\hat{\Psi}(\textbf{r},t)=[\hat{\Psi}(\textbf{r},t), \hat{H}]$, we finally obtain the GP equation for the time evolution of the condensate field $\Psi_0(\textbf{r},t)$:
\begin{equation}
i\hbar\frac{\partial}{\partial t}\Psi_0(\textbf{r},t)=\bigg(-\frac{\hbar^2\nabla^2}{2m}+V_{\text{ext}}(\textbf{r})+g|\Psi(\textbf{r},t)|^2\bigg)\Psi_0(\textbf{r},t).\label{GP}
\end{equation}

One may proceed with the GP equation (\ref{GP}) and obtain its static form by considering static solutions $\tilde{\Psi}_0 (\textbf{r})$ so that one may consider harmonic time-dependent wave function
\begin{equation}
\Psi(\textbf{r},t)=\tilde{\Psi}_0 (\textbf{r})exp\left(\frac{-i\mu t}{\hbar}\right),\label{psi}
\end{equation}
where $\mu$ is the chemical potential defined as $\frac{\partial E}{\partial N}$. After plugging Eq.(\ref{psi}) into Eq.(\ref{GP}), the form of GP reduces to 
\begin{equation}
\bigg(-\frac{\hbar^2\nabla^2}{2m}+V_{\text{ext}}(\textbf{r})+g|\tilde{\Psi}_0 (\textbf{r})|^2-\mu \bigg)\tilde{\Psi}_0 (\textbf{r})=0.\label{GP2}
\end{equation}





\section{Derivation of Gross-Pitaevskii Equation from $\phi^4$ Theories}

The theories based on the relativistic dynamics of a scalar field, say $\phi$, with its self-interacting potential energy $\frac{\lambda}{4}|\phi|^4$, where $\lambda$ is the dimensionless coupling constant, are called $\phi$-four theories \cite{Mukhanov:2007zz,Itzykson:1980rh,Peskin:1995ev,Maggiore:2005qv}. The dynamics of the scalar field is described by the Klein-Gordon (KG) equation:
\begin{equation}
\bigg[-\square+m^2+\lambda |\phi|^2 \bigg]\phi=0,\label{KG2}
\end{equation}
where $\square \phi=g^{\mu\nu}\nabla_\mu \nabla_\nu\phi=\frac{1}{\sqrt{-g}}\partial_\mu[\sqrt{-g}g^{\mu\nu}\partial_\nu\phi]$ in curved space-time \cite{Mukhanov:2007zz} (i.e in the presence of a gravitational background) and it reduces to the form $\square \phi=g^{\mu\nu}\partial_\mu \partial_\nu\phi$ in flat space-time (i.e., in the absence of a gravitational background), and we use $\hbar=c=1$ by convention. The Klein-Gordon Eq.(\ref{KG2}) is obtained by the variation of the action
\begin{equation}
S_\phi=\int dx^4 \sqrt{-g}\bigg(-\frac{1}{2} g^{\mu\nu}\partial_\mu \phi^\star \partial_\nu \phi-\frac{1}{2} m^2 |\phi|^2-\frac{\lambda}{4}|\phi|^4 \bigg)\label{L}
\end{equation}
with respect to the dynamical complex scalar field $\phi^\star$.

To study the implications of the BEC in the context of cosmology (i.e., in the presence of the gravitational background), one should understand the analogy between the KG equation for the scalar field and the GP equation for Bose-Einstein condensates. To this end, we first begin with the scalar field satisfying KG equation in the flat spacetime background and see how this KG equation corresponds to the GP equation. Then, we continue with the same analysis for the curved spacetime. 

\subsection{Reducing Klein-Gordon Equation in Flat Spacetime to Gross-Pitaevskii Equation}
We begin with analysing a scalar field described by a KG equation in the absence of a gravitational background. Considering metric in flat spacetime
\begin{equation}
ds^2=-dt^2+dx^2+dy^2+dz^2
\end{equation}
and $\square \phi=g^{\mu\nu}\partial_\mu \partial_\nu\phi$, the KG equation (\ref{KG2}) is written as
\begin{equation}
\ddot{\phi}-\triangle\phi+m^2 \phi + \lambda |\phi|^2 \phi = 0, \label{KG33}
\end{equation}
where the dot denotes a derivative with respect to $t$ and $\triangle$ is the 3-dimensional Laplace operator. If we consider the transformation \cite{Erdem:2016hqw}:
\begin{equation}
\phi=e^{-iwt}\chi,
\end{equation}
where $w=\sqrt{\vec{p}^2+m_\phi^2}$ is the energy of the field $\phi$, Eq.(\ref{KG33}) then becomes
\begin{equation}
\ddot{\chi}-2iw\dot{\chi}-\triangle\chi+(m_\phi^2-w^2)\chi+\lambda|\chi|^2\chi=0. \label{KG333}
\end{equation}
Since the GP equation is considered as the non-relativistic limit of $\phi ^4$ theories, we here assume that $\phi$ is non-relativistic for a long time then its time evolution should be slow. This implies that $|\frac{\ddot{\chi}}{\dot{\chi}}|\ll 2 m_\phi$ and $w^2\simeq m_\phi^2$. Then Eq.(\ref{KG333}) reduces to
\begin{equation}
i\dot{\chi}=(-\frac{\triangle}{2m_\phi}+\frac{\lambda}{m_\phi}|\chi|^2)\chi,
\end{equation}
which is the GP equation (\ref{GP}) without external potential.

We now give an example \cite{Castellanos:2013ena} where a specific solution of $\phi$ in one-dimensional case (for simplicity) is considered:
\begin{equation}
\phi(t,x)=e^{iwt}\chi(x),\label{SS}
\end{equation}
which is the harmonic time solution of $\phi$. The field here is considered to be confined in an infinite potential well. Plugging the solution (\ref{SS}) into Eq.(\ref{KG33}), we get
\begin{equation}
\chi^{\prime \prime}-(m^2-w^2)\chi-\lambda |\chi|^2 \chi=0, \label{KG4}
\end{equation}
where the prime denotes a derivative with respect to $x$. Eq.(\ref{KG4}) is in the specific form of the GP equation given in Eq.(\ref{GP2}) (the three-dimensional analysis of the study can be found in \cite{Erdem:2016hqw,Matos:2011kn,Matos:2012qu,Castellanos:2013kga,T. Matos and E. Castellanos}).

Further, one may search for a solution to the KG equation for the field $\chi(x)$. The integration of Eq.(\ref{KG4}) leads to 
\begin{equation}
\frac{1}{2}\chi^{\prime 2}-\frac{\lambda}{4}\left(\chi^2+\frac{m^2-w^2}{\lambda}\right)^2=C,
\end{equation}
where $C$ is a constant. The last equation can be solved by a kink solution \cite{Castellanos:2013ena,Kink}
\begin{equation}
\chi(x)=|\chi_0|\text{tanh}\left(x \sqrt{\frac{w^2-m^2}{2}} \right), \label{K1}
\end{equation}
with a constant $\chi_0$, which represents the wave function far away from the wall of the potential well. The solution (\ref{K1}) has the same form as the one derived for a Bose-Einstein condensate confined in a box \cite{pethick_smith_2008}.

Hence, it is shown that in the absence of a gravitational background, one may relate the classical scalar field (satisfying the KG equation) to a Bose-Einstein condensate (satisfying the GP equation).

\subsection{Reducing Klein-Gordon Equation in Curved Spacetime to Gross-Pitaevskii Equation}
We now proceed with the analysis for a scalar field  satisfying the KG equation in the presence of a gravitational background \cite{Erdem:2016hqw}. In this respect, we consider the FRW metric (\ref{met.}). Using the definition $\square \phi=g^{\mu\nu}\nabla_\mu \nabla_\nu\phi=\frac{1}{\sqrt{-g}}\partial_\mu[\sqrt{-g}g^{\mu\nu}\partial_\nu\phi]$ in curved space-time with the metric (\ref{met.}), the KG equation (\ref{KG2}) takes the following form:
\begin{equation}
\ddot{\phi}+3H\dot{\phi}-\frac{1}{a^2}\triangle\phi+m^2 \phi + \lambda |\phi|^2 \phi = 0. \label{KGCU}
\end{equation}
Using the transformation (\ref{SS}) in Eq.(\ref{KGCU}), we get
\begin{equation}
\ddot{\chi}-2iw\dot{\chi}+3H\dot{\chi}-\triangle\chi+(m_\phi^2-w^2)\chi-3iHw\chi+\lambda|\chi|^2\chi=0. \label{KGCU3}
\end{equation}
We know that the GP equation is the equation for a non-relativistic particle in a condensate at atomic scale, where the time evolution of the cosmic scale factor $a$ is negligible \cite{Erdem:2016hqw}. Therefore, the terms with the Hubble parameter in Eq.(\ref{KGCU3}) are negligible compared to the other terms. Further, considering the non-relativistic limit, one may assume (as we have noticed in the previous subsection) $|\frac{\ddot{\chi}}{\dot{\chi}}|\ll 2 m_\phi$ and $w^2\simeq m_\phi^2$. Then Eq.(\ref{KGCU3}) reduces to
\begin{equation}
i\dot{\chi}=(-\frac{\triangle}{2a^2 m_\phi}+\frac{\lambda}{m_\phi}|\chi|^2)\chi,
\end{equation}
which is the GP equation (\ref{GP}) without external potential.

As an example for the GP equation in curved spacetime \cite{Castellanos:2013ena}, we now consider the spherically symmetric and static metric
\begin{equation}
ds^2=-A(r)dt^2+\frac{1}{A(r)}dr^2+r^2d\theta^2+r^2 sin^2 \theta d\varphi^2.\label{cm}
\end{equation}
Using the definition $\square \phi=g^{\mu\nu}\nabla_\mu \nabla_\nu\phi=\frac{1}{\sqrt{-g}}\partial_\mu[\sqrt{-g}g^{\mu\nu}\partial_\nu\phi]$ in curved space-time with the metric (\ref{cm}), the KG equation (\ref{KG2}) takes the following form:
\begin{equation}
\frac{\ddot{\phi}}{A}-A\phi^{\prime\prime}-\left(\frac{2A}{r}+A^{\prime}\right)\phi^{\prime}+m^2\phi+\lambda|\phi|^2\phi=0,\label{cm2}
\end{equation}
where the prime denotes a derivative with respect to $r$. If we particularly use the monopolar and the harmonic time solution of the scalar field 
\begin{equation}
\phi(t,r)=e^{iwt}\frac{u(r)}{r},
\end{equation}
Eq.(\ref{cm2}) becomes
\begin{equation}
u^{\prime\prime}-\left[A\left(m^2+\frac{A^\prime}{r^\star}\right)-w^2 \right]u-\lambda\frac{A}{r^2}|u|^2 u=0,
\end{equation}
which is in the specific form of the GP equation (\ref{GP2}). Here, we introduce a new coordinate $r^\star=\int\frac{dr}{A}$ and the prime denotes a derivative with respect to $r^\star$.  The field is here trapped by the effective potential 
\begin{equation}
V_\text{eff}=A \left( m^2+\frac{A^\prime}{r^\star} \right),\label{veff}
\end{equation}
which is identified with the external (or trapping) potential $V_\text{ext}(\textbf{r})$ of the GP equation (\ref{GP2}). In contrast to the case of flat spacetime, we here do not need to introduce an external trapping potential. The curvature of a spacetime associated with some $A(r)$ can naturally provide trapping potential by Eq.(\ref{veff}) to confine the scalar field in some region of gravitational background. As examples of such backgrounds, Schwarzschild spacetime with $A(r)=1-\frac{2MG}{r}$ and Schwarzschild-de Sitter spacetime with $A(r)=1-\frac{2MG}{r}-\frac{\Lambda}{3}$ (where $M$ is the mass of the blackhole and $\Lambda$ is the cosmological constant) may be given. With some conditions, they can confine the scalar field in some region of their backgrounds so that stationary (or quasi-stationary) scalar field distributions can be obtained. In this regard, such  distributions reinforce the idea that there may be close relations between such scalar fields in cosmology and Bose-Einstein condensates in atomic physics. Detailed discussions on these examples are given in \cite{Castellanos:2013ena,Barranco:2011eyw,Barranco:2012qs}.





\section{Some Cosmological Bose-Einstein Condensate Scalar Field Models}

As we have clarified in the previous chapter, Bose-Einstein condensates (BEC) of scalar fields at the cosmological level may be good candidates for DM and DE. In this regard, there are some studies on BEC scalar field models for DE and DM \cite{Fukuyama:2007sx, Erdem:2016hqw, UrenaLopez:2008zh, Silverman:2001gx, Silverman:2002qx,Harko:2015nua, Nishiyama:2004ju, Besprosvany:2015ura, Das:2014agf, Takeshi:2009cy, Das:2015dca}. In one of them \cite{Das:2014agf} it is shown that if a gas of bosons all in the condensate phase is identified with DM, the critical temperature (\ref{BEC1}) (below which the bosons start to form a BEC) reduces to
\begin{equation}
T_c=\frac{6\times 10^{-12}}{m^{1/3}a} \ K,
\end{equation}
where $a$ is the scale factor and m is in the unit of $kg$. Thus, for $m<1 \text{eV}$, $T_c > 2.7/a$. Because $2.7/a$ is the background temperature of the universe at all times, the BEC phase of a tiny mass of bosons can start at the very early era of the universe. Further, the bosons in this phase have approximately zero momenta and zero pressure, meaning that they can be considered CDM.

Moreover, we have seen that a BEC is a macroscopic quantum state, which is represented by a scalar field in the form of $\phi=Re^{iS}$, where $R$ and $S$ are the real functions of space and time, as clarified in the previous sections. Considering quantal (Bohmian) trajectories \cite{PhysRev.85.166, BOHM1987321, holland_1993} where the velocity field is defined by $u_a=\hslash \partial_\alpha S/m$ and also defining the induced metric $h_{ab}=g_{ab}-u_a u_b$,
the quantum mechanical contribution to the cosmological constant in the Friedmann equation is obtained by \cite{Das:2013oda, Das:2014sia, Ali:2014qla}
\begin{equation}
\Lambda_Q=\frac{\hslash^2}{m^2 c^2}h^{ab}\nabla_a \nabla_b\left(\frac{\square R}{R}\right).
\end{equation}
Here $\Lambda_Q$ is related to the relativistic quantum potential $V_Q=\frac{\hslash^2}{m^2 c^2}\left(\frac{\square R}{R}\right)$ \cite{Das:2013oda}. Because our universe is homogeneous and isotropic, the wavefunction $R$ is expected to disperse uniformly over the range of the Hubble radius, say $L_0$,  of the universe. Due to the condensate nature of the field $\phi$, $R$ is also taken to be time-independent. With all these physical points of view, the most suitable choices of $R$ may be given as $R=R_0e^{-r^2/L_0^2}$ \cite{Brack} (which describes the ground state of the harmonic oscillator) or $R=R_0\text{tanh}(r/L_0\sqrt{2})$ for $g>0$ or $R=\sqrt{2}R_0 \text{sech}(r/L_0)$ for $g<0$ \cite{RogelSalazar2013, article}, where $g$ is the interaction of strength. All these choices give rise to the same expression as $\nabla_a \nabla_b \left(\frac{\square R}{R}\right)\simeq 1/L_0^4$. Moreover, $L_0$ may also be considered as the Compton wavelength of the bosons of mass $m$, that is, $L_0=h/mc$ \cite{Wachter}. Then by using the present value of Hubble radius $L_0=1.4\times 10^{26} \text{meter} $, the mass of the bosons is obtained as $m\simeq 10^{-68} kg$ or $10^{-32} eV$, which gives rise to the observed value of DE
\begin{equation}
\Lambda_Q=10^{-123}l_{\text{Pl}}^{-2},
\end{equation}
where $l_\text{Pl}$ denotes the Planck units. In conclusion, it is shown that DM can be explained by the density of the bosons of mass $m$ all in BEC, while DE can be explained by the quantum potential of the BEC macroscopic wavefunction.

There is another study \cite{Nishiyama:2004ju} on BEC employed as DE and DM. In the model, it is first introduced that at the very beginning of the universe, the energy density of the excited Bose gas, $\rho_g$, identified as DM is assumed to dominate the energy density of the condensation of the bosons, $\rho_c$ employed as DE.  Next, by the expansion of the universe, the amount of $\rho_g$ is decreased while $\rho_c$ is unchanged (This idea is thermodynamically well-explained in \cite{Nishiyama:2004ju}). Therefore, $\rho_c$ eventually dominates $\rho_g$. This results in the accelerated expansion of the universe at the large scale and the collapse of the condensation at the smallest scale, yielding the localized compact objects as DM. The new energy density $\rho_l$ is then introduced for the localized energy density due to the collapse of the condensation at small scales.

According to the $\Lambda$CDM, the evolution of all these different forms of energy densities are given by the following equations:
\begin{equation}
\dot{\rho}_c=\Gamma \rho_g, \ \ \ \dot{\rho}_g=-3H\rho_g-\Gamma \rho_g, \ \ \ \dot{\rho}_l=-3H\rho_l,
\end{equation}
where $\rho=\rho_c+\rho_g+\rho_l$ and $H\equiv\frac{\dot{a}}{a}=\sqrt{8\pi G \rho / 3}$. These equations are valid when the energy density of the condensate has the condition $\rho_c<\rho_g+\rho_l$. After the time $\Gamma^{-1}$, it becomes $\rho_c>\rho_g+\rho_l$ and the inhomogeneous part of the condensate would collapse and contributes to $\rho_l$. Then the condition $\rho_c<\rho_g+\rho_l$ is realized again. In each time scale $\Gamma^{-1}$, this situation, i.e., 'chase and collapse', is repeated until $\rho_c \approx \rho_g+\rho_l$, that is, until the ratio of the energy densities of DE and DM becomes approximately order of one as predicted by $\Lambda\text{CDM}$. This kind of dynamics is known as Self Organized Criticality (SOC) as discussed in \cite{Nishiyama:2004ju,https://doi.org/10.48550/arxiv.cond-mat/9906077}.

\newpage
\chapter{A Mechanism for the Formation of Bose-Einstein Condensation in Cosmology}

\section{An Effective Minkowski Space Formulation }
In this mechanism, based on the paper \cite{Erdem:2021mrw}, we aim to realize BEC scalar field in cosmology with particular emphasis on its microscopic description in particle physics. To this end, by taking the FRW metric
\begin{equation}
ds^2=-dt^2+a^2(t)[dr^2+r^2(d\theta^2+\text{sin}^2\theta d\phi^2)],\label{frw}
\end{equation}
we consider the following action
\begin{equation}
S=\int d^4x \sqrt{-g} \frac{1}{2} \lbrace -g_{\mu\nu}[\partial_\mu \phi \partial_\nu \phi + \partial_\mu \chi \partial_\nu \chi ]-m_\phi^2 \phi^2- m_\chi^2 \chi^2 -\mu \phi^2\chi\rbrace
\end{equation}
with its effectively Minkowskian form \cite{Mukhanov:2007zz}
\begin{equation}
S=\int d^3xd\eta\frac{1}{2}\lbrace \tilde{\phi}^{\prime \ 2} - (\vec{\nabla}\tilde{\phi})^2+\tilde{\chi}^{\prime \ 2} - (\vec{\nabla}\tilde{\chi})^2-\tilde{m}_{\phi}^2 \tilde{\phi}^2- \tilde{m}_\chi^2 \tilde{\chi}^2 -\tilde{\mu} \tilde{\phi}^2 \tilde{\chi}\rbrace,\label{eff}
\end{equation}
which is obtained by use of the following transformations and derivatives:
\begin{align}
d\eta &=\frac{dt}{a(t)}, \ \ \ \tilde{\phi}=a\phi, \ \ \ \tilde{\chi}=a\chi, \ \ \ a^\prime=\frac{da}{d\eta}, \ \ \  \dot{a}=\frac{da}{dt}, \ \ \  a^{\prime\prime}=\frac{d^2a}{d\eta^2}, \ \ \  \ddot{a}=\frac{d^2a}{dt^2},
\\ &
\tilde{\mu}=a\mu, \ \ \ \tilde{m}_i^2=m_i^2a^2-\frac{a^{\prime\prime}}{a}=a^2\bigg(m_i^2-\frac{\ddot{a}}{a}-\frac{\dot{a}^2}{a^2}\bigg), \label{effm}
\end{align}
where prime denotes the derivative with respect to conformal time $\eta$ and a dot denotes the derivative with respect to ordinary time $t$ and the subscript $i$ takes the values, $i=\phi, \chi$.

It is evident from Eq.(\ref{eff}) that locally (i.e., at each time $\eta$) we may consider the spacetime effectively Minkowskian. This, in turn, implies that for sufficiently small intervals of $\eta$, $\Delta \eta$, spacetime may be approximately considered to be effectively Minkowskian. In this section, we will clarify how small $\Delta \eta$ should be taken so that an effective Minkowski formulation of spacetime is reliable, and we will show that the effective Minkowski formulation is cosmologically relevant. Then, we consider the problem of formation of scalar field condensation in the framework of Minkowski space formulation in the following section. In this context, we use the interaction term $\phi^2\chi$ in the Lagrangian, which shall induce the evolution and the formation of the condensation via the particle physics processes $\chi\chi\rightarrow \phi\phi$, during the time $\triangle t=\frac{1}{n_\chi \beta\sigma v}$ for each process. Here, we assume that initially, there are only relativistic $\chi$ particles causing the production of non-relativistic $\phi$ particles, i.e., $\tilde{m}_\chi\ll \tilde{m}_\phi$. However, being in an effectively Minkowskian space in $\triangle t$  is not enough to proceed with our model to realize the condensation. We should also be able to use the tools of the usual perturbative quantum field theory \cite{Itzykson:1980rh, Peskin:1995ev, Maggiore:2005qv} for the calculation of the rates and the cross sections in an effective Minkowski space for each process given in Figure 5.1, where the leading order contributions to the production of $\phi$ particles are given. To this end, we need the following two conditions:
\begin{itemize}
\item The first condition is the assumption that the rate of each process $\chi\chi\rightarrow \phi\phi$ is much larger than the Hubble parameter, $H\equiv \frac{\dot{a}}{a}$, during the so small time interval $\triangle t$. Then one may take the masses $\tilde{m}_\chi$ and $\tilde{m}_\phi$ constant during each process; that is to say, there is no time dependence of the effective masses in $\triangle t$, i.e, in microscopic scales, where each particle process occurs while the time dependence is observed only at cosmological scales. A concrete expression of this condition is
\begin{equation}
\Bigg\lvert\frac{\triangle\tilde{m}^2}{\tilde{m}^2}\Bigg\rvert=\Bigg\lvert\frac{\triangle t \bigg(\frac{da^2(m^2-\dot{H}-2H^2)}{dt}\bigg)}{a^2(m^2-\dot{H}-2H^2)}\Bigg\rvert\ll 1.\label{con1}
\end{equation}
\item The next condition requires the constancy of the effective coupling constant $\tilde{\mu}$ in each time interval $\triangle t$ which is imposed by the expression
\begin{equation}
\bigg\lvert\frac{\triangle \tilde{\mu}}{\tilde{\mu}}\bigg\rvert=|H\triangle t|\ll 1.\label{con2}
\end{equation}
\end{itemize}
In Appendix A, it is shown that by letting $H=\xi a^{-s}$, which includes all simple interesting cases e.g. radiation, matter, stiff matter, cosmological constant dominated universes, these two conditions (\ref{con1}) and (\ref{con2}) are satisfied for a considerable range of parameters.
\begin{figure}
\begin{center}
\includegraphics[scale=0.5]{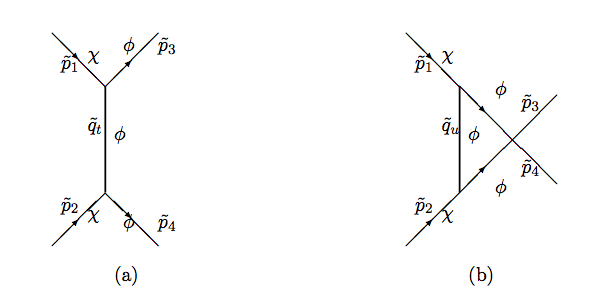} 
\end{center}
\caption{The leading order Feynman diagrams contributing to the production of $\phi$ particles. Here $\tilde{q}_t=\tilde{p}_1-\tilde{p}_3=\tilde{p}_4-\tilde{p}_2,\ \tilde{q}_u=\tilde{p}_1-\tilde{p}_4=\tilde{p}_3-\tilde{p}_2 $ are the 4-momenta carried in the internal lines. Note that the s-channel is forbidden in this case by kinematics.}\label{Feyn1}
\end{figure}

The derivation in Appendix B shows that the quantized expression of any field $\chi$ (satisfying the conditions in $\triangle t$ mentioned above) is given by
\begin{equation}
\tilde{\chi}^{(i)}(\vec{r},\eta)\simeq \int \frac{d^3\tilde{p}}{(2\pi)^{\frac{3}{2}}\sqrt{2w_p^{(i)}}}\bigg[a_p^{(i)-}e^{i\big(\vec{\tilde{p}}.\vec{r}-w_p^{(i)}(\eta-\eta_i)\big)}+a_p^{(i)+}e^{i\big(-\vec{\tilde{p}}.\vec{r}+w_p^{(i)}(\eta-\eta_i)\big)}\bigg],\label{chii}
\end{equation}
where $\eta_i<\eta<\eta_{i+1}$, and the superscript $(i)$ refers to the $i$th time interval between the $i$th and $(i+1)$th processes. In fact, Eq.(\ref{chii}) is a consequence of the previous conditions.

Thus, by considering the conditions (\ref{con1}) and (\ref{con2}) together with the form of Eq.(\ref{chii}) we have
\begin{equation}
d\tilde{s}^2=-d\eta^2+d\tilde{x}_1^2+d\tilde{x}_2^2+d\tilde{x}_3^2\label{met1}
\end{equation}
in each interval $\eta_i<\eta<\eta_{i+1}$, for $\chi$, where the masses and the coupling constant of the particles are constant and they alter as one passes from one interval to the other, and $\tilde{x}_i$ are related to Eq.(\ref{frw}) by $d\tilde{x}_1^2+d\tilde{x}_2^2+d\tilde{x}_3^2=dr^2+r^2(d\theta^2+\text{sin}^2\theta d\phi^2)$. Therefore, one may employ the usual perturbative quantum field theory tools in each interval in Eq.(\ref{met1}).

\section{Realizing the Conditions for Condensation}
Having introduced the effective flat space formulation with the constancies of the particle's effective masses and effective coupling constant by Eq.(\ref{con1}) and Eq.(\ref{con2}) for each process $\chi\chi\rightarrow \phi\phi$ during $\triangle t$ , we are now in a position to observe the tendency of the $\phi$'s towards Bose-Einstein condensation via the curved space effects in cosmology \cite{Erdem:2019gpk}. To this end, we need to realize the coherence and correlation for the de Broglie wavelengths of the particles, and achieve large, finite number density of the particles.

\subsection{Achieving Coherence and Correlation}

To begin with, for each process occurring in the effective flat space, we consider the center of mass frame where the conservation of energy amounts to $\vec{\tilde{p}}^2+\tilde{m}_\chi^2=\vec{\tilde{k}}^2+\tilde{m}_\phi^2 $. Here $\vec{\tilde{p}}=\vec{\tilde{p}}_1$ and $\vec{\tilde{k}}=\vec{\tilde{p}}_3$ (see Figure 5.1). Then, by the definition of the effective masses given in Eq.(\ref{effm}), we get
\begin{equation}
\vec{\tilde{p}}^2-\vec{\tilde{k}}^2=a^2(m_\phi^2-m_\chi^2).\label{rbec}
\end{equation}
To proceed with our study on condensation using Eq.(\ref{rbec}), one should see whether the effective momenta are time dependent or not. To this end, we should compare the effective momentum (corresponding to the metric (\ref{met1})) with the physical momentum (corresponding to the metric (\ref{frw})) as follows:
\begin{equation}
|\vec{\tilde{p}}|\,=\,\tilde{m}_\chi\sqrt{\tilde{g}_{ij}\frac{d\tilde{x}^i}{d\tilde{\tau}}\frac{d\tilde{x}^j}{d\tilde{\tau}}}\,=\,
a\,m_\chi\sqrt{g_{ij}\frac{dx^i}{d\tau}\frac{dx^j}{d\tau}}\,=\,a |\vec{p}| \;. \label{t1y}
\end{equation}
In the above, tilde $``\ \tilde{}\ "$ over a quantity refers to its form for the metric (\ref{met1}) while the quantities without a tilde refer to its form for the metric (\ref{frw}), and $d\tau^2=-ds^2$, $d\tilde{\tau}^2=-d\tilde{s}^2$. Eq.(\ref{t1y}) is obtained by use of $g_{ij}= a^2\tilde{g}_{ij}$ and $\frac{d\tilde{x}^i}{d\tilde{\tau}}=a^2\left(\frac{m}{\tilde{m}}\right)\frac{dx^i}{d\tau}$ which, in turn, follows from the geodesic equations for the actions $\int\,\tilde{m}_\chi\,\sqrt{-\tilde{g}_{\mu\nu}\frac{d\tilde{x}^\mu}{d\tilde{\tau}}\frac{d\tilde{x}^\nu}{d\tilde{\tau}}}\,d\tilde{\tau}$ and
$\int\,m_\chi\,\sqrt{-g_{\mu\nu}\frac{dx^\mu}{d\tau}\frac{dx^\nu}{d\tau}}\,d\tau$ for the metrics (\ref{met1}) and (\ref{frw}), respectively, after requiring that $|\vec{\tilde{p}}|$ and $|\vec{p}|$ coincide for $a=\mbox{constant}=1$. We then observe that $\lvert \vec{\tilde{p}}\rvert$ and $\lvert \vec{p}\rvert$ are related to each other by $\lvert \vec{\tilde{p}}\rvert=a\lvert \vec{p}\rvert$. This relation implies that $\lvert \vec{\tilde{p}}\rvert=\text{constant}$ because the redshift dependence of the physical momentum is given by $\lvert \vec{p}\rvert\propto \frac{1}{a}$.
Thus the constancies of $\lvert \vec{\tilde{p}}\rvert$ and $\lvert \vec{\tilde{k}}\rvert$ imply that if Eq.(\ref{rbec}) is satisfied for a value of $\vec{\tilde{p}}^2-\vec{\tilde{k}}^2$ (at the moment of some transition $\chi\chi\rightarrow \phi\phi$ ) then, in general, it is not satisfied at a later time by the same $\vec{\tilde{p}}^2-\vec{\tilde{k}}^2$. This means that during a process $\chi\chi\rightarrow \phi\phi$, the scale factor $a$ is unchanged. However, during another process (being a later time), $a$ is again unchanged except that it has a new increased constant value with respect to the previous process. Then for each next process, $a$ increases while it behaves as a constant value at the moment of these processes. In this regard, for a given value of $\lvert \vec{\tilde{p}}\rvert$ in Eq.(\ref{rbec}), the corresponding $\lvert \vec{\tilde{k}}\rvert$ will get smaller and smaller over time via the scale factor $a$. Here the curved space effect is then provided by the scale factor $a$. Therefore, this observation is one of the key points for deriving the tendency of the $\phi$'s towards Bose-Einstein condensation in the level of cosmology.

Another approach to the tendency for the condensation of $\phi$ particles can also be given by considering the phase space evolution of the particles by the evolution of the distribution function for one of the final state particles \cite{Zhou:2017zql} through the equation
\begin{align}
\frac{d\tilde{f}(\vec{\tilde{p_4}},\eta)}{d\eta}=&\frac{1}{32(2\pi)^5\tilde{E}_4}\int\int\int \delta^{(4)}(\tilde{p_1}+\tilde{p_2}-\tilde{p_3}-\tilde{p_4})|\tilde{M}|^2\nonumber
\\&
\times\lbrace \tilde{f}_1\tilde{f}_2(1+\tilde{f}_3)(1+\tilde{f}_4)-\tilde{f}_3\tilde{f}_4(1+\tilde{f}_1)(1+\tilde{f}_2)\rbrace\frac{d^3 \vec{\tilde{p_1}}}{\tilde{E}_1}\frac{d^3 \vec{\tilde{p_2}}}{\tilde{E}_2}\frac{d^3 \vec{\tilde{p_3}}}{\tilde{E}_3},\label{dist}
\end{align}
which is written for the effective Minkowski space as clarified by $``\ \tilde{}\ "$ over the quantities. In Eq.(\ref{dist}), $\tilde{M}$ denotes the transition matrix element for $\chi\chi\rightarrow\phi\phi$ dominated by the tree-level diagrams in Figure 5.1 and $\tilde{f}_i=\tilde{f}(\vec{\tilde{p_i}},\eta)$ is the number density in phase space. For the distribution of $\phi$ particles in phase space at early times after the start of the process $\chi\chi\rightarrow\phi\phi$, we consider the initial times for the sake of simplicity. Then we may write
\begin{align}
\tilde{f}^{(f)}(\vec{\tilde{p_j}},\eta)\simeq 0 \ \ \ \text{with} \ \ \ j=3,4\label{f1}
\end{align}
\begin{align}
\tilde{f}^{(i)}(\vec{\tilde{p_j}},\eta)\simeq & \tilde{n}_\chi\frac{[\Theta (|\vec{\tilde{p}}|_\text{max}-|\vec{\tilde{p}}|_\text{min})-\Theta (|\vec{\tilde{p}}|_\text{min}-|\vec{\tilde{p}}|_\text{max})]}{4\pi(|\vec{\tilde{p}}|_\text{max}-|\vec{\tilde{p}}|_\text{min})|\vec{\tilde{p}}_j|^2}\nonumber
\\&
\times [\Theta (|\vec{\tilde{p}}_\text{j}|-|\vec{\tilde{p}}|_\text{min})-\Theta (|\vec{\tilde{p}}_\text{j}|-|\vec{\tilde{p}}|_\text{max})] \ \ \ \ \ \ \ \ \ \ \ \ \text{with} \ \ \ j=1,2,\label{f2}
\end{align}
where $\Theta$ denotes the Heaviside function (i.e., unit step function). For obtaining Eq.(\ref{f2}), we get use of the expression $\tilde{n}_\chi(\eta)=\int d^3\tilde{p}\tilde{f}^{(i)}(\vec{\tilde{p}},\eta)$ (where $\tilde{n}_\chi$ is the number density of $\chi$ in the co-moving frame defined by Eq.(\ref{met1})). Here we also assume that the momentum of the $\chi$ particles (that have enough energies to induce $\chi\chi\rightarrow\phi\phi$ processes) satisfies $0\leq |\vec{\tilde{p}}|_\text{min}=\sqrt{\tilde{m}_\phi^2-\tilde{m}_\chi^2}<|\vec{\tilde{p}}|<|\vec{\tilde{p}}|_\text{max}=\sqrt{\vec{\tilde{k}}^2_\text{max}+\tilde{m}_\phi^2-\tilde{m}_\chi^2}$, and the spatial distributions of the $\chi$ particles are homogeneous and isotropic. Then, after using the expressions (\ref{f1}) and (\ref{f2}) in Eq.(\ref{dist}), we get
\begin{equation}
\frac{d\tilde{f}^{(f)}(\vec{\tilde{p}}_4, \eta)}{d\eta}\simeq \tilde{n}^2_\chi\frac{\delta (\sqrt{\frac{1}{4}\vec{\tilde{p}}^2_4+\tilde{m}_\phi^2-\tilde{m}_\chi^2}-\sqrt{\tilde{m}_\phi^2-\tilde{m}_\chi^2})}{(\sqrt{\frac{1}{4}\vec{\tilde{p}}^2_4+\tilde{m}_\phi^2-\tilde{m}_\chi^2}-\sqrt{\tilde{m}_\phi^2-\tilde{m}_\chi^2})}\mathcal{B}(|\vec{\tilde{p}}_4|),\label{rbec1}
\end{equation}
where
\begin{eqnarray}
{\cal B}(|\vec{\tilde{p}}_4|)
&=&
\frac{1}{128(2\pi)^7\tilde{E}_4}\int\int\int\,
\frac{d^3\vec{\tilde{p}}_1}{\tilde{E}_1}\frac{d^3\vec{\tilde{p}}_2}{\tilde{E}_2}\frac{d^3\vec{\tilde{p}}_3}{\tilde{E}_3}\,
\delta^{(4)}(\tilde{p}_1+\tilde{p}_2-\tilde{p}_3-\tilde{p}_4)
\,|\tilde{M}|^2\nonumber \\
&&\times \Bigg\lbrace \,\frac{\left[\Theta\left(|\vec{\tilde{p}}_1|-|\vec{\tilde{p}}|_{min}\right)-
\Theta\left(|\vec{\tilde{p}}_1|-|\vec{\tilde{p}}|_{max}\right)\right]}{\vec{\tilde{p}}_1^2 \,\vec{\tilde{p}}_2^2} \nonumber \\
&& \, \, \, \, \, \, \, \, \, \times \,\frac{\left[\Theta\left(|\vec{\tilde{p}}_2|-|\vec{\tilde{p}}|_{min}\right)-
\Theta\left(|\vec{\tilde{p}}_2|-|\vec{\tilde{p}}|_{max}\right)\right]}{\vec{\tilde{p}}_1^2 \,\vec{\tilde{p}}_2^2} \, \Bigg\rbrace .
\label{ab2xxa3}
\label{x1}
\end{eqnarray}
and $\delta$ denotes Dirac delta function for which we have used that $\text{lim}_{x\rightarrow y}\frac{\Theta(y)-\Theta(x)}{y-x}=\delta(x)$. 

(Note that the $|\vec{\tilde{k}}|_{max}$ dependence in Eq.(\ref{x1}) may be eliminated (in favor of $|\vec{\tilde{p}}_4|$ and a constant $|\alpha|_{max}$) by making use of
\begin{eqnarray}
&&\vec{\tilde{p}}_3=\tilde{m}_3\frac{\vec{\tilde{P}}}{\tilde{m}_3+\tilde{m}_4}+\vec{\tilde{k}}=\frac{\vec{\tilde{P}}}{2}+\vec{\tilde{k}}~,~~
\vec{\tilde{p}}_4=\tilde{m}_4\frac{\vec{\tilde{P}}}{\tilde{m}_3+\tilde{m}_4}-\vec{\tilde{k}}=\frac{\vec{\tilde{P}}}{2}-\vec{\tilde{k}}
\label{ab1a} \\
&&\vec{\tilde{P}}=\vec{\tilde{p}}_3+\vec{\tilde{p}}_4~,~~
\vec{\tilde{k}}=\frac{(\tilde{m}_2\vec{\tilde{p}}_3-\tilde{m}_1\vec{\tilde{p}}_4)}{\tilde{m}_3+\tilde{m}_4}=\frac{\vec{\tilde{p}}_3-\vec{\tilde{p}}_4}{2} \;.
\label{ab1b}
\end{eqnarray}
Eq.(\ref{ab1b}) implies that, for a fixed $\vec{\tilde{p}}_4$, there exist the largest number $|\alpha|_{max}$ that maximizes $|\vec{\tilde{k}}|$ with $\vec{\tilde{p}}_3=-|\alpha|_{max}\vec{\tilde{p}}_4$, i.e., $|\vec{\tilde{k}}|_{max}=\frac{1}{2}\left(1+|\alpha|_{max}\right)|\vec{\tilde{p}}_4|$.)

The final form Eq.(\ref{rbec1}) of the phase space evolution of $\phi$ particles indicates that the evolution of their momenta are towards $|\vec{\tilde{p}}_4|=0$. This is the main indication of the Bose-Eeinstein condensation because this gives rise to coherence (necessary for the $\phi$ particles being at the same de Broglie wavelengths) and correlation by $\lambda \gg  n_\phi^{-1/3} $  (necessary for the overlap of the de Broglie wavelengths of the particles) of the system.

One should here notice that while Eq.(\ref{rbec1}) is an important equation to indicate the formation of BEC, it does not show its complete formation. For the derivation of Eq.(\ref{rbec1}) we have used Eq.(\ref{f2}), which is given for the initial time when  $\tilde{f}^{(f)}(\vec{\tilde{p}}_j,\eta)\ll\,1$ (when the formation of BEC has not been realised yet); on the other hand, Eq.(\ref{rbec1}) is given for much later times when $|\vec{\tilde{k}}|_{max}\,\rightarrow\,0$. This implies that we have not given proof but a hint towards the formation of BEC at later times. Additionally, even while evolution towards  $|\vec{\tilde{p}}_4|=0$ is a crucial point towards forming of BEC, it is insufficient \cite{Semikoz}. In fact, essentially, Eq.(\ref{rbec}) also indicates $|\vec{\tilde{p}}_4|\,\rightarrow\,0$ by time. Eq.(\ref{rbec1}) reaffirms and supports this finding and encourages a more thorough investigation. Therefore, we have only shown the tendency of the $\phi$'s towards Bose-Einstein condensation instead of proving its formation. For a thorough examination of condensation formation, the same procedures must be followed at all times, including the times when Bose statistics' impact cannot be disregarded (i.e., considering the times when $\tilde{f}^{(f)}(\vec{\tilde{p}}_j,\eta)$ cannot be disregarded on the right hand side of Eq.(\ref{dist})) and then solve the equation for $\tilde{f}^{(f)}(\vec{\tilde{p}}_4,\eta)$ and show that it has a delta function of the form of Eq.(\ref{rbec1}). This requires separate work in future. It is also important to note that there are also other processes  $\phi\phi\,\rightarrow\,\chi\chi$ and $\chi\phi\,\rightarrow\,\chi\phi$. However, we assume that we have only $\chi$ particles at initial times, i.e., $n_\phi$ in phase space is small. Therefore, the rates of these processes can be ignored. Additionally, the process $\chi\phi\,\rightarrow\,\chi\phi$ does not change the number density of the particles, so the condensation is not affected by this process. While it seems that the process  $\phi\phi\,\rightarrow\,\chi\chi$ changes the number density of $\phi$, there is a general rise in $n_\phi$ in case of considering both of $\chi\chi\,\rightarrow\,\phi\phi$ and $\phi\phi\,\rightarrow\,\chi\chi$. As a result, the general evolution, at least until chemical equilibrium, is towards forming BEC. All these factors must be thoroughly considered in future research to obtain a more complete and accurate picture of the evolution of condensation.

\subsection{Achieving Finite Number Density of $\phi$ Particles} 

The other condition for BEC is to realize the macroscopic nature of BEC. To this end, the number density of $\phi$ particles, i.e., $n_\phi$, should achieve a remarkable finite value in the presence of cosmological expansion.
Let us clarify this point as follows: By use of the relation $\tilde{n}=\int \tilde{f}d^3\tilde{p}$ by using Eq.(\ref{dist}) for the initial times of the transition $\chi\chi\rightarrow \phi\phi$, where $\tilde{f}_3$ and $\tilde{f}_4$ are negligible, we obtain
\begin{equation}
\frac{d\tilde{\eta}_4(\eta)}{d\eta}=\tilde{\beta}\tilde{n}_1\tilde{n}_2\tilde{\sigma}\tilde{v},\label{a}
\end{equation}
where $\tilde{\beta}$ is a constant that corresponds to average effective depth of the collisions, $\tilde{\sigma}$ is the total cross-section of the process, $\tilde{v}$ is the average velocity of two inital particles in the spaced defined by ($\ref{met1}$). In Eq.(\ref{a}), the total effective cross-section for the transitions with $\tilde{m}_\chi\ll \tilde{m}_\phi$ is given by
\begin{equation}
\tilde{\sigma}=\frac{(2\pi)^4}{4\sqrt{(\tilde{p}_1.\tilde{p}_2)^2-\tilde{m}_\chi^4}}\int\int \delta^{(4)}(\tilde{p_1}+\tilde{p_2}-\tilde{p_3}-\tilde{p_4})|\tilde{M}|^2\frac{d^3 \vec{\tilde{p_3}}}{\tilde{E}_3}\frac{d^3 \vec{\tilde{p_4}}}{\tilde{E}_4}\simeq\bigg(\frac{\tilde{\mu}}{\tilde{m}_\phi}\bigg)^4\frac{|\vec{\tilde{k}}|}{64 \pi |\vec{\tilde{p}}|^2 \tilde{m}_\phi},\label{cs}
\end{equation}
where we have used the definition of the transition matrix to find
\begin{equation}
\tilde{M}=\tilde{\mu}^2\bigg[\frac{1}{(\tilde{p}_1-\tilde{p}_3)^2+\tilde{m}_\phi^2}+\frac{1}{(\tilde{p}_1-\tilde{p}_4)^2+\tilde{m}_\phi^2}\bigg]\simeq\bigg(\frac{\tilde{\mu}}{\tilde{m}_\phi}\bigg)^2,
\end{equation}
which is obtained by considering the center of mass frame with the assumptions $\vec{\tilde{p}}^2\gg\tilde{m}_\chi^2$ and  $\vec{\tilde{k}}^2\ll\tilde{m}_\phi^2$.

To see the curved space effect on the number density, it is needed to convert the form of Eq.(\ref{a}) into the form defined by the space (\ref{frw}) as follows. Because we have $\tilde{f}_i=f_i$ (where i=$\phi, \chi$) followed by 
\begin{equation}
\tilde{f}_i\,=\,\tilde{f}(\vec{\tilde{p}}_i,\eta)\,=\,\frac{d\tilde{N}(\eta)}{d^3\tilde{p}_{(i)}d^3\tilde{x}_{(i)}}\,=\,\frac{d\tilde{N}(\eta)}{d^3p_{(i)}\,d^3x_{(i)}}
\,=\,\frac{dN(t)}{d^3p_{(i)}\,d^3x_{(i)}}\,=\,f(\vec{p}_i,t)\,=\,f_i\;, \label{s1}
\end{equation}
we get $\tilde{n}_i=a^3 n_i$ by the definition of number density just given above (here the sub-index $(i)$ refers to the i'th particle, and we have used in Eq.(\ref{s1}) $\vec{p}=\frac{1}{a}\vec{\tilde{p}}$, $\vec{x}=a\vec{\tilde{x}}$ where $x$ is the physical length scale). 

At this point, we should notice that there are three different possible cases for $m_\chi^2$ in Eq.(\ref{effm}):
\\
i)  $\frac{\dot{a}^2}{a^2}+\frac{\ddot{a}}{a}\,\leq\,m_\chi^2$ and $\frac{\dot{a}^2}{a^2}+\frac{\ddot{a}}{a}\,\geq\,0$,\\
ii)  $\frac{\dot{a}^2}{a^2}+\frac{\ddot{a}}{a}\,\leq\,m_\chi^2$ and $\frac{\dot{a}^2}{a^2}+\frac{\ddot{a}}{a}\,<\,0$,\\
iii) $\frac{\dot{a}^2}{a^2}+\frac{\ddot{a}}{a}\,>\,m_\chi^2$ 
\\
The case iii) above should be excluded to get rid of troublesome tachyons. Therefore the cases i) and ii) remain as the only safe choices. In the case of Eq.(\ref{t9a3}), i) implies that $\frac{\dot{a}^2}{a^2}+\frac{\ddot{a}}{a}\,=\,\xi^2(2-s)\,a^{-2s}\,\geq\,0$, i.e., $s\,\leq\,2$ while ii) implies that $\xi^2(2-s)\,a^{-2s)}\,<\,0$, i.e., $s\,>\,2$. Most of the straightforward cosmic periods that are physically significant - namely, the radiation, matter, and cosmological constant eras - correspond to $s\,\leq\,2$ while the only physically interesting era for the case $s\,>\,2$ is a possible stiff matter dominated era where $s=3$. Therefore we consider the case i) here while we study the extreme case $|\frac{\dot{a}^2}{a^2}+\frac{\ddot{a}}{a}|\,\gg\,m_\phi^2$ of $s\,>\,2$ in the Appendix C to see the basic implications of ii).

For the case i) above we have $|\frac{\dot{a}^2}{a^2}+\frac{\ddot{a}}{a}|\,\ll\,m_\phi^2$ since $m_\chi^2\,\ll\,m_\phi^2$. This, in turn, implies that
$\sigma\,=\,a^2\tilde{\sigma}\simeq\,a^2\left(\frac{\tilde{\mu}}{\tilde{m}_\phi}\right)^4\frac{|\vec{\tilde{k}}|}{64\pi\,\vec{\tilde{p}}^2\tilde{m}_\phi}\,\propto\,a$ since $\tilde{m}_\phi\,\simeq\,a\,m_\phi$, $\tilde{\mu}\,\simeq\,a\,\mu$,  and $|\vec{\tilde{p}}|$, $|\vec{\tilde{k}}|$ are independent of redshift by Eq.(\ref{t1y}). For the relative velocity of the particles, $\vec{\tilde{v}}$, we have $|\vec{\tilde{v}}|=\frac{|d\vec{\tilde{r}}|}{d\eta}=\frac{|d\tilde{r}|/a}{dt/a}=|\vec{v}|=v_0$. Thus, with these conversions by using $n_i(t)=\frac{C_i (t)}{a^3(t)}$, Eq.(\ref{a}) results in two equations
\begin{equation}
\frac{\dot{C}_\chi}{a^3}=-\beta\bigg(\frac{C_\chi}{a^3}\bigg)\sigma_0 v_0 a \ \ \ \ \ \ \frac{\dot{C}_\phi}{a^3}=\beta\bigg(\frac{C_\chi}{a^3}\bigg)\sigma_0 v_0 a. \label{CCC}
\end{equation}
Integrating out the first equation for $C_\chi$ and using it in the second equation to obtain $C_\phi$ with  considering cosmologically relevant cases defined by $H=\xi a^{-s}$, we get
\begin{equation}
C_\phi=C_1-\frac{C_1}{\frac{C_1\beta \sigma_0 v_0}{(s-2)\xi}(a^{s-2}-a_1^{s-2})+1},\label{C}
\end{equation}
where $C_1=C_\chi (t_1)$ is the value of $C_\chi$ at the start of the conversion of $\chi$s to $\phi$s. Eq.(\ref{C}) can be examined for two different cases:
\begin{itemize}
\item At initial times $t\simeq t_1$, it may be approximated as
\end{itemize}
\begin{equation}
C_\phi\simeq\frac{C_1\beta \sigma_0 v_0}{(|s-2|)\xi}a^{|s-2|}\bigg[1-\bigg(\frac{a_1}{a}\bigg)^{|s-2|} \bigg] \ \ \ \text{for} \ \ \ s-2>0,\label{C1}
\end{equation}
\begin{equation}
C_\phi\simeq\frac{C_1\beta \sigma_0 v_0}{(|s-2|)\xi}a_1^{-|s-2|}\bigg[1-\bigg(\frac{a_1}{a}\bigg)^{|s-2|} \bigg] \ \ \ \text{for} \ \ \ s-2<0\label{C2}.
\end{equation}
Eq.(\ref{C1}) and Eq.(\ref{C2}) imply that $C_\phi$ initially reaches higher values by the increase of the scale factor $a$ for $s<2$ while $C_\phi$ grows faster for $s>2$. 
\begin{itemize}
\item For the late times $t\gg t_1$, Eq.(\ref{C}) is approximated as
\end{itemize}
\begin{equation}
C_\phi \sim C_1-\frac{(|s-2|)\xi}{\beta \sigma_0 v_0}a^{-|s-2|} \ \ \ \text{for} \ \ \ s-2>0 \ \ \ \text{and} \ \ \ \frac{C_1\beta \sigma_0 v_0}{(|s-2|)\xi}a^{|s-2|}\gg 1
\end{equation}
\begin{equation}
C_\phi \sim C_1-\frac{(|s-2|)\xi}{\beta \sigma_0 v_0}a^{|s-2|} \ \ \ \text{for} \ \ \ s-2<0 \ \ \ \text{and} \ \ \ \frac{C_1\beta \sigma_0 v_0}{(|s-2|)\xi}a_1^{-|s-2|}\gg 1.
\end{equation}
The last equations imply that if the process $\chi\chi\rightarrow\phi\phi$ continues till very late times then $C_\phi$ reaches its maximum value of $C_1$ for $s>2$ while it may be smaller than the value in the case $s<2$. However, the study of $C_\phi$ for late times is not very reliable in that the effects of the processes $\phi\phi\rightarrow\chi\chi$ and their statistics are neglected in these times while they can be ignored at initial times. These equations are reliable only if $n_\phi$ has not reached a large value at late times yet.  To better understand the problem at late times, more thorough research will be required in future.

Finally, it is crucial to examine the scope of applicability of Eq.(\ref{con1}) by considering Eq.(\ref{cs})  in such perturbative calculations (where $\left(\frac{\tilde{\mu}}{\tilde{m}_\phi}\right)\,<\,1$) because $\sigma$ becomes smaller for smaller $\frac{\tilde{\mu}}{\tilde{m}_\phi}$. However, $|\vec{\tilde{k}}|$, $|\vec{\tilde{p}}|$, and $\tilde{m}_\phi$ itself also affect the quantity $\tilde{\sigma}$, as well as  $\frac{\tilde{\mu}}{\tilde{m}_\phi}$. Further, instead of $\sigma$ alone, the more relevant quantity in Eq.(\ref{con1}) is $n\sigma$ (given in $\bigtriangleup t$). In this regard, even in the regime of perturbation, there is a sizeable parameter space where such an effective Minkowski space formulation is valid. For example, one may identify $\phi$ by dark matter and let $\frac{\tilde{\mu}}{\tilde{m}_\phi}=0.1$, $\frac{|\vec{\tilde{k}}|}{|\vec{\tilde{p}}|}=0.01$, $|\vec{\tilde{p}}|\sim\,\tilde{m}_\phi\,c$; then $n_0\,\sim\,\frac{10^{-3}\,eV\,cm^{-3}}{\tilde{m}_\phi\,c^2}$, so by Eq.(\ref{cs}), $n_0\tilde{\sigma}_0\simeq\,\frac{10^{-3}\,eV\,cm^{-3}}{\tilde{m}_\phi\,c^2}\left(\frac{\tilde{\mu}}{\tilde{m}_\phi}\right)^4\left(\frac{|\vec{\tilde{k}}|}{|\vec{\tilde{p}}|}\right) \left(\frac{\hbar\,c}{\tilde{m}_\phi\,c^2}\right)^2$ $\sim$ $10^{-19}\times\left(\frac{eV}{\tilde{m}_\phi}\right)^3\,cm^{-1}$ which satisfies  $\frac{H_0}{n_0\sigma_0\,v}\simeq\,\frac{10^{-28}cm^{-1}}{n_0\sigma_0(v/c)}\,\ll\,1$ if $\frac{v}{c}\sim\,1$ and $\tilde{m}_\phi\,\ll\,10^3\,eV$ (including the phenomenologically interesting case of ultra light dark matter). We should here notice that in  the calculations, we cannot use $|\vec{\tilde{k}}|\,\simeq\,0$ because doing so would mean a complete BEC, where the entire system acts as a single quantity, and in this case the standard description of scattering regarding single particles in quantum field theory cannot be considered. In fact, our primary goal in this research is to indicate how curved space effects promote the formation of BEC in a model that incorporates $\phi^2\chi$ kind of interaction terms instead of showing that an effective Minkowski space formulation holds true in every scenario in cosmology. We expect that this formulation and this research may give additional insight to understand the formation of BEC in cosmology.

After these results, we can now conclude that via the construction of the effective Minkowski space formulation at the time scale oparticle physics process $\chi\chi\rightarrow\phi\phi$ (induced by $\phi^2\chi$ type of interaction), we have obtained all necessary conditions, i.e., coherence, correlation, and finite number density, for the condensation of $\phi$ particles at initial times in cosmology, where the curved space effect is provided by the scale factor $a$.
\newpage
\chapter{Curved Space and Particle Physics Effects on the Formation of Bose-Einstein Condensation around a Reissner - Nordstr{\o}m Black Hole}
In the previous chapter, we have shown the evolution of $\phi$ particles is towards condensation in a curved space, particularly in cosmology described by FRW metric. A similar approach can also be applied to other curved spaces. To this end, this chapter \cite{Erdem:2021trk} aims to observe if a condensate of the scalar field is realized around a black hole. 

In particular, we consider an RN black hole \cite{Chandrasekhar:1985kt} (a non-rotating black hole with a non-vanishing electric charge) instead of the simpler case of a Schwarzschild black hole. This is because charged particles of charge $q$ (of the same charge as the black hole) with low enough energies $\omega$ with $0\,<\,\omega\,<q\frac{Q}{r_+}$ (where $r_+=M+\sqrt{M^2-Q^2}$ is the radius of the event horizon) can be scattered by RN black holes \cite{Chandrasekhar:1985kt}. On the other hand, in the case of Schwarzschild black holes, there is no scattering from the black hole. Hence, the particle physics processes essential for the formation of condensation cannot be efficiently realized around a Schwarzschild black hole. Therefore, we consider the problem of scalar field condensation around RN black holes as a simple and efficient framework to study the condensation of scalar fields around black holes.

\section{Framework}
The RN metric is defined as
\begin{equation}
ds^2=-fdt^2+f^{-1}dr^2+r^2(d\theta^2+sin^2\theta d\varphi^2),\label{RN}
\end{equation}
where 
\begin{equation}
f=\bigg(1-\frac{2M}{r}+\frac{Q^2}{r^2}\bigg).\label{f}
\end{equation}
Here $M$ and $Q$ are the black hole's mass and charge. 

We assume that initially, there are a homogeneous distribution of a diluted, relativistic $\chi$ fields in this background, and they are converted into the non-relativistic $\phi$ fields by the particle physics processes $\chi\chi\,\rightarrow\,\phi\phi$. Here we consider an interaction term $\chi^*\chi\phi^*\phi$, which will be introduced in Section 6.1.2. 

Most of the collisions of the $\chi$ particles occur between the ones scattered from the black hole and the others coming from infinity. The relativistic particles extract energy from the black hole after being scattered if they obey the condition $m\,<\,\omega\,<\,q\frac{Q}{r_+}$. This type of extracting energy from a black hole is known as superradiance.

In this study, we focus on the radial head-on collisions of the $\chi$ particles and examine the behaviour of the produced $\phi$ particles. In this respect, we first study the motion of the scalar particles in the radial direction at the level of test particles as given in the following subsection.

\subsection{Motion in Radial Direction}

To proceed, we begin with examining the radial behaviours of the fields. To this end, we consider the 1+1 dimensional subspace of the 3+1 dimensional space (\ref{RN})
\begin{equation}
ds^2\,=\,-f\,dt^2\,+\,f^{-1}dr^2. \label{eq:3}
\end{equation}
The Lagrangian for a test particle of mass $m$ in Eq.(\ref{eq:3}) is
\begin{equation}
L\,=\,m\frac{ds}{d\tau}\,=\,m\sqrt{-f\dot{t}^2\,+\,f^{-1}\dot{r}^2}, \label{eq:4}
\end{equation}
where $\dot{t}=\frac{dt}{d\tau}$, $\dot{r}=\frac{dr}{d\tau}$ with $d\tau=\sqrt{-ds^2}$. The Lagrange equation for the coordinate $t$ results in the conservation of $h$, i.e., the total energy (including the potential energy) per unit mass of a test particle, namely
\begin{equation}
h\,=\,f\,\dot{t}\,=\,\frac{1}{m}\frac{\partial\,L}{\partial\,\dot{t}}\,=\,C\,=\,\mbox{constant}, \label{eq:5}
\end{equation}
where we have used
\begin{equation}
\left(-f\dot{t}^2+f^{-1}\dot{r}^2\right)\,=\,-1 \label{eq:6}
\end{equation}
for massive particles. Eq.(\ref{eq:6}) in combination with Eq.(\ref{eq:5}) results in
\begin{equation}
\dot{r}^2\,=\,C^2\,-\,\left(1\,-\,\frac{2M}{r}\,+\,\frac{Q^2}{r^2}\right).
\label{eq:7}
\end{equation}
As $r\,\rightarrow\,\infty$, Eq.(\ref{eq:7}) becomes
\begin{equation}
\dot{r}_\infty^2\,=\,C^2\,-\,1.
\label{eq:8}
\end{equation}
Here we let $\chi$ particles that are initially at rest fall towards the black hole from a significant distance that may be approximated by infinity, which implies $C_\chi=1$ by Eq.(\ref{eq:8}). Moreover, at the center of mass frame, the conservation of energy for a given process $\chi\chi\,\rightarrow\,\phi\phi$ implies $m_\chi C_\chi=m_\phi C_\phi$ or $C_\phi=C_\chi\frac{m_\chi}{m_\phi}$. Because $m_\chi<m_\phi$, we then get $C_\phi<C_\chi$. This result means that the produced $\phi$ particles scattered by the black hole cannot reach infinity. Therefore, they can reach only a finite distance from the black hole, which is the greater root of $(1-C_\chi^2)r_0^2-2Mr_0+Q^2=0$ for $Q\,<\,M$, which may be found by equating $\dot{r}$ in Eq.(\ref{eq:7}) to zero. In this respect, there will be a belt of $\phi$ particles with zero or almost zero momenta around the black hole. This provides a suitable condition for the formation of Bose-Einstein condensation. This result is obtained at the level of test particles. We now search for the same result at the level of field theory in the following sections.

\subsection{The Field Equations for the Scalars}

We consider the following action with the background of Eq.(\ref{RN}) 
\begin{equation}
S=\int\;d^4x\,\sqrt{-g}\,\{-g^{\mu\nu}\left[D_\mu\phi\,\left(D_\nu\phi\right)^*\,+\,
D_\mu\chi\,\left(D_\nu\chi\right)^*\right]\,-\,m_\phi^2\left|\phi\right|^2\,-\,
m_\chi^2\left|\chi\right|^2\,-\,\lambda\,\phi^*\phi\chi^*\chi\},
\label{eq:9a}
\end{equation}
where $D_\mu\,=\,\partial_\mu\,+\,iq\,A_\mu$ with $q$ being the electric charge of the scalar field and $A_\mu\,=\,\left(\frac{Q}{r}, 0, 0, 0\right)$ denoting the electric field of the black hole. Here we let both $\chi$ and $\phi$ have the same charge q. In Eq.(\ref{eq:9a}), due to the small coupling constant of electromagnetic interactions and we have assumed the low scalar particle density, we have ignored their impact on scalar particle interactions.


If we take $\lambda$ small in comparison with the other terms in Eq.(\ref{eq:9a}), the corresponding field equation for $\phi$ is then given by
\begin{equation}
D_\mu\,D^\mu\,\phi\,-\,m_\phi^2\,\phi\,=\,0.
\label{eq:13}
\end{equation}
The corresponding field equation for $\chi$ may be obtained by replacing $\phi$ in Eq.(\ref{eq:13}) with $\chi$. Although we have neglected the term with the coefficient $\lambda$ in the field equations to obtain approximate solutions, the effect of the interaction is still imposed by the previously obtained expression
\begin{align}
C_\phi=C_\chi\frac{m_\chi}{m_\phi}.\label{con.}
\end{align}
Using the ansatz \cite{RN-superradiance}
\begin{equation}
\phi_\omega\left(t,r,\theta,\varphi\right)\,=\,\sum_{l,m}e^{-i\omega\,t}Y_l^m\left(\theta,\varphi\right)\frac{\psi_\omega\left(r\right)}{r},
\label{c1}
\end{equation}
the field equation (\ref{eq:13}) results in
\begin{equation}
f^2\frac{d^2}{dr^2}\psi_\omega+ff^\prime\frac{d}{dr}\psi_\omega+\left[\left(\omega-\frac{qQ}{r}\right)^2-V\right]\psi_\omega\,=\,0,
\label{c2}
\end{equation}
where $\prime$ denotes derivative with respect to r, and
\begin{equation}
V= f\left(\frac{l(l+1)}{r^2}+\frac{f^\prime}{r}+m_\phi^2\right).
\label{c3}
\end{equation}
In the case of the radial motion for which $l=0$ and by considering $\frac{f^\prime}{r}\,\ll\,m_\chi^2\,<\,m_\phi^2$ (which is satisfied for a wide range of $m_\chi^2$ if $r$ is not close to $r_+$, as will be mentioned in Section 6.2.2), Eq.(\ref{c2}) reduces to
\begin{equation}
\frac{d^2\psi_\omega}{dr_*^2}\,+\,\left[\left(\omega-\frac{qQ}{r}\right)^2\,-\,\tilde{m}_\phi^2\right]\,\psi_\omega\,=\,0,
\label{eq:15}
\end{equation}
where $dr_*\,=\,f^{-1}dr$, $\tilde{m}_\phi^2\,=\,f\,m_\phi^2$. 
Eq.(\ref{eq:15}) is the same expression as the field equation obtained from the corresponding 1+1 dimensional form of the action (\ref{eq:9a}) (see Appendix D).

\section{A Special Solution}
\subsection{Derivation}

In the hope of obtaining a wave-like solution to Eq.(\ref{eq:15}) we consider the following type of solution
\begin{align}
\psi_\omega=e^{isr_*}g(r_*), \label{H1}
\end{align}
where $s^2=\omega^2-m^2$.
Eq.(\ref{eq:15}), after using Eq.(\ref{H1}), becomes
\begin{align}
\frac{g^{\prime\prime}}{g}+2is\frac{g^{\prime}}{g}-\frac{2(qQ\omega-m^2M)}{r}+\frac{(q^2-m^2)Q^2}{r^2}=0,\label{H2}
\end{align}
where $\prime$ denotes the derivative with respect to $r_*$.
We try the following choice
\begin{equation}
\frac{g^{\prime}}{g}=\frac{\alpha_1}{r}+\frac{\beta_1}{r^2}+\frac{\gamma_1}{r^3}, \label{H3a}
\end{equation}
\begin{equation}
\frac{g^{\prime\prime}}{g}=\frac{\alpha_2}{r}+\frac{\beta_2}{r^2}+\frac{\gamma_2}{r^3}, \label{H3b}
\end{equation}
where $\alpha_1$, $\beta_1$, $\gamma_1$, $\alpha_2$, $\beta_2$, $\gamma_2$ are some constants whose explicit forms will be derived below.
Hence, if such a solution exists, then Eq.(\ref{H1}) becomes
\begin{align}
\psi_\omega=e^{isr_*}\,\exp{\left[\int_{r_0}^r{\frac{\alpha_1\,r}{r^2-2Mr+Q^2}dr}+\int_{r_0}^r{\frac{\beta_1}{r^2-2Mr+Q^2}dr}+\int_{r_0}^r{\frac{\gamma_1}{r^3-2Mr^2+Q^2r}dr}\right]}.
\label{H1a}
\end{align}
The explicit form of the solution $\psi_\omega$ for a given $M$ and $Q$ may be obtained as follows. Eq.(\ref{H3a}) and Eq.(\ref{H3b}) solve Eq.(\ref{H2}) if (see Appendix E)
\begin{equation}
\alpha_1=\frac{9}{5}-\frac{9}{20}\bigg(\frac{M}{Q}\bigg)^2 \,, \ \ \ \ \ \ \beta_1=-\frac{9}{2}M \,, \ \ \ \ \ \ \gamma_1=3Q^2\,, \label{H5}
\end{equation}
\begin{align}
\alpha_2=0 \,, \ \ \ \ \ \ \beta_2=\bigg[\frac{9}{5}-\frac{9}{20}\bigg(\frac{M}{Q}\bigg)^2\bigg]\bigg[\frac{4}{5}-\frac{9}{20}\bigg(\frac{M}{Q}\bigg)^2\bigg] \,, \ \ \ \ \ \ \gamma_2=2\bigg[\frac{-9}{5}+\frac{63}{40}\bigg(\frac{M}{Q}\bigg)^2 \bigg]M.  \label{H6}
\end{align}
Inserting Eq.(\ref{H5}) and Eq.(\ref{H6}) into Eq.(\ref{H2}) and using $s^2=\omega^2-m^2$, we get three equations for three unknown quantities $\omega$, $q$, $m$ for a given $M$ and $Q$:
\begin{align}
(m^2-\omega^2)-\frac{M^2}{9Q^4}\bigg(\frac{9}{5}-\frac{63}{40}\frac{M^2}{Q^2}  \bigg)^2=0, \label{H7}
\end{align}
\begin{align}
-\frac{3M^2}{Q^2}\bigg(\frac{9}{5}-\frac{63}{40}\frac{M^2}{Q^2}\bigg)+\bigg(\frac{9}{5}-\frac{9}{20}\frac{M^2}{Q^2} \bigg) \bigg( \frac{4}{5}-\frac{9}{20}\frac{M^2}{Q^2} \bigg) +(q^2-m^2)Q^2=0, \label{H8}
\end{align}
\begin{align}
\frac{2M}{3Q^2}\bigg(\frac{9}{5}-\frac{63}{40}\frac{M^2}{Q^2}\bigg)\bigg(\frac{9}{5}-\frac{9}{20}\frac{M^2}{Q^2}\bigg)-2(qQ\omega-m^2M)=0. \label{H9}
\end{align}
By plugging Eq.(\ref{H7}) and Eq.(\ref{H5}) into Eq.(\ref{H1a}), we obtain the following solution
\begin{small}
\begin{eqnarray}
\psi_\omega&=&
\exp \Bigg\lbrace \left(\frac{\pm\,3 M\left(7 M^2-8 Q^2\right)}{40 Q^4 \sqrt{Q^2-M^2}}\right)
\nonumber \\
&&
\times \, \Bigg[ \left(2 M^2-Q^2\right)\Bigg( \tan ^{-1}\bigg(\frac{M-r}{\sqrt{Q^2-M^2}}\bigg)- \tan ^{-1}\bigg(\frac{M-\text{r}_0}{\sqrt{Q^2-M^2}}\bigg)\Bigg) \nonumber \\
&&-\sqrt{Q^2-M^2} \Big(M \log \left(-2 M r+Q^2+r^2\right)-M \log \left(-2 M \text{r}_0+Q^2+\text{r}_0^2\right)+r-\text{r}_0\Big)\Bigg]\ \nonumber \\
 &&
 + \ \left(\frac{3(6 M^3-4 M Q^2 )}{40 Q^2 \sqrt{Q^2-M^2}}\right)
 \Bigg( \tan ^{-1}\bigg(\frac{M-r}{\sqrt{Q^2-M^2}}\bigg)-\tan ^{-1}\bigg(\frac{M-\text{r}_0}{\sqrt{Q^2-M^2}}\bigg)\Bigg) \nonumber \\
 &&-\frac{3\left(3 M^2+8 Q^2\right)}{40 Q^2}\Big( \log \left(-2 M r+Q^2+r^2\right)-\log \left(-2 M
   \text{r}_0+Q^2+\text{r}_0^2\right)\Big)\nonumber \\
 &&+3\log (r)- 3\log (\text{r}_0)\Bigg\rbrace,
 \label{H1b}
\end{eqnarray} 
\end{small}where the exponential function can be either an increasing or decreasing real exponential in $r$ depending on the values of $M$ and $Q$. The increasing one corresponds to the unphysical case. Therefore, we consider the decreasing one for the implication of the solution $(\ref{H1b})$, as will be examined in the next section. Further, if we solve Eq.(\ref{H7}), Eq.(\ref{H8}), and Eq.(\ref{H9}) for $m$, $\omega$, and $q$ by Mathematica \cite{Mathematica}, we get the following results:
\begin{eqnarray}
m^2_1 =-\frac{9 \left(7 M^4+2 M^2 Q^2\right)}{40 Q^6}, \ \ \ \ \
m^2_2=\frac{9 \left(7 M^2-8 Q^2\right)^2 \left(4 Q^2-3 M^2\right)}{1600 Q^6
   \left(Q^2-M^2\right)},
\label{eq:m2}
\end{eqnarray}
\begin{eqnarray}
\omega_{1,2}=\mp\frac{3 i M \left(7 M^2+12 Q^2\right)}{40 Q^4}, \ \ \ \ \
\omega_{3,4}=\mp\frac{3 i \left(7 M^4-22 M^2 Q^2+16 Q^4\right)}{40 \sqrt{Q^8 \left(M^2-Q^2\right)}},
\label{eq:w2a}
\end{eqnarray}
\begin{eqnarray}
q_{1,2}=\mp\frac{3 i \left(17 M^2-8 Q^2\right)}{20 Q^3}, \ \ \ \ \
q_{3,4}=\pm\frac{3 i M Q \left(27 M^2-28 Q^2\right)}{40 \sqrt{Q^8 \left(M^2-Q^2\right)}},
 \label{eq:q12}
\end{eqnarray}
where $\{m^2_1,\ \omega_{1,2},\ q_{1,2} \}$ and $\{m^2_2,\ \omega_{3,4},\ q_{3,4} \}$ are two different set of solutions.
\subsection{Viability of the Solution}
To examine the viability of our solution, we rescale all the expressions given in Eq.(\ref{eq:15}) in terms of the multiples of $M$ or $M_\circ$ (the mass of the sun) as given below:
\begin{align}
&\overline{r}_*=\frac{c^2}{GM_\circ}r_*,~~ \overline{r}=\frac{c^2}{GM_\circ}r, \label{bar1}
\\&
\overline{\omega}=\omega\frac{GM_\circ}{c^3}=\bigg(\frac{\omega}{s^{-1}}\bigg)5\times 10^{-6},\label{bar2}
\\&
\overline{m}=m\frac{GM_\circ}{\hbar c} =\bigg( \frac{m}{kg}\bigg)\,4.5\times 10^{45}, \label{bar3}
\\&
\overline{M}=\frac{M}{M_\circ}=\bigg(\frac{M}{kg}\bigg)2\times 10^{-30}, \label{bar4}
\\&
\overline{Q}=\frac{Q\sqrt{k}}{M_\circ \sqrt{G}}=\bigg(\frac{Q}{C}\bigg)5,7\times 10^{-21}, \label{bar5}
\\&
\overline{q}=q\frac{M_\circ \sqrt{Gk}}{\hbar c}=\bigg(\frac{q}{C}\bigg) 4,8 \times 10^{55}, \label{bar6}
\end{align}
where $k$ is the Coulomb's constant and we have explicitly written $G$, $c$, $\hbar$, $k$ (that we had set to 1) to see the phenomenological contents of these quantities. With these redefinitions, Eq.(\ref{eq:15}) is given by (see Appendix F)
\begin{align}
\frac{d^2\psi_\omega}{d\overline{r}_*^2}+\bigg[\big(\overline{\omega}-\frac{\overline{q}\overline{Q}}{\overline{r}}\big)^2-\overline{m}_\phi^2\left(1-\frac{2\overline{M}}{\overline{r}} + \frac{\overline{Q}^2}{\overline{r}^2}\right) \bigg]\psi_\omega=0.
\label{scaled_eq}
\end{align}
We should here notice that the rescaled form of our field equation (\ref{scaled_eq}) contains the barred forms of Eqs.(\ref{eq:m2})-(\ref{eq:q12}). In this regard, we have obtained the following results:
\begin{itemize}
\item In Eq.(\ref{eq:q12}) $\bar{q}_{1,2}$ takes the imaginary values. Then the solution set $\{\bar{m}^2_1,\ \bar{\omega}_{1,2},\ \bar{q}_{1,2} \}$ does not correspond to the physical one. Therefore only the other solution set $\{\bar{m}^2_2,\ \bar{\omega}_{3,4},\ \bar{q}_{3,4} \}$ is the physical one if $|\bar{Q}|>\bar{M}$.
\item We also see that there are no physical solutions for $\bar{q}=0$ (or $q=0$) because $\bar{q}=0$ corresponds to the case of $\bar{M}>|\bar{Q}|$, which is not in the physical range of our set of solution. Therefore, in agreement with our model, the solution given in this study is relevant only for $q\neq 0$.
\item The allowed ranges are $1.44<m^2 Q^2<\infty$, $1.2<\omega Q<\infty$, $0<qQ<\infty$, where $m^2 Q^2$ and $\omega Q$ are in the order of $1$ for most of the values of $M$ and $Q$. $qQ$ is in the order of $10^{-1}$ for most of the values of $M$ and $Q$.
\end{itemize}

By taking into account our set of solutions, as discussed above, we are now in a position to check if our wave solution (\ref{H1b}) results in the wave profiles that are expected from a test particle treatment where we examine the tendency of the condensation (the accumulation of the $\phi$ particles with zero momenta) around the black hole. By considering our physically relevant set $|\bar{Q}|>\bar{M}$, we have plotted $\psi_\omega$ plots for various values of its parameters by using a Mathematica code \cite{Mathematica}. We have found mainly two types of behaviours for $|\psi_\omega|$, namely, exponentially increasing with increasing $r$, exponentially decreasing with increasing $r$ with a local peak. We discard the exponentially increasing ones because they are not physical. Finally, as an example we obtain the desired Figure \ref{Wp1} (in agreement with our test particle treatment), as given below, with $\bar{Q}=17$ for a given $\bar{M}=10$. It is also observed that for some $M$, $Q$ pairs the absolute value of $\psi_\omega$ peaks at some $Q$ values, for example, as in Figure \ref{Wp2}. 
We have also checked whether these wave profiles correspond to the $\phi$ particles or not. For a given value $\bar{Q}=17$ and $\bar{M}=10$ we obtain $\frac{\omega}{m}\,<\,1$, which implies the behaviour of the wave profile of the form of Figure \ref{Wp1} for the scalar field $\phi$. Therefore, we conclude that the wave profile in Figure \ref{Wp1} is consistent with the accumulation of the scalar particles at some distance from the black hole suggested by the test particle behaviour as discussed in Section 6.1.1.
\begin{figure}[H]
\begin{center}
\includegraphics[scale=0.8]{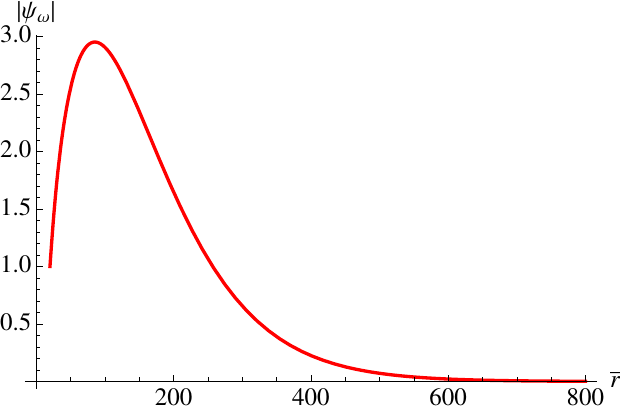} 
\caption{$|\psi_\omega|$ versus $\bar{r}$ graph for $\bar{M}=10$,\ $\bar{Q}=17$,\ $\bar{r}_0=20$ }
\label{Wp1}
\end{center}
\end{figure}
\begin{figure}[H]
\begin{center}
\includegraphics[scale=0.8]{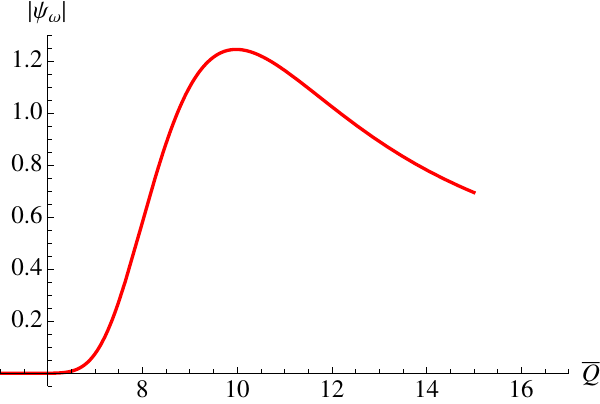} 
\caption{$|\psi_\omega|$ versus $\bar{Q}$ graph for $\bar{M}=10$,\ $\bar{r}=20$,\ $\bar{r}_0=30$ }
\label{Wp2}
\end{center}
\end{figure}

We realize that the $\frac{f^\prime}{r}$ term can be ignored with respect to $m_\chi^2$ for phenomenologically feasible values of the parameters as can be seen below
\begin{equation}
\frac{f^\prime}{\bar{r}}= \frac{2}{\bar{r}^2}\left(\frac{\bar{M}}{\bar{r}}-\frac{\bar{Q}^2}{\bar{r}^2}\right),
\label{c4x}
\end{equation}
where we have used $\bar{r}$, $\bar{M}$, $\bar{Q}$ in place of $r$, $M$, $Q$ by use of the Eqs.(\ref{bar1})-(\ref{bar6}). We here consider the outside of the horizon, i.e., $\bar{r}>\bar{r}_+$. In the case of small values of $\bar{r}$, where $\bar{r}\simeq\bar{M}$, we have $\frac{\bar{M}}{\bar{r}}\simeq 1$. Then the term $\left(\frac{\bar{M}}{\bar{r}}-\frac{\bar{Q}^2}{\bar{r}^2}\right)$ ensures that $\frac{f^\prime}{\bar{r}}\rightarrow 0$  with respect to $m_\chi^2$ as $|\bar{Q}|\rightarrow \bar{M}$. In the case of large values of $\bar{r}$, the term $\frac{2}{\bar{r}^2}$ ensures $\frac{f^\prime}{\bar{r}}\sim\,0$, which is compatible with the wave profile \ref{Wp1}.

For the sake of simplicity, to prevent the fields in this study from altering the space's geometry, we considered their low energy densities. As a result, it is quite challenging to find these dark matter candidates. On the other hand, as long as we are at a sufficiently large distance from the black hole to satisfy $\frac{f^\prime}{r}\,\ll\,m_\chi^2\,<\,m_\phi^2$ and the spherical shape of the wave profile is retained, i.e., $l=0$, we do not anticipate a significant change in the form of the geometry even when the energy density of the fields is raised. In that situation, the RN metric will still characterize the compact object's geometry. In this case, the gravitational effect of these field(s), such as their impact on the rotation curve(s) of their galaxies, can be used to detect the presence of the scalar fields charged with a dark electric charge surrounding an RN black hole (charged with the same dark electric charge) (while such an analysis will have additional, non-trivial points to be addressed). Each of these topics requires a distinct and detailed analysis. All of these aspects must be taken into account in thorough, independent, and in-depth future analysis in order to draw a firm and reliable conclusion regarding the impact of non-negligible energy densities of the scalar fields. In fact, literature has addressed the issue of galaxy rotation curves in the context of Bose-Einstein condensate dark matter, which is made up of a self-gravitating scalar dark matter gas cloud \cite{Boehmer:2007um,Craciun:2020twu}. There are also studies in the literature that consider the source of gravitational as a point source \cite{Chavanis:2019bnu} while they do not discuss the predictions for galaxy rotation curves. Most of these studies are non-relativistic while there are also studies in the context of general relativity \cite{Boehmer:2007um,Castellanos:2017tka}. \cite{Boehmer:2007um} studies the problem through the postulation of a mass density for the scalar field.  On the other hand, \cite{Castellanos:2017tka} studies the problem in the context of charged black holes including RN black holes while the scalar field is taken to be neutral. This paper also finds a localized solution of the scalar field while the explicit form of the solution is derived only at the limits $r_\star \pm \infty$. In other words, some studies have the same topics of interest as the current paper. The current work still includes some novel aspects, such as a model of two interacting scalars that leads to $C_\phi=C_\chi\frac{m_\chi}{m_\phi}$ (which may be considered as the mechanism behind the Bose-Einstein condensation in this context) and the solution given in Section 6.2 is a new analytical solution. 

In this study, it is also possible to consider $\phi$ and/or $\chi$ particles as the only source or as one of the sources of DM in the universe. The idea that this mechanism is the only source of DM cannot be completely ruled out due to the special nature of both the source and DM. An example to a possible scenario may be given that $\chi$ particles, which are non-relativistic at considerable distances from the black hole, may operate as the component of DM/DE at cosmic scales, whereas $\phi$ particles are thought to act in the component of DM at galactic scales. That if such a situation is feasible or not  may be considered in future in a separate study. Another relevant non-trivial point is that a constant pair of $M$, $Q$ gives a constant pair of $m$, $q$ as seen in Eq.(\ref{eq:m2}) and Eq.(\ref{eq:q12}); that is to say, for constant values of $m_\phi$, $q_\phi$ the number of $M$, $Q$ pairs is unchanged. At first sight, one may infer that these fixed numbers of values lead to a particular case of the solitonic solution for the scalar DM behaviour around the black hole, so this study is insufficient to be regarded as an important source of DM in cosmology. Given the following discussions, such a justification may not be valid. The values of $M$ and $Q$ of RN black holes are expected to be changed by accretion, i.e., by swallowing the particles $\chi$ and $\phi$. Because the charge is an additive quantum number, the value of $Q$ increases  proportional to the number of swallowed particles. However, this may not be the same for the value of $M$. The value of $M$ is expected to be smaller than the number of swallowed particles times their masses since the effect of the gravitational potential between the particles should also be considered. At the initial times of the accretion, due to the small number density of the particles, the effect of the gravitational potential between them can be ignored. Therefore, it is expected that  $\frac{\Delta Q}{\Delta M}\simeq \frac{q}{m}$ at these times but $\frac{\Delta Q}{\Delta M} > \frac{q}{m}$ at later times. Hence, by the time the values of $Q$ and $M$ increase and at some time the pair of $Q$ and $M$ reaches some values so that the solitonic solution is formed. Then, $Q$ and $M$ proceed to evolve in order for the formation of the localized wave form. However, at later times this wave profile may be dispersed due to the ongoing accretion and at this time the effect of the gravitational potential between the particles may be significant because of the non-negligible number densities of $\phi$ particles. Thus its mass and charge will effectively change depending on the distance from the black hole. This indicates that at any radial distance the black hole's mass and charge will always meet the criteria required for forming the solitonic solution. This, as well as the gravitational attraction between the particles, tends to prevent the wave profile's dispersion. However, this point requires a more detailed further study to ensure these results. Further, if we let $C_\chi$ take values in an interval about 1, i.e., if we  take $\omega$ to be in an interval about $m_\chi$ rather than taking a sharp value for $\omega$, there will be then more chance that a value of $(m_\phi^2-\omega^2)$ in the interval provides Eq.(\ref{H7}) for a given set of $M$, $Q$. Given these discussions and the novelty of the mechanism and the solution mentioned here, in our opinion, this new mechanism and the solution need in-depth follow-up research that examines all of their characteristics and potential in further studies.

\newpage
\chapter{A Metric for Gravitational Collapse around a Schwarzschild Black Hole}
This chapter aims to be the first step towards a more realistic picture for examining BEC around an accreting black hole - fluid system \cite{Erdem:2020jgi}. To this end, we consider the problem of gravitational collapse of a fluid under the effect of a small Schwarzschild black hole. We postpone the study of the behaviour of BEC in such a background to future studies. In this chapter, we start with the most general spherically symmetric metric and its corresponding non-zero components of the Einstein tensor, where $G^0_{\;1}$ is responsible for the accretion, i.e., radial inward power flux. To ensure a Schwarzschild and Robertson-Walker metrics as limiting cases of our study, we first derive a condition in the case of non-accretion, i.e., $G^0_{\;1}=0$. We next propose the simplest, most plausible mass function that describes twice the total energy in a spherically collapsing shell at some distance and time. Considering the condition derived in limiting cases and the proposed mass function, we derive a metric for gravitational collapse around a Schwarzschild black hole. We ensure that this metric corresponds to the case of $G^0_{\;1}\neq 0$. In the limiting cases of $r\,\rightarrow\infty$ or $M\,\rightarrow 0$, we observe that $G^0_{\;1}\,\rightarrow 0$ and the metric derived is reduced to the Robertson-Walker metric. Our discussion on these limits shows that the accretion is due to the presence of the black hole instead of the behaviour of the fluid itself. We also discuss the apparent horizon and the singularities of our solution in a particular case of dust. We see that all the singularities are hidden behind the apparent horizon. Additionally, the collapse of dust allows negative pressures implying that the collapsing dust may induce dark energy-like behaviour. 

We organize this chapter as follows. We first introduce our framework, where we derive a condition that ensures Robertson-Walker and Schwarzschild's metrics are particular solutions. In this section, we also propose a mass function. With the condition and mass function, we derive a metric for accreting Schwarzschild black hole-fluid system. In the next section, we discuss some implications of the metric where we examine the general behaviour of our solution with its apparent horizons and singularities in the case of dust. In the final section, we comment on the energy density and the pressures for dust collapse.



\section{Framework}
In this section, we first quickly review the background information that is crucial to understanding this research. Then, by requiring consistency between two distinct formulations of the Misner-Sharp mass function for a marginally bound dust, we derive a requirement crucial to understanding the remaining sections of the study. 

\subsection{Preliminaries}
The most general spherically symmetric metric corresponds to \cite{book,Joshi,Bambi}
\begin{equation}
ds^2=-e^{2\lambda}dt^2+e^{2\psi}dr^2+R^2(d\theta^2+\sin^2\theta\, d\phi^2), \label{eq:1}
\end{equation}
where $\lambda=\lambda(r,t)$, $\psi=\psi(r,t)$, $R=R(r,t)$. The non-zero components of Einstein tensor for the metric (\ref{eq:1}) are given by \cite{book,Bambi}
\begin{eqnarray}
&& G^0_{\;0}\,=-\,\frac{F^\prime}{R^2 R^\prime}\,+\,\frac{2\dot{R}\,e^{-2\lambda}}{R\,R^\prime}\left(\dot{R} ^\prime-\dot R \lambda^\prime-\dot \psi R^\prime\right), \label{eq:1a} \\
&& G^1_{\;1}\,=-\,\frac{\dot F}{R^2\dot R }\,-\,\frac{2R^\prime\,e^{-2\psi}}{R\,\dot{R}}\left(\dot{R}^\prime-\dot R \lambda^\prime-\dot \psi R^\prime\right), \label{eq:1b} \\
&& G^0_{\;1}\,=-\,e^{2\psi-2\lambda}\,G^1_{\;0}\,=\,
\frac{2e^{-2\lambda}}{R}\left(\dot{R} ^\prime-\dot R \lambda^\prime-\dot \psi R^\prime\right), \label{eq:1c} \\
&& G^2_{\;2}\,=\,G^3_{\;3}\,=\,\frac{e^{-2\psi}}{R}\,\left[\left(\lambda^{\prime\prime}+{\lambda^\prime}^2- \lambda^\prime\psi^\prime\right)R\,+\,R^{\prime\prime}+ R^\prime\lambda^\prime-R^\prime\psi^\prime\right] \nonumber\\
&&-\,
\frac{e^{-2\lambda}}{R}\,\left[\left(\ddot{\psi}+\dot{\psi}^2- \dot{\lambda}\dot{\psi}\right)R\,+\,\ddot{R}+ \dot{R}\dot{\psi}-\dot{R}\dot{\lambda}\right],
\label{eq:1d}
\end{eqnarray}
where the indices $0,1,2,3$ corresponds to $t,r,\theta,\phi$, and
\begin{equation}
F(r,t)=R(1-e^{-2\psi}R'^2+e^{-2\lambda}\dot R^2) \label{eq:4}
\end{equation}
is the Misner-Sharp mass function \cite{Misner,book}.

\subsection{Derivations of a Condition that Insures Robertson-Walker and Schwarzschild Metrics being Particular Solutions}
In the following, we first calculate the dust's Misner-Sharp mass function. Despite the fact that this mass function and its derivation are widely known \cite{Misner,book}, we shall repeat them in order to get one of our fundamental requirements, namely, Eq.(\ref{eq:11}). Additionally, we shall also generalize the mass function to various fluids.

In case of the non-accreting black hole, i.e., in case of $G^0_{\;1}\,=\,0$ (e.g. as in Lema{\^i}tre-Tolman-Bondi metric), the energy momentum tensor $T^\mu_\nu$ of the 
Einstein equations may be then considered to be of type I in the Hawking-Ellis classification \cite{types}, i.e., of the form of a perfect fluid
\begin{equation}
T^\mu_\nu=\text{diag}(-\rho,P_1,P_2,P_3) \label{eq:2}
\end{equation}
in comoving coordinates.
$G^0_{\;0}$ component in Eq.(\ref{eq:1a}) for a perfect fluid is then written as \cite{Joshi,Bambi,Joshi5} 
\begin{equation}
\frac{F^{'}}{R^2 R^{'}}\,=\, 8\pi\rho \label{eq:3a},
\end{equation}
which leads to the following mass function \cite{Bambi,Misner}
\begin{equation}
F=8\pi\int_0^{R}\rho\,\tilde{R}_t^2\,d\tilde{R}_t. \label{eq:n5}
\end{equation}
Here $R_t$ corresponds to $R(r,t)$ with $t$ being fixed in the integration. Eq.(\ref{eq:n5}) suggests that, for a perfect fluid, $F(r,t)$ in general, describes twice the total energy inside a sphere of radius $R(r,t)$ in the local Minkowski space at a given $r$ and $t$. For this reason, we must define $\frac{1}{2}F(r,t)$ as the entire energy contained within a collapsing shell at $r$ and $t$ if no power flow occurs in the radial direction.

We assume that at first, a small black hole seed (of mass M) is embedded in a dust cloud. This may be considered to model a small primordial black hole in dust with initial approximate uniform distribution \cite{PBH-matter-dominated}. We demand that the effect of the black hole disappears in the limit of vanishing black hole mass and at infinitely great distances from the black hole. In other words, as $r\,\rightarrow\,\infty$ or $M\,\rightarrow\,0$ we expect that the corresponding metric should become Robertson Walker metric. Moreover, we consider the spherically symmetric gravitational collapse which means that
\begin{equation}
R(r,t)\,=\,f(t)\,r \label{n5a},
\end{equation}
where $f(t)$ is some function of time. The dust initially, almost everywhere, may be well approximated by a homogenous isotropic energy density with $\rho=m\frac{n_0}{f^3(t)}$ (i.e., we consider the dust as gravitationally marginally bound \cite{book}) where $m$ is the rest mass energy of each dust particle, $n_0$ is the number density for $f(t)=1$. The mass function for dust may be obtained from Eq.(\ref{eq:n5}) as \cite{Malafarina,Kopteva1,Kopteva2}
\begin{equation}
F\,=\,m\,n_0\,\frac{8\pi}{3}r^3, \label{n6}
\end{equation}
which is the twice of the total rest mass energy inside a sphere of radius $r$.

Eq.(\ref{n6}) may be also expressed as
\begin{equation}
F\,=\,2m\,N\,=\,2m\int_0^r n\, d^3V, \label{eq:12}
\end{equation}
where
\begin{equation}
d^3V=4\pi R^2 e^\psi dr \label{eq:7}
\end{equation}
is the infinitesimal volume element for Eq.(\ref{eq:1}) \cite{Misner}.

On the other hand
the mechanical energy of a test particle of mass m, excluding its potential energy, is
\begin{equation}
E=m\,e^\lambda \frac{dt}{ds} \label{eq:8}.
\end{equation}
This expression may be obtained by dividing both sides of Eq.(\ref{eq:1}) by $ds^2$, and then identifying the result as the local Minkowski expression $m^2=E^2-\vec{p}^2$ for unit mass.
Note that this is the relevant quantity in an energy-momentum tensor rather than the total mechanical energy of the particle $m\,e^{2\lambda} \frac{dt}{ds}$ (that may be obtained from $\frac{\partial\,L}{\partial\,t^\prime}$, where $L=-m\frac{\sqrt{-ds^2}}{dt}$, $t^\prime=\frac{dt}{ds}$). The relevance of this identification may be better seen by considering the non-relativistic limit of Einstein equations for weak fields, namely, Poisson equation, $\vec{\nabla}^2\phi=4\pi\,G\,\sum\,m_i\delta(\vec{r}-\vec{r}_i)$ for a set of point masses. The Poisson equation implies that the gravitational potential is not a part of the source term (i.e., the energy-momentum tensor part) instead it is a part of the left hand side of the equation (i.e., the Einstein tensor part)
If we assume that the fluid is made of particles of mass m and number density $n$, then the energy density of the fluid is
\begin{equation}
\rho=E\,n, \label{eq:9}
\end{equation}
where $E$ is given by Eq.(\ref{eq:8}). After using Eq.(\ref{eq:7}), Eq.(\ref{eq:8}), Eq.(\ref{eq:9}); Eq.(\ref{eq:n5}) becomes
\begin{equation}
F=2\int_0^r R^{'}e^{-\psi}\rho\, d^3V=2\int_0^r R^{'}e^{-\psi}E\,n\, d^3V\,=\,2m\int_0^r n\,R^{'}\frac{dt}{ds}e^{\lambda-\psi}\, d^3V \label{eq:10-}.
\end{equation}
Equating Eq.(\ref{eq:10-}) and Eq.(\ref{eq:12}) requires the condition
\begin{equation}
R^{'}\frac{dt}{ds}e^{\lambda-\psi}=1 \label{eq:11},
\end{equation}
which is satisfied, in particular, by the Schwarzschild metric and the Robertson-Walker metric for dust. Thus, the condition (\ref{eq:11}) makes the framework consistent since the metric corresponding to the present study should reduce to the Schwarzschild and Robertson-Walker metrics within the appropriate limits. 
Hence, Eq.(\ref{eq:10-}) (after the use of Eq.(\ref{n6}), Eq.(\ref{eq:8}) and Eq.(\ref{eq:11})) becomes
\begin{equation}
F(r)=2m \int_0^r n\, d^3V \,=\, m \, n_0 \,\frac{8\pi}{3}r^3   \label{eq:12x}
\end{equation}
since the dust particles move with the Hubble flow in comoving coordinates (that is ensured by the choice (\ref{eq:2})), i.e., the particles keep their position at the same $r$ during their whole story. It is clear that with (18) it is impossible to get an accretion; therefore, a black hole that is accreting (i.e., has increasing inhomogeneity) cannot be considered when $G^0_{\;1}\,=\,0$. Rather than Eq.(\ref{eq:12x}), making $F$ time and $r$ dependent with increasing inhomogeneity so that at the beginning of the accretion it reduces to Eq.(\ref{eq:12x}) for $M\rightarrow\,0$ may be one method of obtaining accretion. However, it doesn't appear tractable to achieve this goal. The simplest option for time-dependent F that reduces to Eq.(\ref{eq:12x}) as $M\rightarrow\,0$ is given by
\begin{equation}
F(r,t)\, =\,m\,n_0\,\frac{8\pi}{3}r^3\,b(t)  \label{eq:12xx},
\end{equation}
where $b(t)$ is a function to be determined by Einstein equations. Thus, after plugging the contribution of the black hole into the last equation, we obtain
\begin{equation}
	F\,=\,F\left(r_c\right)\,=\,2M\,+\,m\,n_0\,\frac{8\pi}{3}r_c^3 ,
	\label{eq:16}
\end{equation}
where the first term is due to the black hole's contribution, and 
\begin{equation}
r_c\,\equiv \,b(t)^{1/3}\,r.  \label{eq:16ax}
\end{equation}
However, the true accretion (i.e., a growing inhomogeneity) can be achieved by Eq.(\ref{eq:16}). The rate of change of $F\left(r_c\right)$ is the same for all $r$. For instance, for $b(t)=a^4$ (i.e., for a phantom field) $F(r_c)$ grows for all $r$ simultaneously. In fact, we shall later clarify that using time dependent $b(t)$ in place of $b(t)=1$ corresponds to another fluid in place of dust, respectively. Then replacing $b(t)=1$ by $b(t)$ in Eq.(\ref{eq:16}) with Eq.(\ref{eq:16ax}) is not an actual method of achieving accretion (due to a local gravitational collapse). Nevertheless, we maintain the generic form in Eq.(\ref{eq:16ax}) to keep the study as general as possible unless we consider the specific example of dust. We should here add that an inhomogeneous metric by itself is insufficient for an accreting black hole. For an accretion, the inhomogeneity must also be growing. The original McVities metric, for instance, is considered for an inhomogeneous distribution of perfect fluids without an accretion \cite{Faraoni:2021nhi}. 


In summary, we have first introduced Eq.(\ref{eq:16}), the most logical, simple option for the mass function of the gravitational collapse of dust with $G^0_{\;1}\,=\,0$ in the presence of a black hole. We have also observed that condition (\ref{eq:11}) is required for our consistent formulation. However, as it is shown in Appendix G, Eq.(\ref{eq:16}) and Eq.(\ref{eq:11}) together do not satisfy $G^0_{\;1}\,=\,0$, so one must leave one of the conditions $G^0_{\;1}\,=\,0$ or  Eq.(\ref{eq:16}) or Eq.(\ref{eq:11}) for the consistency of the framework. We continue to use Eq.(\ref{eq:16}) because it is the most straightforward and logical option for an accreting gravitational collapse. We also maintain Eq.(\ref{eq:11}) because it ensures that a Schwarzschild black hole and homogeneous isotropic dust are included as the limiting cases of the obtained metric. In this regard, we shall leave $G^0_{\;1}\,=\,0$ in the following section and we shall obtain a metric satisfying Eq.(\ref{eq:11}) and Eq.(\ref{eq:16}).


\section{Derivation of a Metric for Gravitational Collapse}

In this section, we aim to derive a metric for an accreting black hole-fluid system that initially consists of a Schwarzschild black hole and a homogeneous and isotropic dust. We start with taking the derivative of Eq.(\ref{eq:11}) with respect to time and dividing it by itself gives 
\begin{equation}
\dot R ^\prime + \dot \lambda R^\prime-\dot{\psi} R^\prime=0 \;\Rightarrow~ \dot{R}^\prime - \dot{\psi} R^\prime = - \dot{\lambda} R^\prime.
\label{eq:24}
\end{equation}
If we use Eq.(\ref{n5a}) in the last equation, we get
\begin{equation}
\frac{\dot{f}}{f}\,=\,\frac{\partial\,\left(\psi-\lambda\right)}{\partial\,t},
\label{eq:28c}
\end{equation}
which after integration gives
\begin{equation}
e^{-2\lambda}\,=\,Q(r)\,f^2\,e^{-2\psi}\, , \label{eq:28d}
\end{equation}
where $Q$ is an arbitrary function of $r$ (whose form to be determined by initial conditions). Eq.(\ref{eq:4}) may be rewritten as
\begin{equation}
e^{2\psi}\,=\,\left[1+e^{-2\lambda}\dot{R}^2-\frac{\,F(r)}{R}\right]^{-1}R^{\prime\;2}.
\label{eq:28e}
\end{equation}
Eq.(\ref{eq:28e}) may be solved for $e^{2\psi}$ after using Eq.(\ref{eq:28d}),  Eq.(\ref{eq:16}) and Eq.(\ref{n5a}) as
\begin{equation}
e^{2\psi}\,=\,\frac{f^2\left(1-r^2Q\,\dot{f}^2\right)}{\left[1-\frac{\,F(r_c)}{R}\right]},
\label{eq:28f}
\end{equation}
where $F(r_c)$ is given by Eq.(\ref{eq:16}). Then one may substitute Eq.(\ref{eq:28f}) in Eq.(\ref{eq:28d}) to obtain
\begin{equation}
e^{2\lambda}\,=\,\frac{\left(1-r^2Q\,\dot{f}^2\right)}{Q\left(1-\frac{\,F(r_c)}{R}\right)}.
\label{eq:28g}
\end{equation}

After leaving the condition $G^0_{\;1}\,=\,0$, one may here expect a non-zero $G^0_{\;1}\,=\,0$ in general. By use of Mathematica \cite{Collapse.nb}, we observe that this is really the case. Then, in Einstein field equations, a non-zero $T^0_{\;1}$ term must be added to the energy-momentum tensor. This can be achieved by altering the energy momentum tensor's type from type I to type II.The canonical form of type II energy-momentum tensors is given by \cite{types}
\begin{eqnarray}
\left(T^{\mu\nu}\right)\,=\,\left(\begin{array}{cccc}
\xi+\sigma&\sigma&0&0\\
\sigma&-\xi+\sigma&0&0\\
0&0&p_2&0\\
0&0&0&p_3\end{array}\right),
\label{eq:28r}
\end{eqnarray}
where $\xi$, $\sigma$, $p_2$, $p_3$ are some functions of the coordinates. An arbitrary type II energy-momentum tensor may be put into its canonical form given in Eq.(\ref{eq:28r}) by using a Lorentz transformation in the Minkowski space as discussed in \cite{types,types2}. Again by use of Mathematica \cite{Collapse.nb}, we have verified that the Einstein tensor here is of type II \cite{Erdem:2020jgi}.

We are now in a position to conclude that after using Eq.(\ref{eq:16}) in Eq.(\ref{eq:28f}) and Eq.(\ref{eq:28g}), we obtain the metric for gravitational collapse as
\begin{eqnarray}
ds^2&=&-\frac{\left(1-r^2Q(r)\,\dot{f}^2(t)\right)}{Q(r)\,\left(1-\frac{2\,\left(M+mn_0\frac{4\pi}{3}\,b(t)\,r^3\right)}{f(t)\,r}\right)}
dt^2 \nonumber \,+\,\frac{f^2(t)\,\left(1-r^2Q(r)\,\dot{f}^2(t)\,\right)}{\left(1-\frac{2\,\left(M+mn_0\frac{4\pi}{3}\,b(t)\,r^3)\right)}{f(t)\,r}\right)}
dr^2
\nonumber \\
&&
\,+\,f^2(t)\,r^2(d\theta^2+\sin^2\theta d\phi^2). \label{eq:31a}
\end{eqnarray}
Different choices of $f(t)$ are expected to give different evolutions of the system.
We require that Eq.(\ref{eq:31a}) must reduce to the Schwarzschild metric if the dust is removed (i.e., when $n_0=0$). This condition specifies $Q(r)$ as
\begin{equation}
Q(r)\,=\,\frac{1}{\left(1-\frac{2M}{r}\right)^2}.
\label{eq:31aa}
\end{equation}

\section{Some General Implications of the Metric}
\subsection{General Behaviour of the Solution}

After using a Mathematica code \cite{Collapse.nb}, we have observed that the components of the Einstein tensor for the metric (\ref{eq:31a}) have very complicated expressions. However, to understand some characteristics of the metric, it is possible to get more concrete general observations regarding the components $ G^0_{\;0}$ and $G^1_{\;1}$ as follows. If we use Eq.(\ref{n5a}), Eq.(\ref{eq:16}), and Eq.(\ref{eq:31a}) (i.e., Eq.(\ref{eq:28f}) and Eq.(\ref{eq:28g}) in the Eq.(\ref{eq:1a}) and Eq.(\ref{eq:1b}), we obtain that
\begin{eqnarray}
&& G^0_{\;0}\,=-\,\frac{8\pi\,m\,n_0\,\,b(t)}{f^3\left(t\right)}+\,\frac{r\dot{f}(t)}{f(t)}\,G^0_{\;1}\,=\, -8 \pi \rho,
  \label{eq:1aa} \\
&& G^1_{\;1}\,=-\frac{8\pi\,m\,n_0\,\,\dot{b}(t)}{3f^2\left(t\right)\dot{f}}-\,
\frac{\left(1-\frac{2\,M}{r}\right)^2}{r\,f(t)\dot{f}(t)}\,G^0_{\;1}\, =\, 8 \pi P_r.
\label{eq:1ba} 
\end{eqnarray}
When we are at large distances from the black hole (i.e., as $r\,\rightarrow\infty$) or when there is no local gravitational collapse due to the black hole (i.e., as $M\,\rightarrow 0$), one may expect that the metric (\ref{eq:31a}) should reduce to the Robertson-Walker metric and $G^0_{\;1}\,\rightarrow\,0$ in Eq. (\ref{eq:1aa}) and Eq.(\ref{eq:1ba}). This is also physically reasonable because in these limits we expect not only the homogeneous and isotropic structure of the universe but also no accretion. By Mathematica \cite{Collapse.nb}, we have shown that this is the case, i.e., as $r\,\rightarrow\infty$ or $M\,\rightarrow 0$, we get $G^0_{\;1}\,\rightarrow\,0$ , so the last equations become
\begin{equation}
G^0_{\;0}\,=-\,\frac{8\pi\,m\,n_0\,\,b(t)}{f^3\left(t\right)}\,=\, -8 \pi \rho \label{eq:21a}
\end{equation}
\begin{equation}
G^1_{\;1}\,=-\frac{8\pi\,m\,n_0\,\,\dot{b}(t)}{3f^2\left(t\right)\dot{f}}\,=\, 8 \pi P_r. \label{eq:21b}
\end{equation}
The corresponding equation of state is
\begin{equation}
\omega_r\,=\,\frac{P_r}{\rho}\,=\,-\frac{1}{3}\left(\frac{\dot{b}(t)}{b(t)}\right)\left(\frac{f(t)}{\dot{f}(t)}\right).
\label{eq:33ap}	
\end{equation}
Eq.(\ref{eq:33ap}) indicates that $\omega_r\,<\,0$ if black hole is accreting with cosmic expansion, i.e., $\dot{b}\,>\,0$ , $\dot{f}\,>\,0$.
The fractional change in the active gravitational mass \cite{book} of the system during a Hubble time $\left(\frac{\dot{f}(t)}{f(t)}\right)^{-1}$ is
\begin{equation}
\frac{\Delta\,F}{F}\,=\,F^{-1}(r_0)\,\dot{F}(r_0)\left(\frac{f(t)}{\dot{f}(t)}\right)\,=\,\left(\frac{\dot{b}(t)}{b(t)}\right)\left(\frac{f(t)}{\dot{f}(t)}\right)
=-3\omega_r.  \label{apa2}
\end{equation}
Eq.(\ref{apa2}) indicates that the accretion of the black hole is not possible for Eq.(\ref{n6}). An accretion for $G^0_{\;1}\,=\,0$ is possible for $\dot{b}\,>\,0$. However, rather than being caused by the gravitational field, such accretion is caused by the fluid's behaviour. It appears that a true gravitational collapse may occur only for $G^0_{\;1}\,\neq\,0$. A time-dependent $b(t)$ corresponds to a fluid other than dust. This point may be seen more clearly as follows. Consider an energy density of the form $\rho(t)\,=\,\frac{\rho_0}{f^s(t)}\,=\,\frac{\rho_0}{f^3(t)}\,b(t)$ where $s$, $\rho_0$ are some constants, and $b(t)\,\equiv\,\frac{f^3(t)}{f^s(t)}$. This energy density includes the cases of radiation dominated (i.e., $s=4$), dust dominated (i.e., $s=3$), cosmological constant dominated (i.e., $s=0$) universes as subcases. It is seen that $b$ is a time-dependent function except in the case of dust. One obtains Eq.(\ref{eq:16}) after substituting $\rho(t)$ in Eq.(\ref{eq:n5}). This point may also be seen in the following way. As $r\,\rightarrow\infty$, Eq.(\ref{eq:31a}) becomes
\begin{eqnarray}
ds^2&=&c(t)\,\left[-dt^2\,+\,f^2(t)\,dr^2\right]
\,+\,f^2(t)\,r^2\left[d\theta^2+\sin^2{\theta}\,d\phi^2\right]
\label{eq:31xx}
\end{eqnarray}
where $c(t)\ \equiv \,f(t)\,\frac{\dot{f}^2(t)}{\frac{8\pi}{3}mn_0\,b(t)}$. Eq.(\ref{eq:31xx}) is actually the same as Robertson Walker metric, because $c(t)=1$ gives the fundamental Friedmann equation for a perfect fluid $\left(\frac{\dot{f}}{f}\right)^2=\frac{8\pi}{3}\rho=\frac{8\pi}{3}\frac{\rho_0}{f^3(t)}b(t)$ or vice versa. In other words, $c(t)=1$ if the fluid is perfect fluid. Hence
Eq.(\ref{eq:31xx}) (or Eq.(\ref{eq:31a}) as $r\,\rightarrow\infty$) is the same as the Robertson-Walker metric
\begin{eqnarray}
ds^2=-dt^2\,+\,f^2(t)\,dr^2
\,+\,f^2(t)\,r^2\left[d\theta^2+\sin^2{\theta}\,d\phi^2\right]\label{eq:31xxx}
\end{eqnarray}
for a cosmological fluid, in particular for dust. A similar result holds for the $M\,\rightarrow\,0$ limit too. In this limit, $c(t)=\frac{f(t)-r^2f(t)\dot{f}(t)^2}{f(t)-\frac{8\pi}{3}mn_0 b(t) r^2}$ in Eq.(\ref{eq:31xx}). After employing the same approach as in the case of $r\rightarrow\infty$ given just above, Eq.(\ref{eq:31xx}) (or Eq.(\ref{eq:31a})) as $M\,\rightarrow\,0$) reduces to Eq.(\ref{eq:31xxx}) for a cosmological fluid. This tells us that the source of the gravitational collapse is the presence of the black hole. Moreover, $G^0_{\;1}\,\rightarrow\,0$ for Eq.(\ref{eq:31a}) as $r\,\rightarrow\infty$ or as $M\,\rightarrow\,0$ as we have mentioned before. This suggests that the source of a true accretion, i.e., $G^0_{\;1}\,\neq\,0$ is the gravitational attraction of the black hole. These observations together suggest that Eq.(\ref{eq:31a}) with Eq.(\ref{eq:31aa}) describes the gravitational collapse of a perfect fluid (including dust, i.e., $b(t)=1$ as a subcase). We mainly focus on the case of dust collapse (with $G^0_{\;1}\,\neq\,0$) in the following analysis.

\subsection{A Brief Overview of the Apparent Horizons and the Singularities of the Solution}

To have a better understanding of the evolution of the resulting black hole - fluid system, we must point out the other characteristics of the metric (\ref{eq:31a}) such as its apparent horizons and singularities. We first give some discussions on the apparent horizon(s) of the solution as follows. Because the metric is spherically symmetric, the apparent horizon(s) may be obtained through one of the conditions $g^{\mu\nu}\left(\partial_\mu\,R\right)\left(\partial_\nu\,R\right)\,=\,0$ or $g^{RR}\,=\,0$ after converting the coordinates of Eq.(\ref{eq:31a}) (i.e., the comoving gauge) to the Kodama gauge \cite{Faraoni2}. Using one of these methods, we get the following equation for the apparent horizon(s)
\begin{equation}
\frac{8\pi}{3}\,m\,n_0\,r^3\,b(t)\,-\,f(t)\,r\,+\,2M\,=\,0.
\label{eq:40}	
\end{equation}
The roots of Eq.(\ref{eq:40}) are \cite{galaxies}
\begin{equation}
r_{k+1}\,=\,2\,\sqrt{\frac{f(t)}{8\pi\,mn_0\,b(t)}}\;\sin{\left[\frac{1}{3}\,\sin^{-1}{\left(6M\sqrt{2\pi\,mn_0}\;
\left(\frac{b^\frac{1}{2}(t)}{f^\frac{3}{2}(t)}\right)\right)}
+\frac{2\pi\,k}{3}\right]}
~~,~~~~k=0,1,2.
 \label{eq:41a}	
\end{equation}
$r_3$ is irrelevant from a physical point of view since radial coordinates cannot be negative. $r_1$ corresponds to the apparent horizon of the black hole - fluid system. This may be easily seen as follows. In most of the cases $3M\,\sqrt{8\pi\,m\,n_0}\,\ll\,1$
since $\sqrt{8\pi\,m\,n_0}$ is some multiple of $H_0\,\sim\,\frac{1}{\left(3\times\,10^8\,meters/second\right)\times\,3\times\,10^{17}\,seconds}$ while $M$ is in the order of the Schwarzschild event horizon that is at the order of kilometers for stellar black holes and smaller than $\sim\,10^{14}\,kilometers$ for known supermassive black holes. In the limit of $3M\,\sqrt{8\pi\,m\,n_0}\,\ll\,1$ for $b(t)\,=\,1$, $r_1$ may be approximated by
\begin{equation}
r_1\simeq\frac{2M}{f(t)}\left[1+\frac{4}{27}\left(3M\sqrt{8\pi m n_0}\right)^2\frac{1}{f^3(t)}+\frac{73}{972}\left(3M\sqrt{8\pi m n_0}\right)^4\frac{1}{f^6(t)}\right],
\end{equation}
where we have used approximate forms of sine and arcsine for tiny arguments. After ignoring all the terms (being much smaller than $1$ as stated above) in parenthesis of the last expression, we get $r_1\,\simeq\,\frac{2M}{f(t)}$ (which is the Schwarzschild event horizon of the black hole for $f(t)=1$). In a similar way we obtain $r_2\,\sim\,\frac{\sqrt{f(t)}}{H_0}$, i.e., the cosmological horizon
for $f(t)=1$ (in the same limit as $r_1$).

\begin{figure}[H]
\centerline{\includegraphics[scale=0.6]{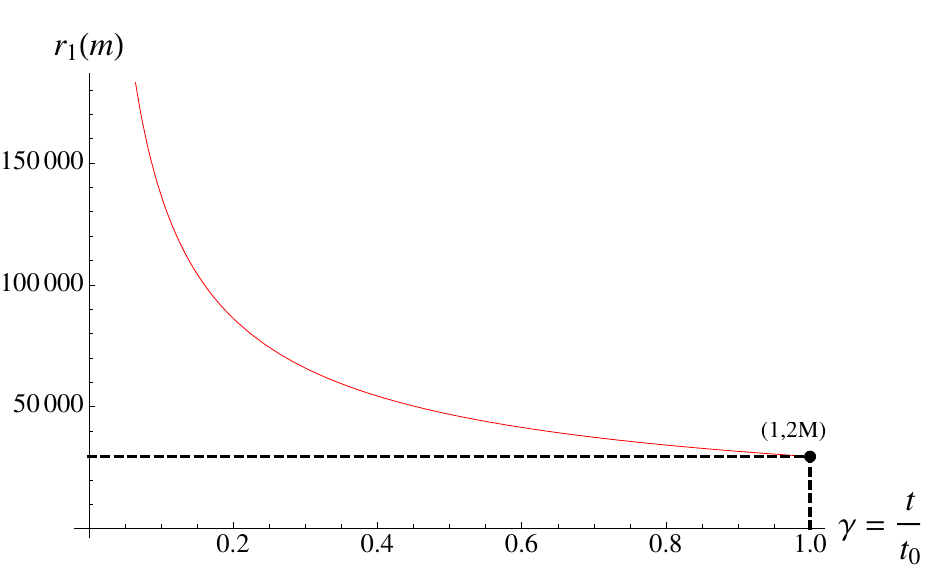}}
\caption{$\gamma=\frac{t}{t_0}$ (where $t$ is the time of any epoch of the universe and $t_0$ is the present time) versus the location of apparent horizon for $M=10M_\odot$, where $M_\odot$ is the solar mass.}
\label{nfig:1}
\end{figure}

After discussing the apparent horizon, the other important point is to determine the singularity structure of the metric (\ref{eq:31a}), to determine whether the singularities are naked or hidden behind a horizon. This issue may be studied by calculating (using a Mathematica code \cite{Collapse.nb}) the curvature invariants, the Ricci scalar $R$, $R_{\mu\nu}\,R^{\mu\nu}$, and the Kretschmann scalar $R_{\mu\nu\sigma\kappa}R^{\mu\nu\sigma\kappa}$ of Eq.(\ref{eq:31a}). We have observed that all the corresponding curvature invariants, even for the particular case of cosmological dust $f(t)\,=\,a(t)\,\propto\,t^\frac{2}{3}$, are rather complicated. We particularly consider the cosmological dust. By factoring out the expressions and then determining the values of the parameters that make the denominators zero, we can determine the following singularities 
\begin{eqnarray}
&&i)~~r=0,~~~ii)~~t=0,~~~iii) ~~-2M\,+\,r\,=\,0, \label{s1a} \\
&&iv)~~2M\,+\,\frac{4r^3}{9t_0^2}\,-\,r\left(\frac{t}{t_0}\right)^\frac{2}{3}\,=\,0,~~\mbox{i.e.,}~~
\left(\frac{t}{t_0}\right)^\frac{2}{3}\,=\,\frac{1}{r}\left(2M\,+\,\frac{4r^3}{9t_0^2}\right), \label{s1b} \\
&&v)~~4r^4\left(\frac{t}{t_0}\right)^\frac{1}{3}\,-\,9\left(-2M\,+\,r\right)^2t\,t_0\,=\,0,~~\mbox{i.e.,}~~
\left(\frac{t}{t_0}\right)^\frac{2}{3}\,=\,\left(\frac{4r^4}{9(-2M\,+\,r)^2t_0^2}\right),\nonumber \\
\label{s1c}
\end{eqnarray}
where the scale factor is expressed as $f(t)=\left(\frac{t}{t_0}\right)^\frac{2}{3}$ with $t_0$ being the age of the universe. The first two singularities correspond to the well-known Schwarzschild black hole and the cosmological big bang singularity, respectively. The third one corresponds to the coordinate singularity of the Schwarzschild black hole while it is a true singularity here. The fourth one is for the apparent horizon itself given in Eq.(\ref{eq:41a}) in the case of cosmological dust (i.e. in the case of $\frac{\dot{f}^2}{f^2}=\frac{8\pi}{3}m\,n_0\left(\frac{a(t_i)}{a(t)}\right)^3$). The only new additional singularity particular to our solution is the last one. This singularity starts at $r=0$, $t=0$, and $r$ increases as $t$ increases. However, this singularity never reaches $r\,=\,2M$, so it remains hidden behind $r\,=\,2M$. In case of the fourth singularity, the minimum value of the time is $t_\text{min}=2\sqrt{3} M$ when $r=\left(\frac{9M}{4t_0}\right)^\frac{1}{3}t_0=2M\left(\frac{9}{32}\right)^\frac{1}{3}\left(\frac{t_0}{M}\right)^\frac{2}{3}$. After singularity starts at $t_\text{min}=2\sqrt{3} M$, $r$ may either increase or decrease in time. For generic values of $M$, $t_0\,\gg\,M$, so $r\,\gg\,2M$. For the generic values of $r$, $t_0\,\gg\,r$ (note that $t_0$ is in the order of $10^{26}\,meters$). Then, $2M\,\gg\,\frac{4r^3}{9t_0^2}$ in Eq.(\ref{s1b}), so $r$ takes values smaller than its starting value in time, and it finally reaches $r=2M$ at $t\simeq\,t_0$, and then disappears. The singularity (\ref{s1b}) corresponds to the increase in the size of the horizon of an isolated Schwarzschild black hole  (at $r=2M$) due to the effect of the presence of dust. Another point worth to mention is that $r=2M$ singularity in Eq.(\ref{s1a}) is a naked singularity. This is not as problematic as it seems at first sight. It is hidden behind the singularity in Eq.(\ref{s1b}) which is also the apparent horizon $r_1$ of Eq.(\ref{eq:41a}) for most of the time between $t=0$ and the present time $t_0$ as may be seen, for example, in Figure \ref{nfig:1}. The singularity in Eq.(\ref{s1b}) does not form till $\left(6M\sqrt{2\pi\,mn_0}\;
\left(\frac{b^\frac{1}{2}(t)}{f^\frac{3}{2}(t)}\right)\right)\leq 1$. Until the formation of this singularity, the $r = 2M$ singularity remains wholly naked. However, it takes some time after the big bang till the primordial density perturbations to grow enough to form the seed black holes.  Thus, it is possible that the $r = 2M$ singularity forms after the formation of the singularity in Eq.(\ref{s1b}). The singularity in Eq.(\ref{s1b}) becomes smaller than $r = 2M$ after  $t > t_0$. However, the formation of the singularities may be related to the completely pressureless nature of dust (although realistic gas has thermodynamical pressure).
The fact that the apparent horizon of the Schwarzschild black hole at $r=2M$ turns into the singularity of Eq.(\ref{s1b}) may be well-explained by the presence of dust. As clarified before, in the absence of the dust, the metric reduces to the Schwarzschild metric, in which case all the singularities discussed above reduces to coordinate singularity of the isolated Schwarzschild black hole at $r=2M$. The reason that  the coordinate singularity of Schwarzschild black hole at $r=2M$ turns into the true singularities at $r=2M$ and Eq.(\ref{s1b}) and Eq.(\ref{s1c}) after consideration of dust may be clarified as follows. When the collapse starts, the collapsing dust accumulates near the horizon. In the absence of an internal repulsive pressure in the dust near the horizon, the energy density of the dust approaches to infinity after some time after the beginning of the collapse of the dust. Conversion of the coordinate singularity of the Schwarzschild metric at $r=2M$ to the singularity cannot be realised in the absence of the dust. In conclusion, all singularities of our solution are hidden behind the singularity given in Eq.(\ref{s1b}) which corresponds to the conversion of the event horizon of the Schwarzschild metric to a singularity because of the accumulation of the (zero internal pressure) dust near the new horizon given by Eq.(\ref{s1b}). Despite being a naked singularity, it is conceivable to predict that, in the case of a more realistic dust, this singularity changes into a coordinate singularity (which acts as an effective event horizon).
\section{Some Additional Comments about the Energy Density and the Pressures in the Dust Collapse}

\begin{figure}[H]
\centerline{\includegraphics[scale=1]{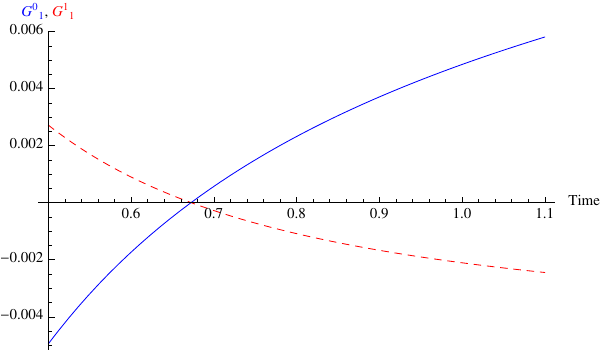}} 
\caption{Time versus $G^0_{\;1}$ (blue) and $G^1_{\;1}$ (dashed red) graphs for $c=1$, $M=0.1$, $r =3$.}
\label{fig:1}
\end{figure}

\begin{figure}[H]
\centerline{\includegraphics[scale=1]{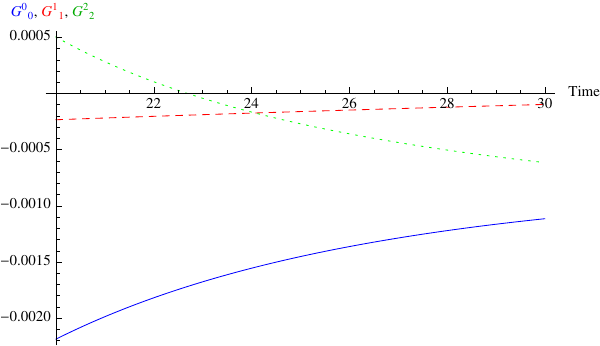}}
\caption{Time versus $G^1_{\;1}$ (dashed red), $G^0_{\;0}$ (blue), $G^2_{\;2}$ (dotted green) graphs for $c=1$, $M=1$, $r =3$.}
\label{fig:2}
\end{figure}

\begin{figure}[H]
\centerline{\includegraphics[scale=1]{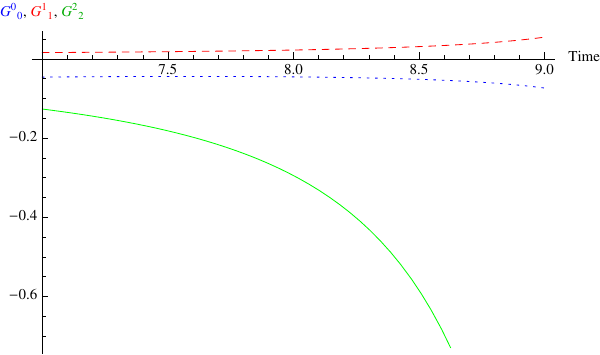}}
\caption{Time versus $G^1_{\;1}$ (dashed red), $G^0_{\;0}$ (dotted blue), $G^2_{\;2}$ (green) graphs for $c=0.1$, $M=1$, $r =30$.}
\label{fig:3}
\end{figure}

\begin{figure}[H]
\centerline{\includegraphics[scale=1]{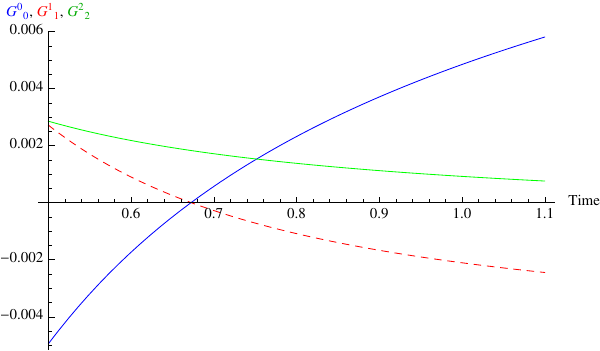}}
\caption{Figure 2 with the function $H=\left(\frac{\dot{a}}{a}\right)^2\,a^3\,r^3-a\,r+2M$ is included. The zeros of $H$ correspond to Eq.(\ref{eq:40}) (where $\left(\frac{\dot{a}}{a}\right)^2\,a^3\,=\,\frac{8\pi}{3}m\,n_0$ is used). The green solid line here corresponds to $\frac{H}{10^4}$.}
\label{fig:4}
\end{figure}

To understand the behaviour of the metric (\ref{eq:31a}) more clearly and explicitly, our aim in this section is to focus on the situation where the fluid is at first a cosmological dust, i.e., dust in an expanding dust dominated universe, which is described by
 \begin{equation}
 f(t)\,=\,a(t)\,\,=\,c\,t^\frac{2}{3},
 \label{i3}
 \end{equation}
 where $c$ is some constant. As we previously pointed out in the absence of gravitational collapse (i.e., without the radial power flux or when $G^0_{\;1}=0$), the pressure of the dust is determined to be zero (see Eq.(\ref{eq:1ba})) and the energy density is $\rho=\frac{\rho_0}{a^3}$ (see Eq.(\ref{eq:1aa})) as in the case of the Robertson-Walker metric for dust. However, in the presence of gravitational collapse due to the black hole (i.e., with the radial power flux or when $G^0_{\;1}\neq 0$), the pressure of the dust becomes non-zero and the energy density of the dust given above is modified by the inclusion of $G^0_{\;1}$. In this regard, it is important to examine some characteristics of $G^0_{\;1}$ for collapsing dust. By use of Mathematica \cite{Collapse.nb}, the corresponding $G^0_{\;1}$ is obtained as
 \begin{eqnarray}
 G^0_{\;1}=
-A\,
 \left[4 c^3 r^2 t^{4/3} (3 M-2 r)+6 c^2 r^3 t^{2/3}+9 c t^2 (r-2 M)+9 t^{4/3} (2 M-r)\right]
         \label{G01}
 \end{eqnarray}
 with
 \begin{equation}
 A\,=\,\frac{24 M r}{c}\left(4 c^2 r^4-9 t^{2/3} (r-2 M)^2\right)^{-2}t^{-\frac{5}{3}}. \label{G01a}
 \end{equation}
 It is clear that either $G^0_{\;1}>0$ or $G^0_{\;1}<0$ is determined by the relative magnitudes of $c$, $t$ and $r$. To understand the situation more easily, one may write $a(t)\,=\,c\,t_0^\frac{2}{3}\gamma^\frac{2}{3}$ where $\gamma= \frac{t}{t_0}$ and $t_0$ denotes the present time. Then, the common convention $a(t_0)=1$ implies $c=1$. In this unit system (i.e., in the case of setting $c=1$), Eq.(\ref{G01}) and Eq.(\ref{G01a}) now becomes
  \begin{eqnarray}
 G^0_{\;1}=
-A\,
 \left[4 r^2 t^{4/3} (3 M-2 r)+6 r^3 t^{2/3}+9 t^2 (r-2 M)+9 t^{4/3} (2 M-r)\right]
         \label{G01b}
 \end{eqnarray}
 with
 \begin{equation}
 A\,=\,24 M r \left(4 r^4-9 t^{2/3} (r-2 M)^2\right)^{-2}t^{-\frac{5}{3}}. \label{G01ab}
 \end{equation}
The last equations now depend only on $t$ and $r$. In these equations, $t=t_0=1$ corresponds to present time, i.e., to the age of the universe or to the size of the observable universe, so $r\ll t_0$ (in geometric units). Now, $G^0_{\;1}>0$ or $G^0_{\;1}<0$ is determined by only the relative values of $r$ and $t$. For instance, in the case of $r\gg M$, $G^0_{\;1}>0$ for $t\simeq\,t_0=1$, and $G^0_{\;1}<0$ for $t\gg t_0$. For small $t$, $G^0_{\;1}>0$ or $G^0_{\;1}<0$ depending on how small $t$ is. In the case of small $r$ with $r\,>\,2M$, again $G^0_{\;1}>0$ or $G^0_{\;1}<0$ depending on values of the parameters. For instance, $G^0_{\;1}<0$ for $r=3M$ and $t\sim\,t_0$. This, in the light of Eq.(\ref{eq:1ba}) indicates that the collapsing dust may lead to a positive or negative pressure in the radial direction. This situation may be better seen by the plots of $G^0_{\;1}$ and $G^1_{\;1}$ drawn by Mathematica \cite{Collapse.nb} e.g. as given in Figure \ref{fig:1}. There is no straightforward relation between $G^0_{\;1}$ and $G^2_{\;2}=G^3_{\;3}$. This case may be better seen by the plots of $G^1_{\;1}$ and $G^2_{\;2}$ which show that both may have the same or different signs e.g. as given in Figures \ref{fig:2} and \ref{fig:3}. In the plots, $G^0_{\;0}<0$ when $r$ is at the outside horizon (as expected) since it corresponds to $-\rho$. Note that one must be careful about the location of the horizon since it increases with time as may be seen from Eq.(\ref{eq:41a}) for $f(t)\,=\,a(t)$ and $G^0_{\;0}$ should be negative for the values of $r$ outside the horizon. We have also observed that the plots we have considered, for instance in Figures \ref{fig:1}, \ref{fig:2}, \ref{fig:3}, mostly correspond to the regions outside the horizon. For instance, for the values of the parameters in Figure \ref{fig:1}, the situation is shown in Figure \ref{fig:4}.
  
There are also some other interesting points of our solution worth to mention at this point. As emphasized previously, for the collapsing dust solution, we have non-zero $G^0_{\;1}$. This implies that after the start of the collapse, we have non-zero $G^1_{\;1}$ that leads to the emergence of a radial pressure as opposed to the zero pressure for our initial case of the fluid, where  we consider homogeneous and isotropic dust. This situation may also be seen as follows. The energy density of dust considered for the present work may be given as
\begin{equation}
\rho\,=\,m\,n\,=\,m\,\frac{dN}{4\pi\,R^2\,e^\psi\,dr}\,=\,\frac{m\,e^{-\psi(r,t)}}{a^2(t)}\,\frac{dN}{4\pi\,r^2\,dr}
\,=\,\frac{e^{-\psi(r,t)}\rho_0}{a^2(t)},
 \label{eq:7x}
\end{equation}
where $n$ denotes the number density of $dN$ particles in a spherical shell of radius $r$ and width $dr$, i.e., in a volume $d^3V\,=\,4\pi\,R^2\,e^\psi\,dr$ as given in Eq.(\ref{eq:7}), and $\rho_0\,=\,m\,n_0\,=\,m\frac{dN}{4\pi\,r^2\,dr}$ is the energy density of homogeneous  dust at $a(t)\,=\,1$. The last equation implies that if 
 $e^{\psi(r,t)}$ differs from $\,a(t)$, the dust particles can no longer be homogeneous dust and particles deviate from the comoving frame, which is, in general, the case after gravitational collapse starts. In this case, the equation of state of the dust can be different from zero. Further, we have observed that when we leave the limit $M\,\rightarrow\,0$, our solution for dust induces non-trivial pressure. This indicates that this non-trivial pressure is entirely due to gravitational collapse rather than due to an exotic energy density.

\newpage

\newpage
\chapter{Concluding Remarks}

In this thesis, we particularly pay attention to Bose-Einstein condensate scalar fields because cosmological scalar fields may be considered as a result of Bose-Einstein condensation (BEC) so that natural time dependence of the fields is realised. To this end, the study of the formation of BEC in cosmology is a valid point. In this sense, our first goal is to show the curved space and particle physics effects on forming Bose-Einstein condensates in cosmology. In this part, to use the techniques and the relative simplicity of the usual (flat) Minkowski space quantum field theory, we introduce an effective Minkowski formulation. By doing this, we eliminate the contributions of gravitational particle production to transition amplitudes so that we may more easily identify the transition amplitudes that are directly due to particle physics processes. 

Moreover, the calculation of transition amplitudes and cross-sections becomes easier so that we see cosmological evolution more easily. This evolution shows us that all necessary conditions, i.e., coherence, correlation, and finite number density, are realised for the initial phase of the condensation of the produced particles. In this study, the curved space effect on the condensation is provided by the scale factor $a$. Thus, we conclude that the formation of the BEC scalar field in cosmology with particular emphasis on its microscopic description in particle physics is promoted. In this part, we have considered the initial stages of the evolution of Bose-Einstein condensation. To demonstrate the same behaviour in subsequent stages of the evolution of Bose-Einstein condensation, a more thorough investigation of the effect of Bose-Einstein statics using numerical calculations must be conducted in future as a separate study.

Secondly, we have shown that a condensate scalar field may also be realized in the spacetime around a Reissner - Nordstr{\o}m black hole. To better understand the scenario, we first analyse the problem in a completely classical setting at the level of the test particle. We have noticed that the produced particles in particle physics processes tend to gather away from the black hole, which creates an ideal environment for condensation. After then, we looked at the issue within the context of field theory. We have found approximate solitonic-like solutions for the scalar fields where the heavier outgoing particles appear more localized relative to the lighter ingoing particles. This wave profile appears to indicate favourable circumstances for the condensation, similar to those in the fully classical treatment. Our discussions on the phenomenological viability of the model also show that the black hole and the scalar particles should have a dark charge rather than the usual electromagnetic charge. We also indicate that the $\chi$ particles are non-relativistic when they are very far away from the black hole, so they may play the role of DM/DE at cosmic scales while $\phi$ particles are thought to act in the component of DM at galactic scales. A more detailed investigation on this point (for instance, the effect of these particles on the rotation curve(s) of their galax(ies)) may be studied in future. Moreover, in this work, we have ignored the effect of electromagnetic interactions between the scalar particles and the density of the scalar particles is taken to be small. However, a separate investigation in the future for a more general framework that does not ignore the electromagnetic interactions between the particles and considers the non-negligible energy density of the produced particles in the derivation of the wave profile  would be helpful to comprehend the study fully. The other point is that for the sake of simplicity, we have considered only the radial motion of the particles, i.e., $l=0$, so only radial interactions of the particles are considered. To make the results as general as possible, the non-trivial case of $l\neq 0$ (i.e when we leave the spherical shape of the wave profile) will be also examined in future in a separate study.

We finally study the gravitational collapse of a fluid (initially considered a homogeneous dust) by the effect of a small Schwarzschild black hole. We derive a corresponding metric that describes the behaviour of this collapsing fluid around the black hole. We have observed that the metric induces the non-vanishing inward radial power flux (i.e $G^0_{\;1}\neq 0$), which means that the black hole is accreting. We have also verified that as $M\rightarrow 0$ or $r\rightarrow \infty$, the corresponding metric reduces to the Robertson-Walker metric, so inward radial power flux vanishes (i.e., $G^0_{\;1}=0$). Therefore, the source of the inhomogeneities and pressures may be explained by the local gravitational collapse due to the black hole. We give particular attention to the case of collapsing dust rather than more general fluid because in this case, the solutions become clearer and simpler, so the behaviour of the metric is seen more easily. This is also physically acceptable due to the larger ratio of matter content (i.e dark matter and baryonic matter) of the universe. The results of this case have shown that the collapsing dust may lead to positive or negative pressures depending on the relative values of the parameters. Even if negative pressure is acceptable here, it does not mean that it is the source of the accelerated expansion of the universe since our metric is different from the Robertson-Walker metric. However, this dark energy-like behaviour deserves to be addressed in future study because it may have some characteristics of cosmological dark energy. We have also discussed some important characteristics of the singularities and the apparent horizons of the metric. In this present work, we have initially assumed almost homogeneous and gravitationally bound fluid. Some other initial cases, for instance, initially inhomogeneous and gravitationally bound and unbound fluids in the framework of the present study, may also be examined in future in a separate study. Another comment is that in this work, we have considered the cosmologically relevant expanding dust, where  $\dot{R}(r,t)\,>\,0$. As an extension to this study, the case of contracting dust (that corresponds to local gravitational collapse of self gravitating dust), where $\dot{R}(r,t)\,<\,0$, i.e., $\rho\,\propto\,a^3(t)$ may be also examined in future. This situation corresponds to considering two distinct scalar factors, the one for the cosmological expansion, i.e., $\dot{f}_1(t)\,>\,0$ and the other one for the local gravitational collapse of self gravitating dust, i.e., $\dot{f}_2(t)\,<\,0$. Another essential point to be addressed in future study is to see the effect of the gravitational collapse of the present work on Bose-Einstein condensation since the metric we derived describes a more realistic framework in the sense that the black hole is accreting in the background of the cosmological fluid.


\newpage
\appendix
\chapter{Condition (\ref{con1}) in the Case of Radiation, Matter, Cosmological Constant, and Stiff Matter Dominated Era}
Using
\begin{equation}
H\,=\,\xi\,a^{-s} \;,\label{t9a3}
\end{equation}
which consists of all the interesting examples, such as those involving radiation, matter, stiff matter, and cosmological constant dominated universes, in Eq.(\ref{con1}), we get
\begin{equation}
\left|\frac{2a^2H\left[m_\chi^2-(s^2-3s+2)H^2\right]}{n_\chi\beta\sigma\,v\,a^2\left(m_\chi^2-(2-s)H^2\right)}\right|
\,=\,\left|\frac{2H}{n_\chi\beta\sigma\,v}\left(1-\frac{s(s-2)H^2}{m_\chi^2+(s-2)H^2}
\right)\right|\ll\,1\;. \label{t1xxx}
\end{equation}

Essentially, Eq.(\ref{t1xxx}) is provided by two distinctive situations: \\
${\bf i})$ $|\left(1-\frac{s(s-2)H^2}{m_\chi^2+(s-2)H^2}\right)|\,\lesssim\,\mathcal{O}(1)$ (i.e.,
$s\simeq\,0$ or $s\simeq\,2$ if $\frac{m_\chi^2}{H^2}$ is not close to $2-s$ or $\frac{m_\chi^2}{(s-2)H^2}+1\,\gg\,s$). For this situation, having $|\frac{2H}{n_\chi\beta\sigma\,v}|\,\ll\,1$ is sufficient to provide Eq.(\ref{t1xxx}). Note that $|\frac{H}{n_\chi\beta\sigma\,v}|\,<\,1$ should already be provided in order for the process to occur. \\
${\bf ii})$ $|\frac{2H}{n_\chi\beta\sigma\,v}|\,\lesssim\,\mathcal{O}(1)$. For this situation having $|\left(1-\frac{s(s-2)H^2}{m_\chi^2+(s-2)H^2}\right)|\,\ll\,1$ is sufficient, i.e., to set $\frac{m_\chi^2}{H^2}\,\simeq\,(s-1)(s-2)$ which in addition means that $s\,>\,2$ or $s\,<\,1$ (additional to $|\frac{m_\chi^2}{H^2}|\,\simeq\,|(s-1)(s-2)|$).

The condition (\ref{t1xxx}) is provided for a wide range of parameters. For instance, the situation i) above can be achieved in the case of radiation dominated universe ($s=2$) well after its beginning (for applicability of $|\frac{2H}{n_\chi\beta\sigma\,v}|\,\ll\,1$) independent of the value of $m_\chi$. The situation i) is provided for the present era of accelerated cosmic expansion as well (where $s\,\simeq\,0$) independent of the value of $m_\chi$ if $\frac{H_0}{n_0\beta\sigma_0\,v}\simeq\,\frac{H_0}{n_0\sigma_0\,v}\simeq\,\frac{10^{-26}m^{-1}}{n_0\sigma_0(v/c)}\,\ll\,1$. For instance, in case $n_0\,\gg\,10^{14}(meter)^{-3}$ with $\sigma_0\sim\,(\frac{c}{v})10^{-40}\,(meter)^2$, and in case $n_0\,\gg\,10^{7}(meter)^{-3}$ with $\sigma_0\sim\,(\frac{c}{v})10^{-33}\,(meter)^2$ (where $c$ is the speed of light) $\frac{H_0}{n_0\beta\sigma_0\,v}\,\ll\,1$ is easily provided. It should be noted that the cross sections for electromagnetic and weak interactions are at the order of $10^{-32}\,(meter)^2$
and $10^{-40}\,(meter)^2$, respectively,  while the current number densities of photons and baryons are $>\,10^8\,(meter)^{-3}$ and $\sim\,10^{-1}\,(meter)^{-3}$, respectively. It is clear that one may cover a considerably larger parameter space than those in the above instances by appropriately altering the values of cross sections and number densities. Therefore, a sufficient number of models may be employed using this method. The situation ii) can be provided in a possible stiff matter dominated epoch ($s=3$) after inflation or in the present accelerated expansion epoch, $s\neq\,0\sim\,0$ if $m_\chi^2\sim\,2H_0^2$.
\newpage
\chapter{Formal Derivation of Equation (\ref{chii})}
Considering the Fourier modes of the field $\tilde{\chi}$, one may write 
\begin{equation}
\tilde{\chi}(\vec{r},\eta)\,=\,\frac{1}{\sqrt{2}}\int\,\frac{d^3\tilde{p}}{(2\pi)^\frac{3}{2}}\left[a_p^-\,v_p^*(\eta)e^{i\vec{\tilde{p}}.\vec{r}}
\,+\,a_p^+\,v_p(\eta)e^{-i\vec{\tilde{p}}.\vec{r}}\right] \;,
\label{e2aaa}
\end{equation}
where $\vec{r}=(\tilde{x}_1,\tilde{x}_2,\tilde{x}_3)$, $a_p$ are the expansion coefficients that are identified by annihilation operators after quantization, $v_p$ are the basic normalized solutions of the equation of motion of $\tilde{\chi}$. $\tilde{\chi}$'s can be considered as being free particles in the time interval between two particle physics processes (for example those in Figure \ref{Feyn1}). Therefore, these particles satisfy
\begin{equation}
v_p^{\prime\prime}\,+\,\omega_p^2(\eta)\,v_p\,=\,0~~~\mbox{where}~~\omega_p\,=\,\sqrt{|\vec{\tilde{p}}|^2+\tilde{m}_\chi^2} ~~,~~~
v_p^\prime\,v_p^*-v_p\,v_p^{*\,\prime}\,=\,2i \;. \label{e2aab1}
\end{equation}
We ensure by the condition (\ref{con1}) that changes in the magnitudes of  $\tilde{m}_\chi^2$ and $\tilde{m}_\phi^2$ are small. Then the variation in  $\omega_p^2$ given in Eq.(\ref{e2aab1}) with time is small because $\vec{\tilde{p}}$ is time-independent. Thus, WKB approximation \cite{Mukhanov:2007zz} can be considered at this point, by which $v_p$ in the time interval between $\eta_i$ and $\eta_{i+1}$ is given by
\begin{equation}
v_p(\eta)\,=\,\frac{1}{\sqrt{W_p(\eta)}}exp{\left(i\int_{\eta_i}^\eta\,W_p(\eta)\,d\eta\right)}, \label{e2c1}
\end{equation}
where $W_p$, by Eq.(\ref{e2aab1}), satisfies
\begin{equation}
W_p^2\,=\,\omega_p^2-\frac{1}{2}\left[\frac{W_p^{\prime\prime}}{W_p}-\frac{3}{2}\left(\frac{W_p^\prime}{W_p}\right)^2\right] \;.\label{e2c2}
\end{equation}
The approximate solution to Eq.(\ref{e2c2}), considering slow variation in $\omega_p$, may be given as
\begin{equation}
^{(0)}W_p=\omega_p~,~~^{(2)}W_p=\omega_p\left(1-\frac{\omega_p^{\prime\prime}}{4\omega_p^3}+\frac{3\omega_p^{\prime\;2}}{8\omega_p^4}\right),~~\mbox{etc.},
\label{e2c3}
\end{equation}
where $^{(0)}W_p$ stands for the zeroth order approximation in $\frac{\Delta\,\omega_p}{\omega_p}$, and $^{(2)}W_p$ stands for the second order approximation in $\frac{\Delta\,\omega_p}{\omega_p}$ and is obtained by substituting  $^{(0)}W_p$ on its left hand side of Eq.(\ref{e2c2}) and then Taylor expanding the square root for slowly varying $\omega_p$. To achive higher order approximations, the similar method is followed. 

We consider the adiabaticity condition because it makes the formulation simpler and it is useful in figuring out the form of the mode function approximately. However, it is not sufficient for the Minkowskian mode function. The condition Eq.(\ref{con1}) should be also imposed so that $^{(0)}W_p\,=\,\omega_p\,\simeq\,constant$ is obtained in the zeroth order approximation. Thus, the approximate Minkowskian space is achieved in each interval between $\eta_i$ and $\eta_{i+1}$. In this situation, the variation in $\omega_p$ is not only slow but, in fact, it extremely slow in each time interval by Eq.(\ref{con1}). In the following derivations, it will be shown that Eq.(\ref{con1}) leads to  $^{(2)}W_p\,\simeq\,^{(0)}W_p\,=\,\omega_p\,\simeq\,constant$.  $\omega_p\,=\,\sqrt{|\vec{\tilde{p}}|^2+\tilde{m}_\chi^2}$ implies that
\begin{equation}
\frac{\omega_p^{\prime}}{\omega_p^2}\,=\,\frac{a}{\omega_p^2}\frac{d\omega_p}{dt}\,=\,\left(\frac{a}{2\omega_p^3}\right)\frac{d\tilde{m}_\chi^2}{dt}
\,=\,\left(\frac{a}{2\omega_p}\right)\left(\frac{\tilde{m}_\chi}{\omega_p}\right)^2(\Delta\,t)^{-1}
\left[\frac{\left(\frac{d\tilde{m}_\chi^2}{dt}\right)(\Delta\,t)}{\tilde{m}_\chi^2}\right]\,\simeq\,0,
\label{rx1}
\end{equation}
where $\Delta\,t\,\sim\,\left(\frac{1}{n_\chi\beta\sigma\,v}\right)$. Via the condition (\ref{con1}), we know that $\left[\frac{\left(\frac{d\tilde{m}_\chi^2}{dt}\right)(\Delta\,t)}{\tilde{m}_\chi^2}\right]\ll 1$ in Eq.(\ref{rx1}). Then, to get $\frac{\omega_p^{\prime}}{\omega_p^2} \simeq 0$, the other terms in Eq.(\ref{rx1}) must not be much larger than 1. We know that $\left(\frac{\tilde{m}_\chi}{\omega_p}\right)^2\,<\,1$ and $\left(\frac{a}{\omega_p}\right)(\Delta\,t)^{-1}\sim\left(\frac{a\,n_\chi\beta\sigma\,v}{\omega_p}\right)\ll 1$ for reasonable values of the parameters. For instance, for the values of parameters discussed after Eq.(\ref{t1xxx}), i.e., for $\hbar\,n_\chi\beta\sigma\,v\,\gg\,\left(6.5\,\times\,10^{-16}\,eV.sec\,\times\,10^{-26}\,sec^{-1}\right)\,\simeq\, 10^{-41}\,eV$, we have  $\left(\frac{a}{\omega_p}\right)(\Delta\,t)^{-1}\sim\left(\frac{a\,n_\chi\beta\sigma\,v}{\omega_p}\right)\,\ll\,1$ if $\hbar\,\omega_p \gg 10^{-41}\,eV$. We will next obtain that for reasonable values of the parameters $\frac{\omega_p^{\prime\prime}}{4\omega_p^3}\,\simeq\,0$. For sake of simplicity, using the case $H=\xi\,a^{-s}$ introduced in the previous appendix and $\frac{d^2\omega_p^2}{dt^2}\,=\,\frac{d^2\tilde{m}_\chi^2}{dt^2}$, and
\begin{equation}
\frac{d^2\omega_p^2}{dt^2}\,=\,2\left(\frac{d\omega_p}{dt}\right)^2+2\omega\frac{d^2\omega_p}{dt^2},~~\mbox{and}~ \frac{d^2\tilde{m}_\chi^2}{dt^2}\,=\,2\left(\frac{d\tilde{m}_\chi}{dt}\right)^2+2\tilde{m}_\chi\frac{d^2\tilde{m}_\chi}{dt^2} \label{q},
\end{equation}
we get that
\begin{eqnarray}
&&\frac{d\tilde{m}_\chi}{dt}\,=\,\frac{1}{2\tilde{m}_\chi}\frac{d\tilde{m}_\chi^2}{dt}\,=\,H\tilde{m}_\chi\,+\,\frac{a^2s(2-s)H^3}{\tilde{m}_\chi},
\nonumber \\
&&\frac{d~^2\tilde{m}_\chi}{dt^2}\,=\,
(1-s)\,H^2\tilde{m}_\chi\,+\,\frac{a^2s(2-s)(2-3s)H^4}{\tilde{m}_\chi}\,-\,\frac{a^4s^2(2-s)^2H^6}{\tilde{m}_\chi^3}\;. \label{rx2}
\end{eqnarray}
Thus we observe
\begin{eqnarray}
&&\frac{\omega_p^{\prime\prime}}{4\omega_p^3}\,=\,\frac{a}{4\omega_p^3}\frac{d\left(a\frac{d\omega_p}{dt}\right)}{dt}
\,=\,
\frac{a^2H\tilde{m}_\chi}{4\omega_p^4}\frac{d\tilde{m}_\chi}{dt}
\,+\,\frac{a^2}{4\omega_p^4}\left(1-\frac{\tilde{m}_\chi^2}{\omega_p^2}\right)\left(\frac{d\tilde{m}_\chi}{dt}\right)^2
\nonumber \\
&&+\,\frac{a^2\tilde{m}_\chi}{4\omega_p^4}\left[(1-s)\,H^2\tilde{m}_\chi\,+\,\frac{a^2s(2-s)(2-3s)H^4}{\tilde{m}_\chi}\,-\,\frac{a^4s^2(2-s)^2H^6}{\tilde{m}_\chi^3}\right]
\nonumber \\
&&=\,
\left(\frac{a}{\omega_p}(\Delta\,t)^{-1}\right)^2\left(\frac{\tilde{m}_\chi}{\omega_p}\right)^2
\{\frac{1}{8}H\Delta\,t\,
\left[\frac{\left(\frac{d\tilde{m}_\chi^2}{dt}\right)(\Delta\,t)}{\tilde{m}_\chi^2}\right]
\,+\,\frac{1}{16}
\left(1-\frac{\tilde{m}_\chi^2}{\omega_p^2}\right)\left[\frac{\left(\frac{d\tilde{m}_\chi^2}{dt}\right)(\Delta\,t)}{\tilde{m}_\chi^2}\right]^2
\nonumber \\
&&+\,\frac{1}{4}(H\Delta\,t)^2\left[
(1-s)\,\,+\,a^2s(2-s)(2-3s)\frac{H^2}{\tilde{m}_\chi^2}\,-\,a^4s^2(2-s)^2\left(\frac{H}{\tilde{m}_\chi}\right)^4\right]\}\,\simeq\,0\;
\label{rx3}
\end{eqnarray}
if Eq.(\ref{con1}) is provided. Both Eq.(\ref{rx1}) and Eq.(\ref{rx2}) implies that $\frac{\omega_p^{\prime\prime}}{4\omega_p^3}\simeq 0$ in  Eq.(\ref{rx3}), so its contribution to Eq.(\ref{e2c3}) can be ignored for reasonable values of the parameters if either $|H\Delta\,t|$ or $|\frac{H}{\tilde{m}_\chi}|$ are not much greater than 1.

By the analysis given above we then observe that in case Eq.(\ref{con1}) is provided, the approximation $\omega_p=\mbox{constant}$ is valid in each time interval $\eta_i\,<\,\eta\,<\,\eta_{i+1}$, so one may consider $v_p\,\simeq\,v_p^{(i)}$ in each interval  where
\begin{equation}
v_p^{(i)}(\eta)\,=\,\frac{1}{\sqrt{\omega_p^{(i)}}}exp{\left(i\omega_p^{(i)}(\eta-\eta_i)\right)}~~\mbox{where}~~\;
\omega_p^{(i)}=\omega_p(\eta_i)~,~~\eta_i\,<\,\eta\,<\,\eta_{i+1} \;. \label{e2c4}
\end{equation}
Thus Eq.(\ref{e2aaa}) may be expressed as
\begin{eqnarray}
\tilde{\chi}^{(i)}(\vec{r},\eta)\,\simeq\,
\int\,
\frac{d^3\tilde{p}}{(2\pi)^\frac{3}{2}\sqrt{2\omega_p^{(i)}}}\left[a_p^{(i)\,-}\,
e^{i\left(\vec{\tilde{p}}.\vec{r}-\omega_p^{(i)}(\eta-\eta_i)\right)}
\,+\,a_p^{(i)\,+}
\,e^{i\left(-\vec{\tilde{p}}.\vec{r}+\omega_p^{(i)}(\eta-\eta_i)\right)}\right]\label{e2aaax} \\
\eta_i\,<\,\eta\,<\,\eta_{i+1}\;, \nonumber
\end{eqnarray}
where $^{(i)}$ refers to the $i$th time interval between the $i$th and $(i+1)$th processes.

A point worth to mention is as follows. For $m_i^2a^2\,<\,\frac{a^{\prime\prime}}{a}\,= \,(2-s)H^2$ (in the case of Eq.(\ref{t9a3})),  $\tilde{m}_i^2$ becomes smaller than $0$ in Eq.(\ref{effm}), which implies that  $\tilde{m}_i$ is tachyonic in this case. However, by time $\frac{a^{\prime\prime}}{a}$ becomes small enough so that $\tilde{m}_i^2$ becomes greater than $0$ for all physically interesting situations as discussed in appendix A excluding the case where strictly $s=0$. Thus, this situation is not an actual problem for the physically relevant cases since the mass either becomes real after a certain amount of time for  $s\,>\,0$ or interaction with other particles can not be taken place  (that makes the tachyonic state safe) for $s\,\leq\,0$ owing to rapid expansion rate. However, the case of tachyonic state can not be handled by this formulation since the would-be ground state (e.g. $\chi=0$, $\phi=0$) will not be the ground state anymore, so the perturbation expansion around the ground state can not be applied. This implies that for the case of $m_i^2a^2\,<\,\frac{a^{\prime\prime}}{a}\,= \,(2-s)H^2$ (in the case of Eq.(\ref{t9a3})), this formulation is not useful. Thus, the case of $s\,\geq\,2$  (e.g. of radiation and stiff matter) is secure in this regard; on the other hand, in the the case of $s\,<\,2$ (e.g. cosmological constant, matter, radiation), $H^2$ should be small enough in comparison to $m_\chi^2$ so that the issue of tachyons does not occur. After combining this limitation with others stated after Eq.(\ref{t1xxx}), one here observes that there is still a remarkable relevant available parameter space remaining. The results given after Eq.(\ref{t1xxx}) remain intact in case of the radiation and stiff matter dominated epochs, and results given in case of the present accelerated expansion epoch still hold if $m_\chi$ is not smaller than $\sim\,H_0\hbar\,\sim\,10^{-33}$ eV.

\newpage
\chapter{The Evolution of the Number Densities in the Case of $s>2$ in Equation (\ref{t9a3})}

Recall that $s>2$ corresponds to the case of ii) (i.e., $\frac{\dot{a}^2}{a^2}+\frac{\ddot{a}}{a}\,\leq\,m_\chi^2$ and $\frac{\dot{a}^2}{a^2}+\frac{\ddot{a}}{a}\,<\,0$) that is introduced just after Eq.(\ref{s1}). We here consider the extreme case $|\frac{\dot{a}^2}{a^2}+\frac{\ddot{a}}{a}|\gg m_\phi^2$ of $s>2$. It seems that BEC of $\phi$ does not play the role of DM in the case of $|\frac{\dot{a}^2}{a^2}+\frac{\ddot{a}}{a}|$=$|\dot{H}+2H^2|\,\gg\,m_\phi^2$ because in this case, we have $|\dot{H}+2H^2_0|\,\sim\,H_0^2\,\sim\,(10^{-31}eV)^2\gg\,m_\phi^2$ which is not relevant to the predicted present value of the mass of DM by observations \cite{ULDM-mass}. However, in this case, BEC of $\phi$ may play the role of DE (if the contribution of the potential term dominates) and of scalars in early universe. 

In this case
\begin{equation}
\tilde{m}^2\,\simeq\,-a^2\left(\dot{H}+2H^2\right) \,=\,\xi^2(s-2)\,a^{2(1-s)}\,>\,0,~~\mbox{i.e.,}~s\,>\,2\label{t10},
\end{equation}
where Eq.(\ref{t9a3}) is considered. Thus, 
\begin{equation}
\tilde{m}^2\,\propto\,a^{2(1-s)}\,;~~\tilde{\mu}\,\propto\,a\,
~;~~~\sigma\,=\,a^2\tilde{\sigma}\,
\simeq\,a^2\left(\frac{\tilde{\mu}}{\tilde{m}_\phi}\right)^4\frac{|\vec{\tilde{k}}|}{64\pi\,\vec{\tilde{p}}^2\tilde{m}_\phi}\,\propto\,a^{5s+1} \;. \label{t11}
\end{equation}
(It is seen from Eq.(\ref{t11}) that $\frac{2H}{n\sigma\,v}=\frac{2H_0}{n_0\sigma_0\,v_0}\,a^{-6s+2}$, so Eq.(\ref{t1xxx}) (i.e., Eq.(\ref{con1})) is always satisfied provided that it is satisfied in the early universe because $s\,>\,2\,>\,\frac{1}{3}$.
Then, from Eq.(\ref{CCC}), we get
\begin{equation}
\frac{\dot{C}_\chi}{a^3}\,=\,-\beta\,C_{\chi}^2\,\sigma_0\,v_0\frac{a^{(5s+1)}}{a^6},
\label{t12}
\end{equation}
which, in turn, results in
\begin{eqnarray}
\frac{dC}{C^2}&=&-\beta\,\sigma_0\,v_0\frac{a^{(5s+1)}}{a^3}dt\,=\,-\beta\,\sigma_0\,v_0\,a^{5s-2}\frac{da}{\xi\,a^{(1-s)}}
\nonumber \\
&&\,=\,-\frac{1}{\xi}\beta\,\sigma_0\,v_0\,a^{6s-3}\,da \;.
\label{t13}
\end{eqnarray}
After integrating out Eq.(\ref{t13}), we have
\begin{equation}
C_\chi\,=\,
\frac{C_1}{\frac{C_1\beta\sigma_0v_0}{(6s-2)\xi}\left(\,a^{6s-2}-a_1^{6s-2}\right)+1} \;.
\label{t15}
\end{equation}
 Thus, we get
\begin{equation}
C_\phi\,=\,C_1\,-\,
\frac{C_1}{\frac{C_1\beta\sigma_0v_0}{(6s-2)\xi}\left(\,a^{6s-2}-a_1^{6s-2}\right)+1} \;.
\label{t16}
\end{equation}
If we consider a similar analysis given for the case of  i) in Section 5.2.2. at initial times, we get  
\begin{equation}
C_\phi\,\simeq\,
\frac{C_1\beta\sigma_0v_0}{(|6s-2|)\xi}\,a^{|6s-2|}\left[\,1-\left(\frac{a_1}{a}\right)^{|6s-2|}\right],
\label{t16a}
\end{equation}
where we impose the condition $s\,>\,2\,>\,\frac{1}{3}$ for excluding the tachyonic case. At late times Eq.(\ref{t16}) may be approximated by 
\begin{equation}
C_\phi\,\sim\,C_1\,-\,
\frac{(|6s-2|)\xi}{\beta\sigma_0v_0}\,a^{-|6s-2|}~~~\mbox{for}~~6s-2\,>\,0~\mbox{if}~\frac{C_1\beta\sigma_0v_0}{(|6s-2|)\xi}\,a^{|6s-2|}\,\gg\,1 \;.
\label{t16d}
\end{equation}
The last equation indicates that $C_\phi$ may get the maximum value $C_1$ if $\chi\chi\,\rightarrow\,\phi\phi$ proceed till very late times. However, because we have ignored the impact of statistics and the impact of the processes $\phi\phi\,\rightarrow\,\chi\chi$ (which cannot be ignored while they can be ignored at initial times), the findings for late times are not entirely valid. Eq.(\ref{t16d}) is safe provided that the number density of $\phi$s has not achieved a large value at late times yet.
\newpage
\chapter{The Scalar Field Equation in 1+1 Dimensions}
In this appendix, we demonstrate that Eq.(\ref{eq:15}), i.e., the approximation of the scalar field equation corresponding to $\frac{f^\prime}{\bar{r}}\simeq\,0$, is the scalar field equation in 1+1 dimensions. 


We take the following action into consideration for $\chi$ and $\phi$ particles
\begin{eqnarray}
S
&=&\int\;dt\,dr\,\{-g^{\mu\nu}\left[D_\mu\phi\,\left(D_\nu\phi\right)^*\,+\,
D_\mu\chi\,\left(D_\nu\chi\right)^*\right]\,-\,m_\phi^2\left|\phi\right|^2\,-\,
m_\chi^2\left|\chi\right|^2\,-\,\lambda\,\phi^*\phi\chi^*\chi\}, \nonumber \\
&&\label{eq:9aap}
\end{eqnarray}
where $D_\mu\,=\,\partial_\mu\,+\,iq\,A_\mu$ with $q$ being the electric charge of the scalar field and $A_\mu\,=\,\left(\frac{Q}{r}, 0, 0, 0\right)$ denoting the electric field of the black hole. We let both $\chi$ and $\phi$ have the same charge q. After the change of variables $dr_*\,=\,f^{-1}\,dr$, Eq.(\ref{eq:9aap}) becomes
\begin{eqnarray}
S\,=\,
\int&dt\,dr_*&\{\left|\frac{\partial\phi}{\partial\,t}+i\frac{qQ}{r}\right|^2-\left|\frac{\partial\phi}{\partial\,r_*}\right|^2 -\tilde{m}_\phi^2\phi^2 \nonumber \\
&&+\,\left|\frac{\partial\chi}{\partial\,t}+i\frac{qQ}{r}\right|^2-\left|\frac{\partial\chi}{\partial\,r_*}\right|^2
-\tilde{m}_\chi^2\chi^2
-\tilde{\lambda}\phi^*\phi\chi^*\chi\},
\label{eq:9bap}
\end{eqnarray}
where
\begin{equation}
\tilde{m}_i^2\,=\,f\,m_i^2~,\tilde{\lambda}\,=\,f\,\lambda~~~~i=\chi,\phi.  \label{eq:10}
\end{equation}
Transforming Eq.(\ref{eq:9aap}) to Eq.(\ref{eq:9bap}) corresponds to changing the metric $ds^2$ into $d\tilde{s}^2$ where $ds^2$ and $d\tilde{s}^2$ are related by
\begin{equation}
ds^2\,=\,f\left(-\,dt^2\,+\,f^{-2}dr^2\right)\,=\,f\,d\tilde{s}^2, \label{eq:11ap}
\end{equation}
where
\begin{equation}
d\tilde{s}^2\,=\, -\,dt^2\,+\,dr_*^2 \label{eq:12ap}.
\end{equation}
In other words, we have moved to an effective Minkowski space denoted by Eq.(\ref{eq:12ap}) at the expense of making the masses and the coupling constant r-dependent (hence, $r_*$-dependent).

In Eq.(\ref{eq:9aap}) and Eq.(\ref{eq:9bap}), due to the low coupling constant of electromagnetic interactions, we have ignored their impact on scalar particle interactions, and the scalar particle density is assumed to be low, and the density of the scalar particles are taken to be small. In a similar way, we take the coupling constant $\lambda$ to be small. Then, the approximate field equation corresponding to Eq.(\ref{eq:9bap}) for $\phi$ is
\begin{equation}
\tilde{D}_\mu\,\tilde{D}^\mu\,\phi\,-\,\tilde{m}_\phi^2\,\phi\,=\,\frac{\partial^2\phi}{\partial\,t^2}\,-\,\frac{\partial^2\phi}{\partial\,r_*^2}
\,+\,\frac{2iqQ}{r}\frac{\partial\phi}{\partial\,t}\,+\,\left(\tilde{m}_\phi^2\,-\,\frac{q^2Q^2}{r^2}\right)\phi\,=\,0,
\label{eq:13ap}
\end{equation}
where $\tilde{D}_\mu=\tilde{\partial}_\mu+iqA_\mu$$=$$\,\left(\frac{\partial}{\partial\,t}+i\frac{qQ}{r}, \frac{\partial}{\partial\,r_*}\right)$.
The corresponding equation for $\chi$ may be obtained by replacing $\phi$ in Eq.(\ref{eq:13ap}) by $\chi$.

The total mechanical energy for the metric (\ref{eq:3}) (i.e., for the local effective Minkowski space) $\tilde{E}$ is equal to the total energy of the particle. In other words,
$\tilde{E}_i^2\,=\,\tilde{p}_i^2+\tilde{m}_i^2\,=\,m_i^2\,C_i^2$. Hence, the oscillatory solutions may be taken as
\begin{equation}
\phi\,=\,R_\phi(r_*)\,e^{-i\omega_\phi\,t},
\label{eq:14ap}
\end{equation}
where $\omega_\phi$ is identified by $\tilde{E}=\hbar\omega_\phi=\hbar\,m_\phi\,C_\phi$. Thus, Eq.(\ref{eq:13ap}) reduces to
\begin{equation}
\frac{d^2R}{dr_*^2}\,+\,\left[\left(\omega-\frac{qQ}{r}\right)^2\,-\,\tilde{m}_\phi^2\right]\,R\,=\,0,
\label{eq:15ap}
\end{equation}
which is the same as Eq.(\ref{eq:15}).
\newpage
\chapter{Derivation of Equations (\ref{H5})-(\ref{H9})}
Consider Eq.(\ref{H2}), namely,
\begin{align}
\frac{g^{\prime\prime}}{g}+2is\frac{g^{\prime}}{g}-\frac{2(qQ\omega-m^2M)}{r}+\frac{(q^2-m^2)Q^2}{r^2}=0,\label{H2ap}
\end{align}
where ``$\ \prime\ $'' denotes the derivative with respect to $r_*$. We let
\begin{align}
\frac{g^{\prime}}{g}=\frac{\alpha_1}{r}+\frac{\beta_1}{r^2}+\frac{\gamma_1}{r^3}, \label{H3apa}
\end{align}
\begin{align}
\frac{g^{\prime\prime}}{g}=\frac{\alpha_2}{r}+\frac{\beta_2}{r^2}+\frac{\gamma_2}{r^3}. \label{H3apb}
\end{align}
Next, we use Eq.(\ref{H3apa}), Eq.(\ref{H3apb}) and $^\prime=\frac{d}{dr_*}=\frac{dr}{dr_*}\frac{d}{dr}$ and the following identity to relate $\alpha_1$, $\beta_1$, $\gamma_1$ and $\alpha_2$, $\beta_2$, $\gamma_2$
\begin{align}
\frac{d}{dr_*}\left(\frac{g^{\prime}}{g}\right)
= \left(\frac{g^{\prime}}{g}\right)^\prime= \frac{g^{\prime\prime}}{g} - \left(\frac{g^{\prime}}{g}\right)^2. \label{H3apx}
\end{align}
Hence, we obtain
\begin{align}
\frac{g^{\prime\prime}}{g}&=\frac{1}{r^2}\alpha_1(\alpha_1-1)+\frac{1}{r^3}2(M\alpha_1-\beta_1+\alpha_1\beta_1)+\frac{1}{r^4}(4\beta_1 M- Q^2\alpha_1 - 3\gamma_1 +2\alpha_1 \gamma_1 +\beta_1^2) \nonumber
\\&
+\frac{1}{r^5}2(3M\gamma_1-Q^2\beta_1+\beta_1\gamma_1)+\gamma_1(\gamma_1-3Q^2)\frac{1}{r^6}=\frac{\alpha_2}{r}
+\frac{\beta_2}{r^2}+\frac{\gamma_2}{r^3}.  \label{H4ap}
\end{align}
Eq.(\ref{H4ap}) implies
\begin{eqnarray}
&&\alpha_2=0\;~,~~\beta_2=\alpha_1(\alpha_1-1)\;~,~~\gamma_2=2(M\alpha_1-\beta_1+\alpha_1\beta_1)\;,\nonumber \\
&&4M\beta_1-Q^2\alpha_1-3\gamma_1+\beta_1^2+2\gamma_1\alpha_1=0 \;~,~~
3M\gamma_1-Q^2\beta_1+\beta_1\gamma_1=0\;~,~\gamma_1-3Q^2=0. \nonumber \\
\label{H4apx}
\end{eqnarray}
Eq.(\ref{H4apx}) reults in
\begin{equation}
\alpha_1=\frac{9}{5}-\frac{9}{20}\bigg(\frac{M}{Q}\bigg)^2, \ \ \ \ \ \ \ \beta_1=-\frac{9}{2}M, \ \ \ \ \ \ \ \gamma_1=3Q^2, \label{H5ap}
\end{equation}
\begin{align}
\alpha_2=0 \,, \ \ \ \ \ \ \beta_2=\bigg[\frac{9}{5}-\frac{9}{20}\bigg(\frac{M}{Q}\bigg)^2\bigg]\bigg[\frac{4}{5}-\frac{9}{20}\bigg(\frac{M}{Q}\bigg)^2\bigg] \,, \ \ \ \ \ \ \gamma_2=2\bigg[\frac{-9}{5}+\frac{63}{40}\bigg(\frac{M}{Q}\bigg)^2 \bigg]M.  \label{H6ap}
\end{align}
Inserting solutions (\ref{H5ap}) and (\ref{H6ap}) into Eq.(\ref{H2}) and using $s^2=\omega^2-m^2$, we get three equations for three unknown quantities $w$, $q$, $m$ for a given $M$ and $Q$:
\begin{align}
(m^2-\omega^2)-\frac{M^2}{9Q^4}\bigg(\frac{9}{5}-\frac{63}{40}\frac{M^2}{Q^2}  \bigg)^2=0,
\end{align}
\begin{align}
-\frac{3M^2}{Q^2}\bigg(\frac{9}{5}-\frac{63}{40}\frac{M^2}{Q^2}\bigg)+\bigg(\frac{9}{5}-\frac{9}{20}\frac{M^2}{Q^2} \bigg) \bigg( \frac{4}{5}-\frac{9}{20}\frac{M^2}{Q^2} \bigg) +(q^2-m^2)Q^2=0,
\end{align}
\begin{align}
\frac{2M}{3Q^2}\bigg(\frac{9}{5}-\frac{63}{40}\frac{M^2}{Q^2}\bigg)\bigg(\frac{9}{5}-\frac{9}{20}\frac{M^2}{Q^2}\bigg)-2(qQ\omega-m^2M)=0.
\end{align}
These three equations solve $m$, $\omega$, and $q$ as given in Eqs.(\ref{eq:m2})-(\ref{eq:q12}).
\newpage
\chapter{Derivation of Equation (\ref{scaled_eq})}
After inserting the Newton's constant $G$, the speed of light $c$, the Planck's constant $\hbar$ into explicitly, the action for a charged free scalar field $\phi$ becomes
\begin{align}
S=\int \pmb{\hbar^2}\bigg[-g^{\mu\nu} (D_\mu \phi) ( D_\mu \phi)^*-\frac{\pmb{m^2c^2}}{\pmb{\hbar^2}}\phi\phi^*\bigg]d^2x.
\end{align}
Because we consider other potential interactions to be small in comparison to the other components in the Lagrangian, it is important to note that we are ignoring possible interaction terms other than electromagnetic interactions. The corresponding equation is
\begin{equation}
\,\frac{\partial^2\phi}{\pmb{c^2}\partial\,t^2}\,-\,\frac{\partial^2\phi}{\partial\,r_*^2}\,+ \frac{2i\pmb{k}}{\pmb{\hbar}\pmb{c^2}}\frac{\pmb{qQ}}{r}\frac{\partial\phi}{\partial\,t}\,+\,\left(\frac{\pmb{c^2}}{\pmb{\hbar^2}}\tilde{\pmb{m}}^2_\phi
\,-\,\frac{\pmb{k^2}}{\pmb{\hbar^2} \pmb{c^2}}\frac{\pmb{q}^2 \pmb{Q}^2}{r^2}\right)\phi\,=\,0,
\label{eq:13apx}
\end{equation}
where $\pmb{k=\frac{1}{4\pi\epsilon_0}}$. After inserting Eq.(\ref{c1}) into Eq.(\ref{eq:13apx}) and using $\frac{f^\prime}{r}\,\ll\,m_\chi^2$, we get
\begin{equation}
\frac{d^2\psi_\omega}{dr_*^2}\,+\,\left[\left(\frac{\pmb{\omega}}{\pmb{c}}-\frac{\pmb{kqQ}}{\pmb{\hbar c} r}\right)^2\,-\,\frac{\pmb{c^2}}{\pmb{\hbar^2}}\pmb{m}^2_\phi\left(1-\frac{2\pmb{GM}}{\pmb{c^2}r}+\frac{\pmb{kGQ^2}}{r^2\pmb{c^4}}\right)\right]\,\psi_\omega\,=\,0.
\label{eq:15'ap}
\end{equation}

To express $r_*$ in terms of the Schwarzschild radius of the sun, we multiply both sides of Eq.(\ref{eq:15'ap}) by $\left(\frac{GM_\circ}{c^2}\right)^2$.
Then, Eq.(\ref{eq:15'ap}) becomes
\begin{align}
\frac{d^2\psi_\omega}{d\overline{r}_*^2}+\bigg[\bigg(\frac{\omega GM_\circ}{c^3}- \frac{kqQ}{\hbar c \overline{r}} \bigg)^2-m_\phi^2\bigg(\frac{GM_\circ}{\hbar c} \bigg)^2\left(1-\frac{2M}{M_\circ \overline{r} } + \frac{Q^2 k}{M_\circ^2 G \overline{r}^2}\right)\bigg]\psi_\omega=0,
\end{align}
where
\begin{align}
\overline{r}_*=\frac{c^2}{GM_\circ}r_*,~~ \overline{r}=\frac{c^2}{GM_\circ}r.
\end{align}
We may define
\begin{align}
&\overline{\omega}=\omega\frac{GM_\circ}{c^3}=\bigg(\frac{\omega}{s^{-1}}\bigg)5\times 10^{-6},
\\&
\overline{m}=m\frac{GM_\circ}{\hbar c} =\bigg( \frac{m}{kg}\bigg)\,4.5\times 10^{45},
\\&
\overline{M}=\frac{M}{M_\circ}=\bigg(\frac{M}{kg}\bigg)2\times 10^{-30},
\\&
\overline{Q}=\frac{Q\sqrt{k}}{M_\circ \sqrt{G}}=\bigg(\frac{Q}{C}\bigg)5,7\times 10^{-21},
\end{align}
\begin{align}
\frac{kqQ}{\hbar c}=\overline{q}\overline{Q}=\frac{q\overline{Q}M_\circ \sqrt{Gk}}{\hbar c } \ \ \ \ \Rightarrow \ \ \ \ \overline{q}=q\frac{M_\circ \sqrt{Gk}}{\hbar c}=\bigg(\frac{q}{C}\bigg) 4,8 \times 10^{55},
\end{align}
where we have used the numerical values of $M_\circ$, $G$, $c$, $\hbar$ in SI unit system.

Then, Eq.(\ref{eq:15'ap}) becomes
\begin{align}
\frac{d^2\psi_\omega}{d\overline{r}_*^2}+\bigg[\big(\overline{\omega}-\frac{\overline{q}\overline{Q}}{\overline{r}}\big)^2-\overline{m}_\phi^2\left(1-\frac{2\overline{M}}{\overline{r}} + \frac{\overline{Q}^2}{\overline{r}^2}\right) \bigg]\psi_\omega=0.
\label{scaled_eqap}
\end{align}



\newpage
\chapter{Incompatibility of the Conditions $G^0_{\;1}\,=\,0$, (\ref{eq:11}), (\ref{eq:16})}

We here show that the simultaneous satisfaction of $G^0_{\;1}\,=\,0$, Eq.(\ref{eq:11}), Eq.(\ref{eq:16}) are not consistent. To this end, we first take the derivative of Eq.(\ref{eq:11}) with respect to time and divide by itself to obtain
\begin{equation}
\dot R ^\prime + \dot \lambda R^\prime-\dot{\psi} R^\prime=0 \;\Rightarrow~ \dot{R}^\prime - \dot{\psi} R^\prime = - \dot{\lambda} R^\prime .
\label{eq:24}
\end{equation}
We then substitute Eq.(\ref{eq:24}) in $G^0_{\;1}\,=\,0$ (i.e., $\dot{R} ^\prime-\dot R \lambda^\prime-\dot \psi R^\prime=0$) which results in
\begin{equation}
 \dot \lambda R^\prime+ \lambda^\prime \dot{R}=0 \Rightarrow \frac{\lambda^\prime}{\dot \lambda}= - \frac{R^\prime}{\dot R }.
\label{eq:25}
\end{equation}
Note that
\begin{eqnarray}
 \frac{\lambda^{'}}{\dot \lambda}=\frac{\big(\frac{\partial e^{2\lambda}}{\partial r}\big) }{\big(\frac{\partial e^{2\lambda}}{\partial t}\big)}~, ~~\frac{R^{'}}{\dot R }= \frac{\big(\frac{\partial R^2}{\partial r}\big)}{\big(\frac{\partial R^2}{\partial t}\big)}.
\end{eqnarray}
One may check that Eq.(\ref{eq:25}) has the following solution,
\begin{eqnarray}
	e^{2\lambda}= \alpha_1 + \alpha_2 \frac{f\left(t\right)}{g\left(r\right)}~, ~~ R^2=\alpha_3 + \alpha_4\left[f\left(t\right)g\left(r\right)\right]^\beta
		\label{eq:26},
\end{eqnarray}
where $\alpha_1,\alpha_2,\alpha_3,\alpha_4, \beta$ are arbitrary constants; and $f\left(t\right),g\left(r\right)$ are arbitrary functions of $t$ and $r$, respectively.

After substituting Eq.(\ref{eq:26}) in Eq.(\ref{eq:24}), we get
\begin{equation}
\frac{\partial \psi}{\partial t}\,=\, \frac{\partial \lambda}{\partial t} +\,\beta \frac{1}{f} \frac{\partial{f}}{\partial t}-\frac{1}{2R^2}\frac{\partial R^2}{ \partial t},
\label{eq:28b}
\end{equation}
which, after integration, results in
\begin{equation}
e^{2\psi}\,=\,Q(r)\,\frac{f^{2\beta}(\alpha_1+\alpha_2fg^{-1})}{(\alpha_3+\alpha_4 f^\beta g^\beta)}.
\label{eq:28b}
\end{equation}
where $Q(r)$ is an arbitrary function of $r$ (whose form to be determined by initial conditions).

The $\lambda$, $\psi$, $R$ given in Eq.(\ref{eq:26}) and Eq.(\ref{eq:28b}) may be substituted in Eq.(\ref{eq:4}) to obtain the corresponding F as
\begin{equation}
F(r,t)=\frac{(\alpha_3+\alpha_4 f^\beta g^\beta)^{1/2}}{(\alpha_1+\alpha_2fg^{-1})}\bigg[(\alpha_1+\alpha_2fg^{-1})-\frac{1}{4Q}(\alpha_4\beta g^\prime g^{\beta-1})^2+\frac{(\alpha_4 \beta f^{\beta-1}\dot{f}g^\beta)^2}{4(\alpha_3+\alpha_4 f^\beta g^\beta)}\bigg].\label{Fgen}
\end{equation}
We find that this F in Eq.(\ref{Fgen}) is not in the form Eq.(\ref{eq:16}). We may see this more explicitly by considering the specific case of $R(r,t)=rf(t)$ in Eq.(\ref{n5a}) that corresponds to $\alpha_3=0$, $\alpha_4=1$, $\beta=2$, $g(r)=r$ in Eq.(\ref{eq:26}). In that case, Eq.(\ref{Fgen}) takes the following form
\begin{equation}
F(r,t)=\bigg(\frac{rf(t)}{\alpha_1+\alpha_2\frac{f(t)}{r}}\bigg)\bigg[(\alpha_1+\alpha_2\frac{f(t)}{r})-\frac{r^2}{Q(r)}+r^2\dot{f}(t)^2\bigg]\label{Fpar},
\end{equation}
which is not in the form of Eq.(\ref{eq:16}) as expexted. This implies that the conditions (\ref{eq:11}) and $G^0_{\;1}\,=\,0$ are not consistent to obtain $F(r,t)$ proposed in Eq.(\ref{eq:16}). Thus, we have to drop one of the conditions $G^0_{\;1}\,=\,0$ or  (\ref{eq:16}) or (\ref{eq:11}) for the consistent framework. In this paper, we choose to drop the condition $G^0_{\;1}\,=\,0$.
\end{document}